\documentclass[12pt, a4paper]{report}

\usepackage{geometry}
\usepackage{comment}
\usepackage{amsfonts}
\usepackage{amssymb}
\usepackage{mathtools}
\usepackage{bm}
\usepackage{bbm}
\usepackage[hang,flushmargin]{footmisc}
\usepackage{natbib}
\usepackage{sidecap}
\usepackage{pdflscape}
\usepackage[nodisplayskipstretch]{setspace}
\usepackage{braket}

\usepackage[T1]{fontenc}
\usepackage[english]{babel}
\usepackage{siunitx}
\usepackage{graphicx}
\usepackage{tipa} 
\usepackage{graphics} 
\usepackage{times} 
\usepackage{amsmath}
\usepackage{latexsym}
\usepackage{amscd}
\usepackage{mathrsfs} 
\usepackage{relsize}
\usepackage{delarray}
\usepackage{pstricks}
\usepackage{theorem}
\usepackage{changepage}
\usepackage{euscript}
\usepackage{textcomp}
\usepackage{esvect}
\usepackage{placeins}
\usepackage{subfigure}
\usepackage{array}
\usepackage{delarray}
\usepackage{stmaryrd}
\usepackage{fancyhdr}
\usepackage{graphpap}
\usepackage{makeidx}
\usepackage{enumerate}
\usepackage{esint}
\usepackage{datetime}
\usepackage{caption}
\usepackage{smartdiagram}
\usesmartdiagramlibrary{additions}
\usepackage{abstract}
\usepackage{color}
\definecolor{igreen}{rgb}{0.0, 0.56, 0.0}
\usepackage{xcolor, colortbl}
\colorlet{gred}{-red!75!green!65!}
\colorlet{mamber}{-red!75!green!15!blue!50!}
\colorlet{grown}{-red!75!blue!20!green}
\colorlet{bled}{-red!85!blue!40!green!45!}
\colorlet{waters}{cyan!25} 
\colorlet{water}{cyan!25!green!20!} 
\definecolor{grin}{HTML}{00F9DE}
\usepackage{rotating}

\usepackage{arydshln}
\usepackage{booktabs}
\usepackage{graphicx}
\usepackage{tikz}
\usepackage{tikz-3dplot}
\usetikzlibrary{
arrows,
arrows.meta,
automata,
backgrounds,
calc,
decorations,
decorations.pathmorphing,
decorations.pathreplacing,
decorations.fractals,
external,
fit,
matrix,
petri,
positioning,
shadows,
shapes,
shapes.multipart,
topaths,
intersections
}
\usepackage{eso-pic}
\def\ba{\begin{array}}
\def\ea{\end{array}}
\def\beann{\begin{eqnarray*}}
\def\eeann{\end{eqnarray*}}
\def\bea{\begin{eqnarray}}
\def\eea{\end{eqnarray}}

\renewcommand{\thesection}{\arabic{section}}
\renewcommand{\thesection}{\thechapter.\number\numexpr\value{section}}

\makeatletter
\newlength\qvec@height
\newlength\qvec@depth
\newlength\qvec@width
\newcommand{\qvec}[2][]{
    \settoheight{\qvec@height}{$#2$}
    \settodepth{\qvec@depth}{$#2$}
    \settowidth{\qvec@width}{$#2$}
  \def\qvec@arg{#1}
  \raisebox{.2ex}{\raisebox{\qvec@height}{\rlap{%
    \kern.05em
    \begin{tikzpicture}[scale=1,shorten >=-3pt,shorten <=-3pt]
    \pgfsetroundcap
    \coordinate (Stx) at (.05em,0) ;
		\coordinate (Arx) at (\qvec@width-.05em,0) ;
    \draw[->](Stx) to[bend left] (Arx);
    \end{tikzpicture}
  }}}
  #2
}
\makeatother
\makeatletter
\newlength\pvec@height
\newlength\pvec@depth
\newlength\pvec@width
\newcommand{\pvec}[2][]{
    \settoheight{\pvec@height}{$#2$}
    \settodepth{\pvec@depth}{$#2$}
    \settowidth{\pvec@width}{$#2$}
  \def\pvec@arg{#1}
  \raisebox{.2ex}{\raisebox{\pvec@height}{\rlap{%
    \kern.05em
    \begin{tikzpicture}[scale=1,shorten >=-3pt,shorten <=-3pt]
    \pgfsetroundcap
    \coordinate (Stx) at (.05em,0) ;
		\coordinate (Arx) at (\pvec@width-.05em,0) ;
    \draw[->](Stx) to[bend right] (Arx);
    \end{tikzpicture}
  }}}
  #2
}
\makeatother
\makeatletter
\newlength\vvec@height%
\newlength\vvec@depth%
\newlength\vvec@width%
\newcommand{\vvec}[2][]{%
  \ifmmode%
    \settoheight{\vvec@height}{$#2$}%
    \settodepth{\vvec@depth}{$#2$}%
    \settowidth{\vvec@width}{$#2$}%
  \else 
    \settoheight{\vvec@height}{#2}%
    \settodepth{\vvec@depth}{#2}%
    \settowidth{\vvec@width}{#2}%
  \fi%
  \def\vvec@arg{#1}%
  \def\vvec@dd{:}%
  \def\vvec@d{.}%
  \raisebox{.2ex}{\raisebox{\vvec@height}{\rlap{%
    \kern.05em%
    \begin{tikzpicture}[scale=1]
    \pgfsetroundcap
    \draw (.05em,0)--(\vvec@width-.05em,0);
    \draw (\vvec@width-.05em,0)--(\vvec@width-.15em, .075em);
    \draw (\vvec@width-.05em,0)--(\vvec@width-.15em,-.075em);
    \ifx\vvec@arg\vvec@d%
      \fill(\vvec@width*.45,.5ex) circle (.5pt);%
    \else\ifx\vvec@arg\vvec@dd%
      \fill(\vvec@width*.30,.5ex) circle (.5pt);%
      \fill(\vvec@width*.65,.5ex) circle (.5pt);%
    \fi\fi%
    \end{tikzpicture}%
  }}}%
  #2%
}
\makeatother
\def\ba{\begin{array}}
\def\ea{\end{array}}
\def\beann{\begin{eqnarray*}}
\def\eeann{\end{eqnarray*}}
\def\bea{\begin{eqnarray}}
\def\eea{\end{eqnarray}}

\usepackage{titlesec}
\usepackage{multirow}
\usepackage{hyperref}
\usepackage{lipsum}
\usepackage{tikz-cd}
\usepackage{float}
\usepackage{titling}
\usepackage{epigraph}
\usepackage[title, titletoc]{appendix}
\titleformat{\chapter}{\normalfont\LARGE}{\thechapter\,$\vert$}{20pt}{\LARGE}{\setcounter{chapter}{0}}
\titlespacing*{\chapter}{0pt}{-20pt}{40pt} 

\titleformat{\section}
   {\normalfont\large\bfseries}{\thesection}{1em}{}

\makeatletter
\newcommand\BackgroundPicturea[3]{
	\setlength{\unitlength}{1pt}
	\put(0,\strip@pt\paperheight){
		\parbox[t]{\paperwidth}{
			\vspace{#2}\hspace{#3}
			\mbox{\includegraphics[scale=0.5]{#1}}
}}}
\newcommand\BackgroundPictureb[3]{
	\setlength{\unitlength}{1pt}
	\put(0,\strip@pt\paperheight){
		\parbox[t]{\paperwidth}{
			\vspace{#2}\hspace{#3}
			\mbox{\includegraphics[scale=0.3]{#1}}
}}}
\makeatother

\usepackage{setspace}
\addto\captionsenglish{
	\renewcommand{\contentsname}%
	{Table of Contents \vspace{-1.0cm}}
	\setcounter{tocdepth}{3}
	\setcounter{secnumdepth}{1}
	}

\hypersetup{pdftitle = A Dynamics-First View of the Spacetime Manifold, pdfauthor = {Daniel Grimmer}, pdfstartview=FitH, pdfkeywords = essay, pdfpagemode=FullScreen, colorlinks, anchorcolor = red, citecolor = blue, urlcolor=blue, filecolor=green, linkcolor=black, plainpages=false}

\fancyheadoffset{9pt}
\fancyfootoffset{9pt}

\newcommand{\Dan}[1]{{\color{red}\bf [Dan: #1]}}

\renewcommand{\d}{\mathrm{d}}
\newcommand{\ii}{\mathrm{i}}
\newcommand{\openone}{\mathbbm{1}}
\newcommand{\surint}{\xrightarrow[\text{int. as}]{\text{survives}}}
\newcommand{\surreint}{\xrightarrow[\text{re-int. as}]{\text{survives}}}
\newcommand{\surext}{\xrightarrow[\text{via ext.}]{\text{becomes}}}
\newcommand{\hra}{\texorpdfstring{$\hookrightarrow$}{-} \ }

\date{}

\title{Deflating Spacetime:\linebreak
A Dynamics-First View\linebreak
of the Spacetime Manifold}
\author{\Large{Daniel Grimmer}
\\ Reuben College
\\
\\
\\
\\
\\ University of Oxford
\\
A 30,000-word Transfer-of-Status Dissertation\\ 
\\ \\
Michaelmas 2022
}
\begin{document}
\AddToShipoutPicture*{\BackgroundPicturea{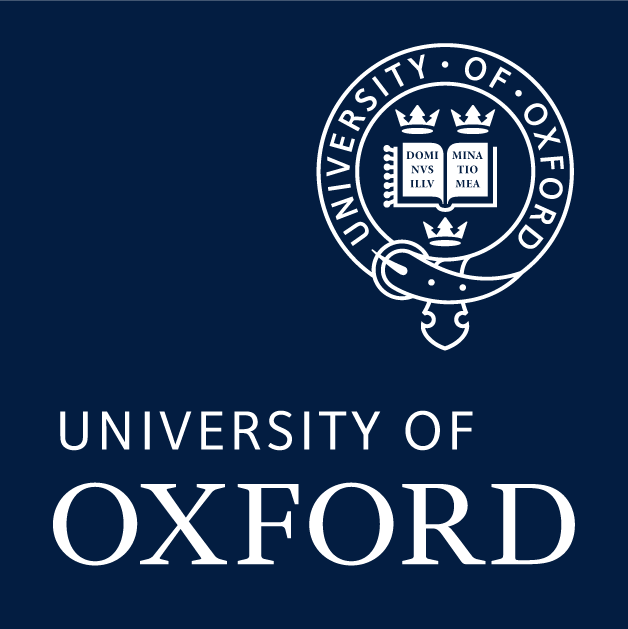}{0.7in}{5.8in}}
\AddToShipoutPicture*{\BackgroundPictureb{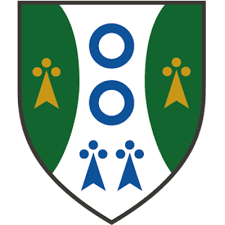}{6.0in}{3.55in}}

\thispagestyle{headings}
{\setstretch{1.5}
    \maketitle}
\FloatBarrier
\pagenumbering{roman}






\thispagestyle{empty}
\begin{abstract}
What, if anything, can help us explain the dynamical behavior of matter? One may be tempted here to appeal to the laws of nature, or to the world's geometric structure, or even to the smooth topological structure of the spacetime manifold itself. Some think, however, that the metaphysics involved in such explanatory strategies is excessively spooky. Indeed, some opt to reverse the arrow of explanation, putting dynamics first. For instance, one can use Lewis's Best Systems Analysis (BSA) to view the laws of nature as merely being codifications of certain patterns in the dynamical behavior of matter. Similarly, one can use Huggett's Regularity Relationism to achieve an analogous view of the world's geometric structure. At present, however, there is no such dynamics-first view of the spacetime manifold itself. This dissertation puts forward such a view: The spacetime manifold is merely a codification of certain patterns in the dynamical behavior of matter.

This dissertation develops powerful mathematical tools for investigating and negotiating between a wide variety of spacetime settings for a wide range of spacetime theories. From here, a competition between different spacetime codifications is invoked analogous to the competition between different law-like codifications in Lewis's BSA. Whereas the BSA judges law-like codifications based on their balance of simplicity and strength, spacetime codifications are to be judged by how well they fit the theory's dynamics and kinematics.
\end{abstract}

{\fontsize{12.0}{12.0}\selectfont \tableofcontents}
\thispagestyle{plain}


\chapter{Introduction}\label{ChapIntro}
\pagenumbering{arabic}
What, if anything, can help us explain the dynamical behavior of matter? For example, one might be interested in answering the following questions: Why do apples fall? Why do like charges repel? Why do light-clocks exhibit time dilation (ticking at different rates when put into relative motion)? Moreover, why does every kind of clock seem to behave this way regardless of its physical composition? Finally, one might ask: Why are the vibrational modes always discrete for a scalar field on a spacetime manifold with spherical topology? Moreover, why do our measuring rods always indicate a positive average Gaussian curvature for such a spacetime manifold? In response to such questions, one may be tempted to appeal to the laws of nature, or to the world's geometric structure, or even to the smooth topological structure of the spacetime manifold itself. Allow me to briefly overview several explanatory strategies based on such appeals.

Beginning with the first three questions, one might try to explain the dynamical behavior of specific apples, charges, and light-clocks by invoking the relevant dynamical laws as revealed by our current best scientific theories (e.g., the laws of gravity, or of electromagnetism). That is, one might try to use laws of nature to explain dynamics. This laws-to-dynamics explanatory strategy typifies what I will call \textit{laws-first views}. For instance, Hempel's covering law model of explanation adopts this strategy.\footnote{See \cite{HempelCover}.} Another important family of laws-first views are the governing views of laws. Roughly, these views claim that things happen as they do ultimately because the laws of nature govern them to that effect.

As I will discuss further in Ch.~\ref{ChapSecondVsFirst}, some find the metaphysical backstories involved in such laws-first explanations excessively spooky. In opposition to laws-first views, a competing strategy is to reverse the arrow of explanation, putting dynamics first. I will call such views \textit{dynamics-first views of laws}. The question of whether the arrow of explanation points from laws to dynamics or vice-versa is at the center of the metaphysics of laws debate.\footnote{For an overview of the metaphysics of laws debate see the SEP article by \cite{sep-laws-of-nature} among many others: \cite{BirdTextbook,BhogalHarjit2020Halo,Lewis1973,Lewis1999,LewisDavid1983Nwfa,DemarestHeather2017PPPL} and \cite{BusinessOfLaws}.} A prominent view on the dynamics-first side of this debate is David Lewis's Best Systems Analysis (BSA).\footnote{See \cite{Lewis1973,Lewis1999}.} Roughly, this view claims that the laws of nature are merely codifications of certain patterns in the dynamical behavior of matter. As this example shows, dynamics-first views can give a metaphysically deflationary account of the laws of nature. The primary cost of such a deflationary move, however, is that we no longer have access to metaphysically-supported laws-based explanations of dynamics. See Sec.~\ref{SecMBE} for further discussion.

Returning to our questions: Why do all kinds of clocks (analog, digital, atomic, biological, etc.) exhibit time dilation? One might see this as a consequence of the fact that all clocks obey dynamical laws which are locally Poincar\'e invariant. But why is this the case? One might try to explain this by invoking the geometric structure of spacetime. Namely, one can point to the fact that our spacetime has a Lorentzian metric and consequently a local Poincar\'e symmetry. Concretely, on this view the geometric structure of spacetime determines the possible symmetries of the dynamical laws and thereby indirectly shapes the dynamical behavior of matter. As this example shows, one might try to use geometry to explain dynamics. This geometry-to-dynamics explanatory strategy typifies what I will call \textit{geometry-first views}.

Just as in the laws context, some find the metaphysical backstories involved in such geometry-first explanations excessively spooky. In opposition to geometry-first views, a competing strategy is to reverse the arrow of explanation, putting dynamics first. I will call such views \textit{dynamics-first views of geometry}. The question of whether the arrow of explanation points from geometry to dynamics or vice-versa is at the center of the geometrical versus dynamical spacetime debate.\footnote{. On the geometrical side, see the work of Tim \cite{MaudlinTim2012Pop:}. For further discussion on either side see: \cite{EarmanJohn1989Weas,TwiceOver,BelotGordon2000GaM,Menon2019,BrownPooley1999,Nonentity,HuggettNick2006TRAo,StevensSyman2014Tdat,DoratoMauro2007RTbS,Norton2008,Pooley2013,BrownRead2018}.} On the dynamics-first side of this debate, see the work of Harvey Brown.\footnote{See \cite{RBrown2005} and \cite{BrownRead2018} for an overview.} Another notable view of this kind is Nick Huggett's Regularity Relationism.\footnote{See \cite{HuggettNick2006TRAo}.} Roughly, this view claims that the world's geometric structure is merely a codification of certain patterns in the dynamical behavior of matter. As this example shows, dynamics-first views can give a metaphysically deflationary account of the the world's geometric structure. The primary cost of such a deflationary move, however, is that we no longer have access to metaphysically-supported geometry-based explanations of dynamics. See Sec.~\ref{SecMBE} for further discussion.

Returning once again to our questions: How can one explain why topologically spherical spacetimes can only ever support discrete vibrational modes? Moreover, why are such spacetimes always measured to have a positive average curvature? One might try to explain either of these by invoking certain topological facts about the spacetime manifold, i.e., the fact that the sphere is topologically compact with Euler characteristic $\chi=2$. Concretely, on this view the compact topology of the spacetime manifold forces the spectrum of any Laplacian operator to be discrete and thereby indirectly shapes the dynamical behavior of matter. Similarly, its Euler characteristic of $\chi=2$ forces the average curvature to be positive by the Gauss-Bonnet theorem. As this example shows, one might try to use the topology of the spacetime manifold to explain dynamics. This topology-to-dynamics explanatory strategy typifies what I will call \textit{topology-first views}.

As is hopefully clear from the above discussion, the laws-first, geometry-first, and topology-first views have some structural similarities. They each take some facet of the dynamical behavior of matter to be the explanandum and something else (e.g., the world's laws, geometry, and/or topology) to be the explanans. Since in each case the arrow of explanation points towards dynamics from elsewhere, I will refer to these views collectively as \textit{dynamics-second views}. The competing views discussed above each reverse the arrow of explanation, putting dynamics first. Consequently, I will refer to these views collectively as \textit{dynamics-first views}.\footnote{It is important to note that these collections of dynamics-first/second views do not need to be accepted or rejected as a package deal. For instance, one could take a dynamics-first view of laws while taking a dynamics-second view of geometry and the spacetime manifold.} In this terminology, both the metaphysics of laws debate and the geometrical vs dynamical spacetime debate are instances of the wider class of dynamics-first vs dynamics-second debates.

Note, however, that the above discussion has neglected to mention the analogous debate regarding the world's topological structure (i.e., the spacetime manifold). Discussion of this debate is missing largely because there is no well-developed alternative to the topology-first view.\footnote{As I will discuss in Ch.~\ref{ChapOverview}, while some existing views in the literature (e.g., spacetime algebraicism, and spacetime functionalism) move in this direction, none of them qualify as dynamics-first views. Ch.~\ref{ChapSecondVsFirst} will review how existing dynamics-first views of laws and geometry are structured. This will lead us in Ch.~\ref{ChapOverview} to demand several Desiderata for an analogous dynamics-first view of the spacetime manifold.} This dissertation seeks to fill this gap by developing a dynamics-first view of the spacetime manifold to the same level of maturity as the corresponding dynamics-first views of laws and geometry. Roughly, the view I put forward is the following: The world's topological structure (i.e., the spacetime manifold) is merely a codification of certain patterns in the dynamical behavior of matter. For brevity, I will call the Dynamics-First view of the Spacetime Manifold put forward in this dissertation the DFSM view.

\section{Connecting Explanation, Metaphysics, and Knowledge Generation}\label{SecMBE}
But why should one care about the dynamics-first vs dynamics-second debates generally? While these debates have been introduced as concerning the direction of certain arrows of explanation, their importance comes from their metaphysical consequences. As I will now discuss, it follows from the fact that dynamics-first and dynamics-second views adopt very different explanatory strategies, that they end up requiring very different metaphysical backstories. 

In particular, if we demand that our explanations be metaphysically supported and metaphysically non-circular (definitions to follow) then dynamics-second views are required take their explanatory aides (e.g., the world's laws, geometry, topology, etc.) to be \textit{metaphysically substantial} in their own right, i.e., more than just a codification of the relevant dynamics. By contrast, dynamics-first views can adopt a deflationary view of laws, geometry, and/or topology and thereby achieve a relatively sparser metaphysics. As mentioned above, the primary cost of adopting this deflated metaphysics is losing access to metaphysically-supported dynamics-second explanations.

In order to establish a connection between explanation and metaphysics, I will now adopt the following two principles. One may wish to demand that our explanations be \textit{metaphysically supported} in the sense that they are backed up by some metaphysical backstory, as follows:\footnote{There is some room available here to claim that at least some kinds of explanations do not require metaphysical support, e.g., explanation by unification. For instance, Newton's theory of gravity gives a unified account of the motions of apples and planets. This theory might have some explanatory merits even if it doesn't give a metaphysical picture of how the instantaneous action at a distance works. One can admit that Newton's explanation is nicely unifying while nonetheless demanding that he owes us a metaphysical backstory. In this dissertation, I am only interested in metaphysically supported explanations.}
\begin{quote}
{\hypertarget{MBE}{\bf Metaphysical Backstory for Explanation (MBE)}} - Any claim that ``X can help us explain Y'' will be metaphysically unsupported (and hence unsatisfying) unless it is backed up with a satisfying metaphysical account of what exactly X and Y are, and moreover how the metaphysical relationship between X and Y facilitates this explanation.
\end{quote}
If one demands that explanations have metaphysical backstories, this is certainly the bare minimum one can ask for. 

The \hyperlink{MBE}{MBE} principle applies equally well to dynamics-first and dynamics-second views; I will demand that both explanatory strategies are supported by \hyperlink{MBE}{MBE}-compliant metaphysical backstories. The following principle, however, distinguishes them. One may additionally wish to demand that our explanations be \textit{metaphysically non-circular} as follows:
\begin{quote}
{\hypertarget{DEM}{\bf Dead-End Metaphysics (DEM)}} - Any claim that ``Z can help us explain certain patterns in the dynamical behavior of matter'' will be metaphysically circular (and hence unsatisfying) if Z itself is merely a codification of those same patterns. To avoid this, Z must be more than just a codification of the relevant dynamical behavior. That is, Z must be metaphysically substantial in its own right (i.e., more than just a codification of the relevant dynamics.).
\end{quote}
Taken together the \hyperlink{MBE}{MBE} and \hyperlink{DEM}{DEM} principles constrain dynamics-second views (but not dynamics-first views) to have a relatively weighty metaphysics. Roughly, \hyperlink{MBE}{MBE} forces our explanations to have metaphysical backstories. Then \hyperlink{DEM}{DEM} requires that when dynamical patterns are the explanandum, the explanans must be metaphysically substantial. Thus, dynamics-second views must take their explanatory aide (e.g., the world's laws, geometry, topology) to be metaphysically substantial. For an extended discussion of this argument, see Ch.~\ref{ChapSecondVsFirst}. Dynamics-first views avoid this conclusion because they take the dynamics to be the explanans not the explanandum. Thus, dynamics-first views are allowed to deflate what dynamics-second views are forced to make metaphysically substantial.

So far I have argued that dynamics-first views \textit{can} achieve a relatively sparse metaphysics. I have not yet argued that dynamics-first views ought to adopt this lighter metaphysics, nor have I argued that there is anything wrong with the heavier metaphysics that dynamics-second views are forced into. Such arguments will be provided in Ch.~\ref{ChapSecondVsFirst}. 

As I will discuss there, the dynamics-first vs dynamics-second debates can be reframed as follows: Are the arrows of explanation and knowledge generations parallel or anti-parallel? In science, we come to know the world's laws, geometric structure, and/or topological structures indirectly by studying and codifying the dynamical behavior of matter. Hence, knowledge generation in science is dynamics-first. 

On dynamics-second views, the arrows of knowledge and explanation are anti-parallel. Studying dynamics teaches us about the world's laws, geometry, and topology. This indirectly known substructure is then taken to be metaphysically substantial in its own right and to subsequently help us explain the more directly known dynamical behavior of matter. This strategy is methodologically fine in principle; Indeed, at heart, this is Plato's methodology. However, as I will discuss in Ch.~\ref{ChapSecondVsFirst} it invites complaints along the lines of Moliere’s dormitive virtue in opium: One cannot explain why opium makes people sleepy by declaring that it has within it the power to make people sleepy (i.e., that it has a dormitive virtue).

By contrast, dynamics-first views align the arrows of explanation and knowledge generation. In doing so, they open up the possibility of a further alignment between metaphysics and knowledge. In particular, as I have discussed above, dynamics-first views can adopt a deflationary metaphysics which parallels scientific practice: The world's law-like, geometric and/or topological structures are merely codifications of the relevant patterns in the dynamical behavior of matter. As I will argue in Ch.~\ref{ChapSecondVsFirst}, this deflationary style of metaphysics is the sparsest metaphysics consistent with scientific practice.

To review: Dynamics-first views \textit{can} benefit from a favorable three-way alignment of explanation, metaphysics, and knowledge. Given the benefits of this three-way alignment, however, I will now restrict the definition of dynamics-first views to those which take full advantage of it. That is, dynamics-first views must adopt both a dynamics-first explanatory strategy as well as a deflationary ``merely a codification of'' style of metaphysics.\footnote{As I will discuss in Ch.~\ref{ChapSecondVsFirst}, further metaphysical posits are optional for dynamics-first views. For instance, dynamics-first views may or may not adopt a Humean style ban on necessary connections between distinct existences.}

\section{Brief Comparison with the Philosophy of Spacetime Literature}
The primary benefit of the DFSM view is that it brings the metaphysical sparsity characteristic of dynamic-first views to the philosophy of space and time. An extended comparison with other views in the philosophy of spacetime literature will be given following an overview of the DFSM view in Ch.~\ref{ChapOverview}. Some brief differences, however, should be noted upfront.

Unlike spacetime substantivalism and spacetime relationalism, the DFSM view denies the existence of metaphysically substantial spacetime points and spacetime relations respectively. Rather the non-spatiotemporal substrate (whose patterns the spacetime manifold is a codification of) is here taken to be algebraic (or, at minimum, vectorial) in nature.\footnote{See Sec.~\ref{SecSurvivingDynamics} for a discussion of the roles that a theory's algebraic and vectorial structures play in this dissertation's methodology.}

Note, however, that despite positing a non-spatiotemporal substrate from which spacetime arises, the DFSM view is not a view of spacetime emergence, at least not in the usual sense. Namely, unlike typical emergence views, the methodology put forward in this dissertation requires no limit-taking, no regime change, and, importantly, no new speculative physics. 

As I will discuss in Ch.~\ref{ChapGenerality}, the DFSM view can be applied to a very wide range of spacetime theories; Namely, it applies to all theories with a sufficient level of \hyperlink{SKC}{spacetime-kinematic compatibility} (defined in Ch.~\ref{ChapGenerality}). Familiar examples include: \hyperlink{QKG}{A non-linear version of the Klein Gordon equation,} \hyperlink{HEDS}{the heat equation on a discrete spacetime,} and \hyperlink{SES}{the Schr\"odinger equation on a sphere}. In these and many other spacetime theories, we can see the spacetime manifold as being merely a codification of patterns in the dynamical behavior of matter. 

Let us take as an example the task of recovering our usual spacetime  description of non-relativistic quantum mechanics (e.g., the Schr\"odinger equation) from a spacetime-neutral Hilbert space description of the theory. Using the results of this dissertation, one can quickly identify a vast range of spacetime description of any given Hilbert space dynamics. In particular, one can identify \textit{every} possible spacetime framing of the theory which has spacetime-kinematic compatibility.\footnote{This scope of all possible spacetime framings for a given theory is captured by the notion of an ISE-equivalence class and is defined in Ch.~\ref{ChapSearch}.} These are in a tight correspondence with what I will call \hyperlink{PSTO}{pre-spacetime translation operations} (defined in Ch.~\ref{ChapExternal}).

In total, we have an over-abundance of possible spacetime settings for the Hilbert space theory. The DFSM view here invokes a competition between different spacetime codifications analogous to the competition between different law-like codifications invokes by Lewis's BSA. There, the rough selection criteria is achieving the balance of simplicity and strength. Here, the rough selection criteria are achieving the best fit with the theory's kinematics and dynamics. 

In this way the DFSM view is a kind of spacetime functionalism:\footnote{For further discussion of spacetime functionalism, see \cite{KNOX2013346,knox2014spacetime,KNOX2019118,LAM201839} and \cite{Lam2020-LAMSFF} among many others.} The theory's spacetime manifold is whatever best plays the role of being a spacetime manifold. A key aspect of the DFSM view, however, is that the full scope of candidates for playing this role is known; They are the pre-spacetime translation operations. Moreover, the story of how we can use these to recover the intuitive spacetime framing of our theory is also known; It is given by the externalization process. 

Despite fixing both the range of candidates and the details of the recovery, the DFSM view offers us a wide range of spacetime framings for a wide variety of spacetime theories. Concretely, as I will discuss in Ch.~\ref{ChapSearch}, given any spacetime theory with spacetime-kinematic compatibility, this dissertation's methodology allows us to re-describe this theory's dynamics in every way possible which maintains this spacetime-kinematic compatibility. As I will argue in Ch.~\ref{ChapGenerality}, spacetime-kinematic compatibility is a very weak condition. The mathematics developed in this dissertation gives us powerful tools for investigating and negotiating between a wide variety of spacetime settings for a wide range of spacetime theories.

Further comparison with the 
philosophy of spacetime literature will be given following an overview of the DFSM view in Ch.~\ref{ChapOverview}. Following this, the remainder of this dissertation will be spent filling in the technical details of the DFSM view and proving that it satisfies various desiderata. Before this, however, in Ch.~\ref{ChapSecondVsFirst} I will provide an extended discussion of the dynamics-first vs dynamics-second debates.

\chapter{The Dynamics-First vs Dynamics-Second Debates}\label{ChapSecondVsFirst}
This chapter will introduce the dynamics-first vs dynamics-second debates in further detail. As mentioned in Ch.~\ref{ChapIntro}, while these debates have been introduced in terms of the direction of certain arrows of explanation, their importance comes from the contrast between the metaphysics backstories which these competing explanatory strategies allow for. As I will discuss, the differences in metaphysical sparsity which these explanatory strategies can achieve is rooted in whether these views take the arrow of explanation to be parallel or anti-parallel to the arrow of knowledge generation. In what follows, particular attention will be paid to how existing dynamics-first views of laws and geometry are structured. Taking these as a model will guide us in creating a dynamics-first view of the spacetime manifold.

\section{Fleshing Out Dynamics-Second Views}\label{SecLawGeometryTopology}
In Chapter~\ref{ChapIntro}, I briefly introduced three dynamics-second views, namely the laws-first, geometry-first, and topology-first views. Roughly, these are the views claiming that world's laws, geometry, and/or topology respectively can help us explain the certain facets of the dynamical behavior of matter. Of interest here, however, are not questions of explanation but rather a question of metaphysics: What sort of metaphysical backstory is needed to achieve a satisfying dynamics-second explanation?

Applying the \hyperlink{MBE}{\bf Metaphysical Backstory for Explanation (MBE)} principle introduced in Ch.~\ref{ChapIntro} to a generic dynamics-second view (of laws/geometry/spacetime) the following metaphysical questions arise:
\begin{enumerate}
    \item[Q1:] What exactly are laws/geometry/spacetime metaphysically speaking?
    \item[Q2:] Moreover, how exactly do laws/geometry/spacetime relate to matter so as to appropriately inform its dynamical behavior? E.g.: How exactly do they govern, interact-with, and/or support matter?
   \item[Q3:] Finally, what sort of a thing is matter such that its dynamical behavior can be shaped in this way by laws/geometry/spacetime?
\end{enumerate}

\noindent In the context of laws, the above-discussed questions can be fleshed out as follows: What kind of a thing is a ``law of nature'' exactly? What properties and relations do they have? How exactly do laws govern matter? By pushing matter around? If so, why can't the particles/fields push back on the laws? What sort of a thing is matter if it can be pushed around by laws, but not push back?\footnote{Such questions are ubiquitous throughout the metaphysics of laws debate. See the SEP article by \cite{sep-laws-of-nature} for an overview among many others: \cite{BirdTextbook,Hall2015,BhogalHarjit2020Halo,Lewis1973,Lewis1999,LewisDavid1983Nwfa,DemarestHeather2017PPPL} and \cite{BusinessOfLaws}.}

To highlight what a successful answer to these questions might look like it's best to briefly consider a failed one. Suppose that one has ample data showing a certain pattern in the dynamical behavior of matter (e.g., falling apples). Suppose that one can write down a set of equations which accurately codify these patterns (e.g., the equations of gravity). One might suggest that the ``laws of nature'' simply means these pattern-codifying equations and moreover that ``being governed'' by these laws simply means satisfying these equations (i.e., fitting these patterns). This explanation is hopelessly circular and has ultimately shown no light on why matter behaves the way it does. This is, of course, a familiar point from the metaphysics of laws debate.

We can see this failure as an instance of the \hyperlink{DEM}{\bf Dead-End Metaphysics (DEM)} principle introduced in Ch.~\ref{ChapIntro}: If the laws of nature are to help us achieve a metaphysically supported explanation of dynamics, they must be more than just a codification of the relevant dynamics. Note, however, that this principle holds even outside of the laws-context. In Ch.~\ref{ChapIntro}, it was concluded that whatever explanatory aide a dynamics-second view invokes (e.g., laws, geometry, the spacetime manifold, etc.) must be metaphysically substantial in its own right. That is, it must be more than just a codification of the relevant dynamical behavior.

For instance, consider a dynamics-second view of geometry which seeks to explain why force-free particles travel in straight lines (i.e., inertia) in terms of the world's geometric substructure. In the context of special relativity, for instance, the relevant geometric structures are the Lorentzian metric, $\eta_\text{ab}$, and the unique covariant derivative, $\nabla_\text{c}$, with $\nabla_\text{c}\eta_\text{ab}=0$. It follows from \hyperlink{DEM}{DEM} that on any successful dynamics-second view of geometry, $\eta_\text{ab}$ and $\nabla_\text{c}$ must metaphysically substantial in their own right. That is, they must be more than just codifications of the relevant dynamical behavior.

Past this, a dynamics-second view of geometry must be able to answer the following sort of questions arising from \hyperlink{MBE}{MBE} regarding the physical correlates of $\eta_\text{ab}$ and $\nabla_\text{c}$: How is a force-free particle supposed to know what a straight line is? Must it consult with some geometric substructure? If so, how? What kind of thing is this geometric substructure anyway? What properties and relations does it have? How does it push matter around? And why can't the particles/fields push back on the geometry?\footnote{Such questions are ubiquitous throughout the dynamical vs geometrical spacetime debate. See \cite{BrownRead2018} for a recent overview among many others: \cite{MaudlinTim2012Pop:,EarmanJohn1989Weas,TwiceOver,BelotGordon2000GaM,Menon2019,BrownPooley1999,Nonentity,HuggettNick2006TRAo,StevensSyman2014Tdat,DoratoMauro2007RTbS,Norton2008,Pooley2013,RBrown2005}.}\footnote{As a historical note, it was this line of questioning which led Einstein to develop his theory of general relativity, see \cite{BrownHarveyR2013Etro}.}

Similar questions arise from \hyperlink{MBE}{MBE} for any dynamics-second view of the spacetime manifold. In this context, \hyperlink{DEM}{DEM} also applies such that for any dynamics-second view to be successful, it must take the spacetime manifold, $\mathcal{M}$, to be metaphysically substantial in its own right. That is, it must be more than just codifications of the relevant dynamical behavior.

In each of these contexts (i.e., laws, geometry, spacetime), the balance to strike here is to invoke enough metaphysics to get these dynamics-second explanations off the ground without excessive hand-waving. As I will now discuss, however, some find the metaphysics involved in dynamics-second explanations to be excessively spooky. In fact, as I will now show, these complaints are rooted in the certain structural features common to all dynamics-second views. To showcase this, it is instructive to compare the arrow of explanation with the arrow of knowledge generation in science.

In practice, how do we come to know the world's laws (or for that matter its geometric or topological structure)? As a matter of scientific practice we get our knowledge of each of these indirectly by studying the dynamical behavior of matter. Namely, by some process of investigating and codifying patterns in the dynamical behavior of matter we come to know the laws of nature. Once we have a few laws of nature established, we may find that certain physical effects are felt by matter in a composition-independent way: All kinds of clocks exhibit time dilation; All kinds of particles have inertia; Moreover, they all follow the same paths under only gravitational forces. The existence of such composition-independent effects suggests that adopting a geometric perspective might be fruitful. Indeed, by writing these laws of nature in a generally covariant way, we can isolate some geometric structures (e.g., $\eta_\text{ab}$ and $\nabla_\text{a}$) common to them all. Past this, there may be other physical effects which are felt by matter in a geometry-independent way. By further investigating these geometry-independent effects, we can isolate certain topological structures common to all dynamical laws, i.e., a spacetime manifold. Thus, our knowledge of laws, geometry, and spacetime topology are each built up from our knowledge of the dynamical behavior of matter. Said differently, the arrow of knowledge generation in science points from dynamics elsewhere. That is, knowledge generation is dynamics-first.

Thus, the dynamics-first vs dynamics-second debates can be recast as follows: Does the arrow of explanation point in the same direction as the arrow of knowledge generation? According to dynamics-second views, the arrows of explanation and knowledge are anti-parallel. The story is roughly as follows: Our dynamical investigations reveal some law-like structure to the world (or relatedly a geometric and topological structure). This now-revealed structure is taken to be metaphysically substantial in its own right (i.e., more than just a codification of the relevant dynamics) and to subsequently help us explain the dynamical behavior of matter. This sounds methodologically fine in principle; Indeed, this is essentially Plato's methodology. 

For example, consider an experiment in which we learn about electrons indirectly via our observations of how certain needles move in the laboratory. One might here adopt a dynamics-second explanatory strategy. That is, one might claim that electrons are metaphysically substantial (i.e., more than just a codifications of the patterned movements of needles) and moreover that they can help us explain our observations. I have no problem adopting such a dynamics-second explanatory strategy in the electron/needle case.

There is always the risk, however, that such dynamics-second explanations are rather hollow along the lines of Moliere’s dormitive virtue in opium. Namely, there is a risk that we are merely restating the yet-to-be-explained dynamical phenomena using new abstract terms: e.g., opium makes people sleepy because ``there is a latent dormitive virtue in opium'' or equivalently because ``opium has within it the power to make people sleepy''. The introduction of such a ``dormitive virtue'' is clearly ad hoc and so surely cannot count as a valid explanation. Such complaints have been explicitly raised against the dynamics-second views of both laws and geometry. See, for instance, \cite{LewisDavid1983Nwfa} against a dynamics-second view of laws, namely  Armstrong's conception of nomic necessity. See also, \cite{Nonentity} against a dynamics-second view of  geometry. The next section will discuss an alternative explanatory strategy available to those who favor a sparser metaphysics. 

\section{Fleshing Out Dynamics-First Views}\label{SecDynamicsFirst}
As the previous section has discussed, dynamics-second views must invoke some (potentially spooky) metaphysics as they take the arrows of knowledge generation and explanation to be anti-parallel. As I will now discuss, dynamics-first views can get by with a much lighter metaphysics by reversing the arrow of explanation, i.e., by taking the dynamical behavior of matter to be the explanans rather than the explanandum.

An immediate complaint against dynamics-first views generally is that they leave the dynamical behavior of matter unexplained (except possibly in terms of more fundamental dynamics). Isn't science supposed to (among other things) explain the fundamental dynamical behavior of matter?\footnote{For a view explicitly denying this, see \cite{VanFraassenBasC.1980Tsie}.} Note, however, that by moving to a dynamics-first we have aligned the arrows of explanation with the arrows of knowledge generation. This is favorable since we are generally on more secure footing when using the known to explain the unknown rather than vice-versa.\footnote{More carefully: We are on more secure footing when using the relatively directly known to explain the relatively indirectly known.} Isn't science supposed to (among other things) explain the unknown in terms of the known? Thus, it may not be so costly after all to leave the fundamental dynamical behavior of matter unexplained. 

As this section will discuss, it is exactly this alignment between the arrows of knowledge and explanation that allows dynamics-first views to achieve a lighter metaphysics (e.g., no dormitive virtues). Specifically, it allows us to additionally parallel the explanation's metaphysical backstory with the process of knowledge generation in science. This three-way three-way alignment between explanation, knowledge, and metaphysics is ultimately what allows dynamics-first views to achieve a sparser metaphysics.

The benefits of aligning metaphysics and scientific practice have already been discussed in the metaphysics of laws literature. As \cite{BhogalHarjit2020Halo} notes:
\begin{quote}
\vspace{-0.6cm}
\singlespacing
One common motivation [for Humeanism about laws] comes from the way that the defender of the BSA claims that their view mirrors scientific practice (e.g., \cite{DemarestHeather2017PPPL}). The BSA sticks very closely to the actual practice of science—in fact, one way to describe the view is that it takes the actual methodology of science and it mirrors that methodology in the account of the metaphysics of laws. \cite{Hall2015} calls this the “unofficial guiding idea” behind Humeanism about laws.\footnote{Ibid. 8} 
\end{quote}
As this section will discuss, the benefits of this three-way alignment extend beyond Humeanism about laws to dynamics-first views generally (about laws, geometry, spacetime, etc.).\footnote{A word of warning is here needed about the conceptual difference between what I am calling ``dynamics-first views'' and a stricter notion of Humeanism. As mentioned at the end of Ch.~\ref{ChapIntro}, dynamics-first views are marked by a ``merely a codification of'' style of metaphysics. As I will discuss further momentarily, dynamics-first views can vary as to the metaphysical details of the to-be-codified substrate. For instance, on a dynamics-first view of laws the generalized Humean mosaic could contain dispositions. For example, see \cite{DemarestHeather2017PPPL}. This is in contrast with more strictly Humean views of laws which enforce Hume's idea that there are no necessary connections between distinct existences. In line with the above quote, dynamics-first views might be called ``unofficially Humean''.}

But how does this paralleling of metaphysics and scientific practice work? Recall the story about how knowledge is generated in science from Sec.~\ref{SecLawGeometryTopology}: Roughly, everything we know comes from studying and codifying the dynamical behavior of matter. Dynamics-first views of laws, geometry, and/or the spacetime manifold take these to merely be codifications of certain patterns in the dynamical behavior of matter. To make things concrete, proponents of the dynamics-first views of laws, geometry, and spacetime manifold might make the following sorts of claims respectively:
\begin{enumerate}
    \item[-] The laws of electromagnetism aren't some metaphysically substantial things in the world, rather they are merely codifications of some dynamical patterns which we have identified;
    \item[-] The Lorentzian metric, $\eta_\text{ab}$, and its associated connection, $\nabla_\text{a}$, aren't some metaphysically substantial things in the world, rather they are merely codifications of some dynamical patterns which we have identified;
    \item[-] The spacetime manifold, $\mathcal{M}$, isn't some metaphysically substantial thing in the world, rather it is merely a codification of some dynamical patterns which we have identified.
\end{enumerate}

\noindent Note that these are exactly the sorts of metaphysical claims which are forbidden to dynamics-second views by the \hyperlink{DEM}{DEM} principle introduced in Sec.~\ref{SecLawGeometryTopology}. Dynamics-first views are allowed to adopt this deflationary style of metaphysics precisely because they don't seek to explain the dynamical behavior of matter. Ultimately, dynamics-first views can achieve a sparse metaphysics because they are allowed to deflate what dynamics-second views are forced to make metaphysically substantial.

Of course, having a sparse metaphysics is not an unqualified virtue. One may worry that this metaphysics is too sparse to do what we require of it. But what exactly is required for a satisfying dynamics-first explanation? According to the \hyperlink{MBE}{MBE} principle introduced in Ch.~\ref{ChapIntro} the salient questions for dynamics-first views (of laws/geometry/spacetime) are:
\begin{enumerate}
    \item[Q1:] Laws/geometry/spacetime are codifications of patterns in what substrate exactly? That is, metaphysically speaking, what are the to-be-codified dynamical behaviors of matter?
    \item[Q2:] Given this substrate, laws/geometry/spacetime are codifications of which of its patterns exactly? 
    \item[Q3:] Given this pattern in that substrate, what does it mean, metaphysically speaking, for laws/geometry/spacetime to be a codification thereof? 
\end{enumerate}
These questions must be addressed for any dynamics-first explanation if it is to be metaphysically supported.

The following two subsections will discuss these questions in the context of laws and geometry respectively. My goal, however, is not to come to any definitive answers or to single out any specific dynamics-first views of laws and/or geometry. Rather, my goal is to better understand the common structure underlying dynamics-first views generally. Consequently, much of the following discussion will be metaphorical and/or pictorial in nature. Once we have a pictorial understanding of how dynamics-first views of laws and geometry are structured, we can then use this understanding as a guide to creating a dynamics-first view of the spacetime manifold.

\subsection{\hra Fleshing Out Dynamics-First Views of Laws}
\begin{figure}[t!]
\begin{center}
\includegraphics[width=0.95\textwidth]{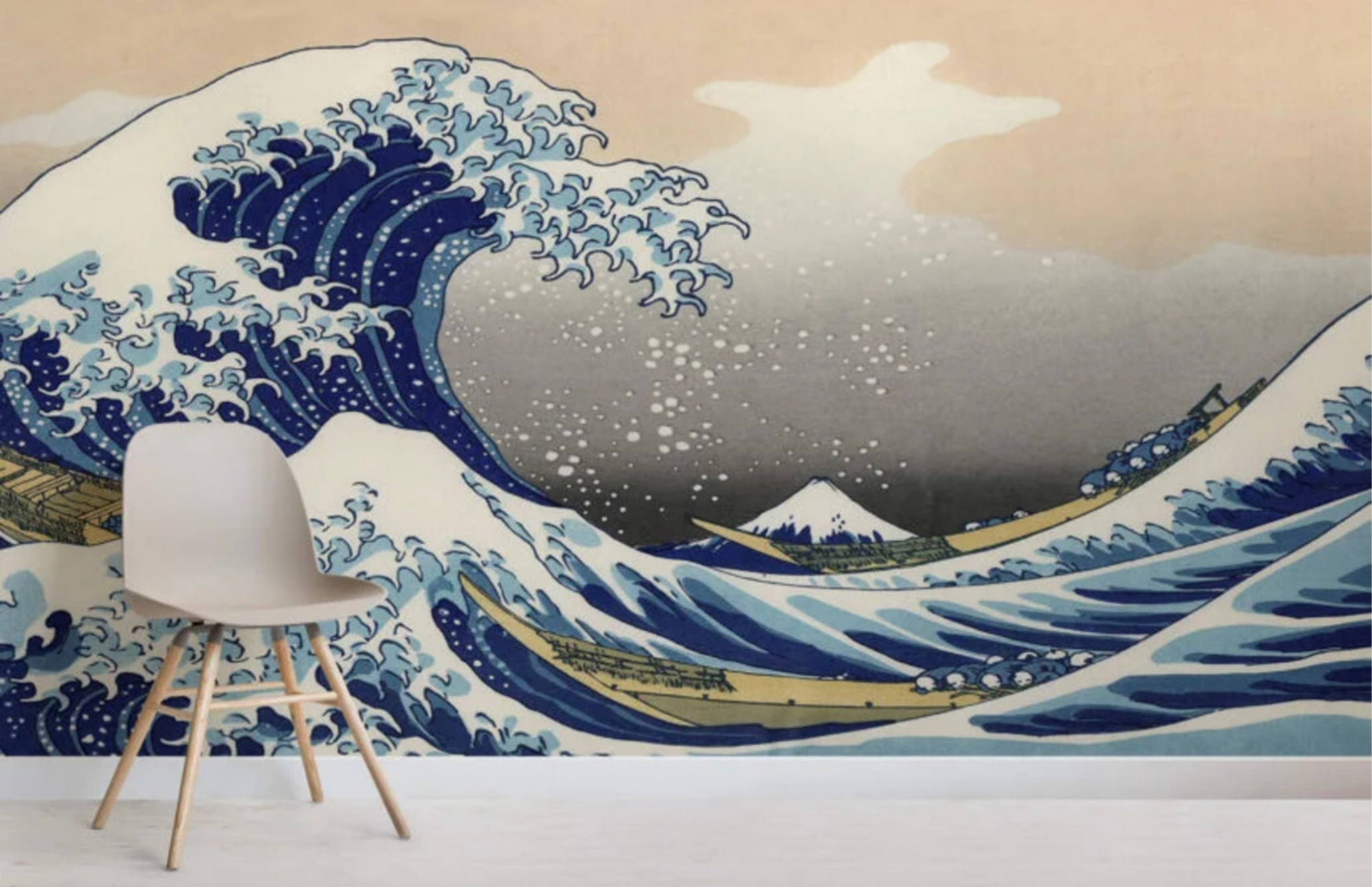}
\caption{The painting ``Under the Wave off Kanagawa'' also known as ``The Great Wave'' by \cite{hokusai_1830} reproduced as a mosaic on a wall.}\label{FigMosaic}    
\end{center}
\end{figure}

Let us first  turn our attention to the dynamics-first vs dynamics-second debate about laws (i.e., the metaphysics of laws debate). Applied to the laws context, Q1 becomes: The laws of nature are codifications of patterns in what substrate exactly? While a wide variety of answers to Q1 are allowed in support of dynamics-first explanations, a prominent example is that given by of Lewis's Best Systems Analysis (BSA).\footnote{See \cite{Lewis1973,Lewis1999}.} Roughly, Lewis views the fundamental ontology of the world to be an arrangement of categorical properties over space and time. On this view, one can imagine the world as a grand painting (i.e., the Humean mosaic) of properties upon a wall. See Fig~\ref{FigMosaic}. Hume himself viewed these bits of paint to be causally independent of each other and consequently reduced causality to ``constant conjunction''. This is a part of Hume's general ban on necessary connections between distinct existences. Fortunately, we can adopt this imagery without agreeing with either Hume or Lewis on the details.

Recall that the goal of this subsection is to come to a pictorial understanding of the general structure of dynamics-first views of laws. To this end, we can generalize the Humean mosaic to include causal relations if we like, as well as any other sorts of relations/properties. For instance, \cite{DemarestHeather2017PPPL} generalizes Lewis's BSA to include dispositions as well as categorical properties. Within the wide umbrella of dynamics-first views, we can take this generalized Humean paint to be as thick or thin as we like, metaphysically speaking. Namely, what I am calling dynamics-first views may or may not enforce a Humean/Lewisian ban on necessary connections between distinct existences.

We can generalize this paint-based metaphor further by imagining a law-like substructure to this painting which somehow governs the generalized Humean paint. We can thereby accommodate primitivism about laws. One might also imagine that this mosaic has some additional geometric or topological substructure (e.g., \cite{HuggettNick2006TRAo} does this). Imagine a painting on a dented sphere. So generalized, this Hume-inspired imagery can also be adapted to fit the markedly non-Humean views. I invite the reader to adopt this generalized Humean imagery: The world is a painting of some kinds of properties/relations/dispositions, possibly with some underlying substructures (law-like, geometric, topological, etc). This imagery will be helpful throughout the rest of this chapter.

The purpose of this wide-scoping paint-based metaphor is to give us a pictorial understanding of how dynamics-first and dynamics-second views of laws differ: Dynamics-first views of laws deny that this generalized Humean mosaic requires any law-like substructure. Further, we can see that dynamics-first views of laws can disagree with each other about the relative metaphysical thickness/thinness of the generalized Humean paint. That is, they can differ in their answers to Q1.

Dynamics-first views of laws can also differ in their answers to Q2: How exactly are the laws supposed to be read off of this generalized Humean mosaic? Once again, there are differing views here. For instance, Lewis's BSA takes the laws of nature to be those which best balance simplicity and strength. Generalizing this, \cite{Cohen2009-COHABB} have modified the BSA account such that one can read the laws off of localized patches of the mosaic.

However it is that dynamics-first views of laws answer Q1 and Q2, any dynamics-first view of laws will put forward a law-like regularity fact of the following form:
\begin{quote}
{\hypertarget{LRF}{\bf Law-Like Regularity Fact (Schema) - }}
\textit{We can conceptualize the dynamical behavior of matter in a law-independent way: Namely, as paint on a generalized Humean mosaic. From this law-neutral perspective we can consider various ways of systematizing the dynamics. One way of doing so in particular gives rise to nice laws.}
\end{quote}
Dynamics-first views of laws can differ on the exact nature of this paint, the technical details of systematizing, as well as what qualifies as a ``nice law''. They all, however, take the laws of nature to be merely codifications of such law-like regularity facts.

Suppose that such a law-like regularity fact has been mathematically proved for a certain spacetime theory. Must a proponent of a dynamics-second view of laws then give up, switching camps? Of course not. They can claim that these sorts of regularity facts require some further metaphysical explanation: Namely, they are explained by the existence of a metaphysically substantial law which governs the dynamical behavior of matter to this effect.

What, then, is the point of proving such law-like regularity facts? While proving such a regularity fact wouldn't end the debate, it certainly brings it to maturity. It is at this point in the discussion that I consider the dynamics-first vs dynamics-second debate about laws to be fully developed. It is tolerably clear what the two sides disagree about (i.e., whether or not these sorts of regularity facts require any further explanation). Moreover, it is tolerably clear what each side of the debate is claiming and how they are to defend their views. Ultimately, however, this laws debate is not the target of this dissertation. For those interested in further discussion of this topic, I refer you to the metaphysics of laws debate.\footnote{See~\cite{sep-laws-of-nature,BirdTextbook,Hall2015,BhogalHarjit2020Halo,Lewis1973,Lewis1999,LewisDavid1983Nwfa,DemarestHeather2017PPPL} and \cite{BusinessOfLaws}.} 

\subsection{\hra Fleshing Out Dynamics-First Views of Geometry}
Let us next turn our attention to the dynamics-first vs dynamics-second debate about geometry (i.e., the geometrical vs dynamical spacetime debate). Applied to the geometry context, Q1 and Q2 become respectively: Geometry is a codification of patterns in what substrate exactly? And moreover, which patterns?

In the geometric context, any answer to Q1 must also answer a more preliminary question Q0: In terms of the above-introduced imagery, how can the generalized Humean paint possibly exist without any geometric underpinnings? The worry associated with Q0 is the following skepticism:\footnote{See, for instance, \cite{DewarNeil2019GC}.} The dynamic-first view may claim that geometry is merely a codification of patterned dynamics, but how do we know that they aren't secretly invoking geometric notions from the wall behind the painting (i.e., the background spacetime)? In order to address this concern, we would need reassurance that the generalized Humean paint can be conceptually divorced from any assumed geometric substructure. Past this, we would then still need an answer to Q2 which reassures us that we can recover our intuitive notions of spacetime geometry from this non-geometric paint. Such reassurances have been achieved by the work of \cite{HuggettNick2006TRAo} on Regularity Relationism.

Recall that the goal of this subsection is to come to a pictorial understanding of the general structure of dynamics-first views of geometry. Consequently, all that is required here is a pictorial account of Regularity Relationism. The full details of this view have been well-discussed elsewhere.\footnote{See~\cite{HuggettNick2006TRAo,Pooley2013}.} Pictorially, Regularity Relationism rises to this challenge by giving a mathematical description of how the following three-step process can be applied to a given spacetime theory:\footnote{Glossing \cite{HuggettNick2006TRAo}'s Regularity Relationism in this way reflects only my own understanding of this view.}
\begin{enumerate}
    \item[1.] First, imagine carefully peeling all of this generalized Humean paint off of the wall in one piece, see Fig.~\ref{FigMosaic}. Moreover, imagine that this detached paint sheet is locally bendable and stretchable but cannot tear or break apart. That is, imagine that this detached paint sheet has only a topological substructure, i.e., no geometric substructure.
    \item[2.] Next, imagine trying out various ways of bending and stretching this detached paint sheet in search of a ``good'' one which fits well with the theory's dynamics and/or kinematics. It is important here that we have multiple options available to us and that one of them is to do nothing.
    \item[3.] Finally, having made our final choice, we then read off the theory's geometry as usual. In the case where we did not bend or stretch the detached paint sheet, this amounts to putting the detached paint sheet straight back on the wall unchanged.
\end{enumerate}
This paint-peeling process serves two purposes: Firstly, it gives us reassurance that the generalized Humean paint can be conceptually divorced from any assumed geometric substructure. Secondly, it gives an indication of how our intuitive notions of spacetime geometry can be recovered from this non-geometric paint.

Consider applying this process to any spacetime theory with nice geometric laws. Applied to such theory, this paint-peeling process will reveal of it the following geometric regularity fact:
\begin{quote}
{\hypertarget{GRF}{\bf Geometric Regularity Fact (Schema) - }}\textit{We can conceptualize the dynamical behavior of matter in a geometry-independent way: Namely as a detached paint sheet. From this geometry-neutral perspective we can consider various ways of bending and stretching this detached paint sheet. One way of doing so in particular gives rise to nice geometric laws.}
\end{quote}
Namely, peeling the paint off the wall and then putting it straight back on yields the theory we started with. Importantly, however, we were not forced to do this. If we can find a better way to bend and stretch, we can find a better geometric description of our theory. Moreover, our ability to bend and stretch the paint sheet helps to demonstrate that it has, in fact, become divorced from its geometric underpinnings.

The exact details of this three-step paint-peeling process and the above-discussed geometric regularity fact can differ between dynamics-first views of geometry. As a motivation for view put forward in this dissertation, only this pictorial gloss is relevant.

Suppose that such geometric regularity fact has been mathematically proved for a certain spacetime theory. Just as in the laws context, such a proof may not convince a proponent of a dynamics-second view of geometry to switch camps. They can respond by claiming that this geometric regularity fact requires some further metaphysical explanation: Namely, it is explained by the existence of a metaphysically substantial geometric substructure to the world. As before, the purpose of proving such a regularity facts is not to end the debate, but rather to bring it to maturity. Ultimately, however, this geometry debate is not the target of this dissertation. For those interested in further discussion of this topic, I refer you to the geometric vs dynamical spacetime debate.\footnote{See~\cite{RBrown2005,EarmanJohn1989Weas,TwiceOver,BelotGordon2000GaM,Menon2019,BrownPooley1999,Nonentity,HuggettNick2006TRAo,StevensSyman2014Tdat,DoratoMauro2007RTbS,Norton2008,Pooley2013,MaudlinTim2012Pop:,BrownRead2018}.}

\subsection{\hra Towards a Dynamics-First View of Spacetime}\label{SecDynamicsFirst3}
Finally, let us turn our attention to the topology-first vs dynamics-first debate (i.e., the dynamics-first vs dynamics-second debate about spacetime). The goal of this dissertation is to develop this debate to the same level of maturity as the above-discussed debates about laws and geometry. What would it take to flesh out a dynamics-first view of the spacetime manifold? Applied to the spacetime  context, Q1 and Q2 become respectively: Spacetime is a codification of patterns in what substrate exactly?\footnote{For a discussion of what non-spatiotemporal substrate might be the supervenience basis for a Humean view of laws, see \cite{LawsBeyondSpacetime}. It should be noted, however, that much of the discussion there assumes a Humean/Lewisian style ban on necessary connections between distinct existences. Dynamics-first views of the spacetime manifold may or may not try to enforce such a ban. \cite{LawsBeyondSpacetime} note that this ban cause tension between Humeanism and naturalism. Where such tensions arise, I tend to favor naturalism.} And moreover, which patterns?

As in the geometric context, any answer here to Q1 must also answer a more preliminary question Q0: In terms of the above-introduced imagery, how can the generalized Humean paint possibly exist without any topological underpinnings?\footnote{\cite{Norton2008} has complained that \cite{HuggettNick2006TRAo}'s Regularity Relationism does not divorce the dynamical behavior of matter from its topological underpinnings. The DFSM view can be seen as a response to Norton's complaint. For further discussion see Sec.~\ref{SecLitComp}.} Past this, we would then still need an answer to Q2 which reassures us that we can recover our intuitive notions of spacetime topology from this non-topological paint.

It is at this point that the paint-based metaphor begins to break down. Fortunately, despite this break-down, the way forward is clear. Taking the above-discussed laws and geometry debates as our guides, the goal of any dynamics-first view of the spacetime manifold should be to prove something like the following topological regularity fact:
\begin{quote}
{\hypertarget{TRFS}{\bf Topological Regularity Fact (Schema) - }}\textit{We can conceptualize the dynamical behavior of matter in a spacetime-independent way: Namely, ``somehow''. From this spacetime-neutral perspective we can consider various ways of ``doing something''. One way of doing so in particular gives rise to nice geometric laws set on a smooth topological manifold.}
\end{quote}
with the quoted phrases somehow made concrete. See Ch.~\ref{ChapOverview} for my proposal. 

Note that proving such a topological regularity fact will give us the mathematical tool relevant for answering questions Q0, Q1, and Q2. Firstly, we will have a method for conceptually divorcing the dynamical behavior of matter from its topological underpinnings. Secondly, we will have a description (at least mathematically) of the non-topological substrate which spacetime is a codification of. Finally, we will have a method for recovering the theory's original spacetime setting (as well as many alternative settings) from this non-topological description of the theory.

Of course, mathematics alone cannot answer all of our philosophical questions. For instance, past its mathematical description, what is this non-topological paint metaphysically speaking? For the DFSM view in particular, the world's non-spatiotemporal substrate is algebraic (or, at minimum, vectorial) in nature.\footnote{See Sec.~\ref{SecSurvivingDynamics} for a discussion of the roles that a theory's algebraic and vectorial structures play in this dissertation's methodology.} Consequently, answering this question requires a theory of algebraic metaphysics.\footnote{For instance, \cite{Geroch1972} and \cite{EarmanJohn1989Weas} advocate for spacetime algebraicism, the view that physical fields exist fundamentally without an underlying spacetime. For further discussion of spacetime algebraicism, see \cite{Chen_2021, Menon2019}.} While the full details of such an algebraic metaphysics are beyond the scope of this dissertation, I do take a significant step in that direction. Whatever this algebraic stuff/goings-on ultimately are the mathematical details the DFSM view are suggestive of how it can be related to our usual descriptions of spacetime physics.

Even with this metaphysical backstory, however, proving this topological regularity fact would not end the debate. As in the previous two debates, any such topological regularity fact can be viewed in two ways. A proponent of the dynamics-first view can claim that the spacetime manifold is merely a codification of this regularity fact. By contrast, a proponent of the topology-first view can claim that this regularity fact requires some further explanation: Namely, it is explained by the existence of some metaphysically substantial topological substructure to the world (i.e., a spacetime manifold). Fortunately, the goal of this dissertation is not to end this debate, but merely to bring it to maturity.

The next chapter will overview the dynamics-first view of spacetime put forward in this dissertation (the DFSM view). Following this, the remainder of this dissertation is spent filling in the technical details of this view, proving that it satisfies various desiderata, and providing illustrative examples.

\chapter{Overview of the DFSM View}\label{ChapOverview}
This chapter will give a skeletal overview of the dynamics-first view of the spacetime manifold put forward in this dissertation (the DFSM view). At a pictorial level, the construction of this view is analogous to the construction of the dynamics-first views of geometry discussed in the previous chapter (e.g., \cite{HuggettNick2006TRAo}'s Regularity Relationism). Concretely, this chapter will put forward a topological regularity fact analogous to the \hyperlink{GRF}{\bf Geometric Regularity Fact} discussed in the previous chapter. This regularity fact will be underwritten by a three-step process analogous to the paint-peeling process of \cite{HuggettNick2006TRAo}'s Regularity Relationism. This new three-step process will serve two purposes:  Firstly, it will give us reassurance that the dynamical behavior of matter can be conceptually divorced from any assumed topological underpinnings. Secondly, it will show how our intuitive notions of spacetime topology can be recovered from this non-topological substrate.

At the outset, however, a word of warning is needed. The overview presented in this chapter, really is skeletal, i.e., it is bare bones. Consequently, four unexplained terms-of-art (i.e., jargon) will appear \underline{underlined}. Explicit definitions of these terms will be given throughout the course of this dissertation.

\section{The Internalize, Search, Externalize (ISE) Methodology}\label{SecISE}
The dynamics-first view of the spacetime manifold put forward in this dissertation (the {\hypertarget{DFSM}{DFSM view}}) is underwritten by the following three-step process. Given a certain kind\footnote{See Ch.~\ref{ChapGenerality} for a discussion of the \hyperlink{Prelims}{preliminary assumptions} which a spacetime theory must satisfy in order to be input to the \hyperlink{ISE}{ISE Methodology}.} of spacetime theory we can apply the following 
{\hypertarget{ISE}{methodology}}:
\begin{enumerate}
    \item[1.] Firstly, we can move to a spacetime-independent view of our theory through a process of \hyperref[ChapInternal]{\underline{internalization}}. This is analogous to the first step in the above-discussed paint-peeling-process. There the generalized Humean paint is divorced from any assumed geometric substructure (although it remains topologically connected). Here the paint is additionally divorced from any assumed topological substructure.
    \item[2.] From this spacetime-neutral perspective we can then search for various  \hyperlink{PSTO}{\underline{pre-spacetime}} \hyperlink{PSTO}{\underline{translation operations}}. This is analogous to the second step of the paint-peeling process where we search for good ways of bending and stretching the detached paint-sheet. As before, the goal is to find a ``good'' one which fits well with the theory's dynamics and/or kinematics.
    \item[3.] Finally, having selected some pre-spacetime translation operations, we can use them to build a new spacetime setting for this theory through a process of \hyperref[ChapExternal]{\underline{externalization}}. Roughly, externalization is the process of taking some hand-picked pre-spacetime translation operations and letting them be honest-to-goodness spacetime translations. This is analogous to the third step of the paint-peeling process where we put the potentially-modified paint sheet back on the wall.
\end{enumerate}
In a single word, these three steps are: Internalize, Search, and Externalize. Consequently, I will call this the {\hyperlink{ISE}{\bf ISE Methodology}}. 

In order to establish the desired topological regularity fact, this methodology must satisfy the following three desiderata.
\begin{quote}
{\hypertarget{D1}{\bf Desiderata \#1 (D1) - }}\textit{Internalization should conceptually divorce the dynamical behavior of matter from any assumed topological underpinnings.}
\end{quote}
\begin{quote}
{\hypertarget{D2}{\bf Desiderata \#2 (D2) - }} \textit{For any spacetime theory which can be produced via externalization, after re-internalizing this theory we should always be able to re-externalize it trivially. That is, for such theories we should always be able to skip the searching step and go right back to the theory's original spacetime setting.}
\end{quote}
\begin{quote}
{\hypertarget{D3}{\bf Desiderata \#3 (D3) - }} \textit{In the searching step, there should be a variety of pre-spacetime translation operations available to choose from. Externalizing these ought to lead us to a variety of inequivalent spacetime settings.}
\end{quote}

Before demonstrating how these desiderata lead to the desired topological regularity fact, the converse of \hyperlink{D2}{\bf D2} should be discussed. From \hyperlink{D2}{\bf D2} it follows that every spacetime theory output by the externalization process should be acceptable as an input to the internalization process. The converse claim is that all spacetime theories which can be input to the internalization process can also be output by the externalization process. For the {\hyperlink{ISE}{ISE Methodology}} put forward in the dissertation, this converse claim is false. In Ch.~\ref{ChapExternal}, I will prove that every spacetime theory producible via externalization has a certain level of \hyperlink{SKC}{ \underline{spacetime-kinematic compatibility}} which is not required for internalization. Moreover, I will there prove that a spacetime theory is producible via externalization if and only if it has this compatibility. I will do so by explicitly demonstrating \hyperlink{D2}{\bf D2} for all such theories.

We are now in a position to see how the above three desiderata come together to prove the desired topological regularity fact. Consider any theory which has the necessary \hyperlink{SKC}{spacetime-kinematic compatibility} to be producible via externalization. Moreover, suppose that this theory has nice geometric laws set on a smooth topological manifold. Applying the {\hyperlink{ISE}{ISE Methodology}} to any such theory will reveal of it the following topological regularity fact:
\begin{quote}
{\hypertarget{TRF}{\bf Topological Regularity Fact - }}
\textit{We can conceptualize the dynamical behavior of matter in a spacetime-independent way: Namely, via internalization. From this spacetime-neutral perspective we can consider various ways of picking out which pre-spacetime translation operations to externalize. One way of doing so in particular gives rise to nice geometric laws set on a smooth topological manifold.}
\end{quote}
Namely, given such a spacetime theory, \hyperlink{D2}{\bf D2} guarantees that after re-internalizing such a theory we can then trivially re-externalize it. This process of re-internalizing and then immediately re-externalizing is analogous to taking the generalized Humean paint off the wall and then putting it right back on. After this re-internalization process, however, we are not forced to go back the way we came (although we are free to). Indeed, \hyperlink{D3}{\bf D3} guarantees that there are alternatives available for us to consider. This is analogous to us being able to consider various ways of bending and stretching the detached paint sheet. Recall that in the paint-peeling process, our ability to bend and stretch the paint sheet demonstrates it has, in fact, become conceptually divorced from its geometric underpinnings. Similarly, our ability to consider various inequivalent spacetime settings demonstrates that we have conceptually divorced the dynamical behavior of matter from any assumed topological underpinnings. That is, \hyperlink{D3}{\bf D3} helps to demonstrate \hyperlink{D1}{\bf D1}.

Now that we have a skeletal overview of the DFSM view put forward in this dissertation, we are in a position to compare it with existing views in the philosophy of space and time literature.

\section{Literature Comparison}\label{SecLitComp}
As I have discussed in Ch.~\ref{ChapIntro} and Ch.~\ref{ChapSecondVsFirst}, the primary benefit of the DFSM view is that it brings the metaphysical sparsity characteristic of dynamics-first views to the philosophy of space and time. In particular, the view put forward in this dissertation is metaphysically deflationary about both spacetime points and spacetime relations. The only metaphysical posit of this view is the dynamical behavior of matter understood as a non-spatiotemporal substrate whose patterns the spacetime manifold is then a codification of. This substrate is here taken to be algebraic (or, at minimum, vectorial) in nature.\footnote{See Sec.~\ref{SecSurvivingDynamics} for a discussion of the roles that a theory's algebraic and vectorial structures play in the \hyperlink{ISE}{ISE Methodology}.}

Despite positing a non-spatiotemporal substrate from which spacetime arises, the DFSM view is not a view of spacetime emergence, at least not in the usual sense. Namely, unlike typical emergence views, the methodology put forward in this dissertation requires no limit-taking, no regime change, and, importantly, no new speculative physics. For any spacetime theory with the necessary spacetime-kinematic compatibility, one can simply prove of it the above-discussed \hyperlink{TRF}{\bf Topological Regularity Fact} and then claim that the theory's spacetime is best understood as a codification of this fact. As I will discuss in Ch.~\ref{ChapGenerality}, this is possible for a very wide range of spacetime theories.

The present work can also be seen as a response to \cite{Norton2008}'s complaint against  \cite{HuggettNick2006TRAo}'s Regularity Relationism. Norton complains that while Huggett may be able to see the world's geometric structure as a codification of some non-geometric substrate, this substrate still has a topological structure. Huggett's view thereby implicitly assumes a spacetime manifold. Norton's complaint can be read as a challenge to extend Regularity Relationism to include the world's topological structure just as it itself extends Lewis's Best System's Analysis to include the world's geometric structure. 

This dissertation rises to that challenge, by putting forward a dynamics-first view of the spacetime manifold (namely, the DFSM view). As I have argued throughout Ch.~\ref{ChapSecondVsFirst} continuing the line of dynamics-first views from laws to geometry to topology involves proving something like the above-discussed \hyperlink{TRFS}{\bf Topological Regularity Fact Schema} for a wide range of spacetime theories. In particular, in analogy with Regularity Relationism's three-step paint peeling process we are lead to the three-step \hyperlink{ISE}{ISE Methodology} and the three above-discussed \hyperlink{D1}{desiderata}. 

To highlight the importance of these desiderata, let us briefly consider another reply to \cite{Norton2008}, namely \cite{Menon2019}.\footnote{For additional comments on \cite{Menon2019} see \cite{Linnemann,Chen_2021}.} Menon replies to Norton citing the \cite{GelfandNaimark} theorem: Roughly, for any Hausdorff compact  manifold, $\mathcal{M}$, knowing only its algebra of smooth scalar functions, $C^\infty(\mathcal{M})$, is enough to uniquely fix $\mathcal{M}$ as a smooth topological manifold. As such, all talk of topological underpinnings might be uniquely replaceable in terms of some equivalent algebraic talk. This point is closely related to \cite{Geroch1972}’s classic paper on Einstein algebras which demonstrates that everything one needs for general relativity (vectors, tensors, connections, etc) can all be characterized algebraically as well.
 
Does \cite{Menon2019}'s view satisfy the (spirit of) the three desiderata discussed above? If not, what issues arise for this view? A key features of Menon's view is the unique recoverability of the theory's topological structure from its algebraic structure. This gives us something like \hyperlink{D2}{\bf D2}: For a certain kind of spacetime theories, their topological spacetime framing can be discarded and re-adopted at will (at least mathematically). However, the unique-ness of this recoverability comes into conflict with the spirit of \hyperlink{D3}{\bf D3}: There ought to be multiple inequivalent spacetime settings available to us. Failing \hyperlink{D3}{\bf D3} also brings \hyperlink{D1}{\bf D1} into question: If the algebraic and spacetime framings of a theory are in tight correspondence then one's view can be accused of not removing the theory's topological commitments but merely hiding it somewhere within the algebra.\footnote{See \cite{Linnemann}'s complaint against \cite{Menon2019}. See also the proof from \cite{ROSENSTOCK2015309} that \cite{Geroch1972}'s Einstein Algebras are category theoretically equivalent to general relativity. Moreover, individual spacetime points can be reconstructed from Einstein Algebras \cite{10.2307/188131}.} By contrast, on the DFSM view theories may have many inequivalent spacetime framings. The only sense in which a particular spacetime framing might be special on the DFSM view is if it is uniquely well-suited to the theory's dynamics or kinematics. As I will discuss in Ch.~\ref{ChapSearch}, whereas Lewis's BSA judges law-codifications based on their simplicity and strength, the DFSM view judges spacetime-codification based on how well the fit the theory's dynamics and kinematics. As I discussed at the end of Ch.~\ref{ChapIntro}, this makes the DFSM view a kind of spacetime functionalism.

The rest of this dissertation will be spent detailing the \hyperlink{ISE}{ISE Methodology} and demonstrating that they satisfy the above-discussed desiderata. In particular, Ch.~\ref{ChapInternal} will discuss internalization and \hyperlink{D1}{\bf Desiderata \#1}. Ch.~\ref{ChapExternal} will discuss externalization and \hyperlink{D2}{\bf Desiderata \#2}. Finally, Ch.~\ref{ChapSearch} will discuss the searching step and \hyperlink{D3}{\bf Desiderata \#3}. Before this, however, Ch.~\ref{ChapGenerality} will give a mathematical characterization of the range of spacetime theories for which one can prove the above-discussed \hyperlink{TRF}{\bf Topological Regularity Fact}. Namely, I will next characterize the range of theories for which we can have a dynamics-first view of their spacetime manifold.

\chapter{Preliminary Assumptions and Example Theories}\label{ChapGenerality}
Chapters~\ref{ChapInternal} -~\ref{ChapSearch} will provide the mathematical details of the \hyperlink{ISE}{\bf ISE Methodology}. Before that, however, it will be useful to know up-front which kinds of spacetime theories can be either input or output by this methodology. This chapter will provide such a characterization. In Sec.~\ref{SecPrelim}, I will state the preliminary assumptions which a spacetime theory must satisfy in order to be a valid input to the internalization process. In Sec.~\ref{SecSKC}, I will state the spacetime-kinematic compatibility conditions which are guaranteed for any theory produced via externalization.\footnote{See Ch.~\ref{ChapExternal} for proof.}

For the sake of concreteness, and to showcase the breadth of spacetime theories considered in this dissertation, I will now introduce five example theories. The following five spacetime theories all satisfy the yet-to-be-introduced preliminary assumptions. However, only the first four satisfy the yet-to-be-discussed spacetime-kinematic compatibility conditions. Thus, while all five of these spacetime theories can be input to the ISE Methodology only the first four theories can be output by this methodology.\footnote{While, the fifth theory regarding \hyperlink{PMNPS}{\bf Periodic Matter on a Non-Periodic Spacetime} does not have spacetime-kinematic compatibility and so cannot be output by the \hyperlink{ISE}{\bf ISE Methodology}, a closely related spacetime theory can. Namely, the fourth theory regarding \hyperlink{MPS}{\bf Matter on a Periodic Spacetime} has spacetime-kinematic compatibility. A brief comparison of these two theories will be given in Ch.~\ref{ChapSearch}. These two spacetime theories are related by a process of kinematic reduction (see Appendix~\ref{AppKinRedSKC}).}

\begin{quote}
{\hypertarget{QKG}{\bf Quartic Klein Gordon - }} Consider a spacetime theory about a field, $\varphi_\text{old}:\mathcal{M}_\text{old}\to\mathcal{V}_\text{old}$, with $\mathcal{M}_\text{old}\cong\mathbb{R}^2$ and $\mathcal{V}_\text{old}\cong\mathbb{R}$. The field states are subject to the following kinematic constraint: In some fixed global coordinate system, $(t,x)$, the field, $\varphi_\text{old}$, must be smooth. In this coordinate system, the field states obey the following dynamics:
\begin{align}\label{QKG}
(\partial_t^2-\partial_x^2+M^2)\,\varphi_\text{old}
+\lambda \, \varphi_\text{old}^3 = 0,
\end{align}    
for some fixed mass parameter, $M\geq0$, and some self-coupling constant, $\lambda\in\mathbb{R}$.
\end{quote}

\begin{quote}
\hypertarget{SES}{{\bf Schr\"odinger Equation on a Sphere - }} Consider a spacetime theory about a field, $\varphi_\text{old}:\mathcal{M}_\text{old}\to\mathcal{V}_\text{old}$, with $\mathcal{M}_\text{old}\cong\mathbb{R}\times S^2$ and $\mathcal{V}_\text{old}\cong\mathbb{C}$. For this theory, there exists an injective map, $C:\mathcal{M}_\text{old}\to \mathbb{R}^4$, which assigns to each point, $p\in \mathcal{M}_\text{old}$, coordinates, $C(p)=(t,x,y,z)\in\mathbb{R}^4$, with \mbox{$x^2+y^2+z^2=1$}. The field states are subject to the following kinematic constraint: In this fixed coordinate system, the field, $\varphi_\text{old}$, must be second-differentiable. In this coordinate system, the field states obey the following dynamics:
\begin{align}
\ii\hbar\,\partial_{t}\varphi_\text{old} &= \frac{-\hbar^2}{2m} \, (L_x^2+L_y^2+L_z^2) \,\varphi_\text{old},
\end{align}
where $L_x$, $L_y$, and $L_z$ are the generators of rotations about the $x$, $y$, and $z$-axis respectively. 
\end{quote}

\begin{quote}
\hypertarget{HEDS}{{\bf Heat Equation on a Discrete Spacetime - }} Consider a spacetime theory about a field, $\varphi_\text{old}:\mathcal{M}_\text{old}\to\mathcal{V}_\text{old}$, with $\mathcal{M}_\text{old}\cong\mathbb{R}\times\mathbb{Z}$  and $\mathcal{V}_\text{old}\cong\mathbb{R}$. The field states are subject to the following kinematic constraint: In some fixed global coordinate system, $(\tau,n)$, at each $\tau$ the vector of field values, $\{\varphi_\text{old}(\tau,n)\}_{n\in\mathbb{Z}}$, must have a finite $L^2$-norm. In this coordinate system, the field states obeys the following dynamics:
\begin{align}
\partial_\tau\varphi(\tau,n)
=\alpha\,[\varphi(\tau,n-1)-2\varphi(\tau,n)+\varphi(\tau,n+1)],
\end{align}
for some $\alpha>0$.
\end{quote}

\begin{quote}
\hypertarget{MPS}{{\bf Matter on a Periodic Spacetime -}} Consider a spacetime theory about a field, $\varphi_\text{old}:\mathcal{M}_\text{old}\to\mathcal{V}_\text{old}$, with $\mathcal{M}_\text{old}\cong \mathbb{R}\times S^1$ and $\mathcal{V}_\text{old}\cong\mathbb{R}$. The field states are subject to no kinematic constraint. The dynamics of this theory is unspecified.
\end{quote}

\begin{quote}
\hypertarget{PMNPS}{{\bf Periodic Matter on a Non-Periodic Spacetime -}} Consider a spacetime theory about a field, $\varphi_\text{old}:\mathcal{M}_\text{old}\to\mathcal{V}_\text{old}$, with $\mathcal{M}_\text{old}\cong\mathbb{R}^2$ and $\mathcal{V}_\text{old}\cong\mathbb{R}$. The field states are subject to the following kinematic constraint: In some fixed global coordinate system, $(t,x)$, the field must be periodic in the $x$-coordinate with some fixed period, $L>0$, as,
\begin{align}
\varphi_\text{old}(t,x+L)=\varphi_\text{old}(t,x).
\end{align}
The dynamics of this theory is unspecified.
\end{quote}

\noindent As these examples show, the spacetime manifold of the theory under consideration is allowed to be curved\footnote{I here mean ``curved'' in the sense that if a metric were to be defined on this spacetime it would have to be curved (by virtue of the Gauss-Bonnet theorem). But spacetimes do not automatically come equipped with metrics. We need to study the theory's dynamics to discover its geometric and spatiotemporal structure.} or disconnected\footnote{In the \hyperlink{HEDS}{\bf Heat Equation on a Discrete Spacetime} example, space is discrete whereas time is continuous. Examples of this kind are discussed at length in ~\cite{DiscreteGenCovPart1}. For an example in which both space and time are discrete, see ~\cite{DiscreteGenCovPart2}.}. Indeed,  $\mathcal{M}_\text{old}$ could, in principle, be \textit{any} smooth manifold.\footnote{More carefully, spacetime theories can satisfy the preliminary assumptions discussed below while being set on any smooth manifold. That is, any smooth manifold can be input to the \hyperlink{ISE}{\bf ISE Methodology}. However, not every smooth manifold can be output by this methodology. In order to satisfy the spacetime-kinematic compatibility assumptions, the theory's spacetime must be homogeneous. See Sec.~\ref{SecSKC} for a definition of homogeneity. See Sec.~\ref{SecRevHomo} for a characterization of homogeneous manifolds.} These example theories also showcase a variety of kinematic constraints (e.g.: smoothness, second-differentiability, normalization, periodicity). Finally, these examples show that the dynamics of the theory in question could be either linear or non-linear.

\section{Preliminary Assumptions (Necessary for Internalization)}\label{SecPrelim}
The five above-discussed spacetime theories all satisfy the following preliminary assumptions.

\vspace{0.25cm}

\noindent {\hypertarget{Prelim}{\bf Preliminary Assumptions:}}
\begin{enumerate}
    \item[PA1)] The value space, $\mathcal{V}_\text{old}$, where the $\varphi_\text{old}$ takes its values must be a vector space over some field, $K$.\footnote{Note that there is no subscript ``old'' on the field, $K$, over which the value space, $\mathcal{V}_\text{old}$, is a vector space. The ISE Methodology will be able to move us into new spacetime settings with new value spaces, $\mathcal{V}_\text{new}\not\cong\mathcal{V}_\text{old}$. However, it is necessary that these values spaces, $\mathcal{V}_\text{new}$ and $\mathcal{V}_\text{old}$ are defined over the same field, $K$.} Let $V_\text{old}^\text{all}$ be the collection of all $\mathcal{V}_\text{old}$-valued functions definable over $\mathcal{M}_\text{old}$.\footnote{All definable functions here really means \textit{all} definable functions. For instance,  $V_\text{old}^\text{all}$ includes functions which are non-smooth and even discontinuous. If any such restrictions are needed (e.g., to state the theory's dynamics) these are to be implemented as kinematic constraints, see PA3 below.} Since addition and scalar multiplication of functions are carried out point-wise, we are thereby guaranteed that $V_\text{old}^\text{all}$ is also a vector space.
    \item[PA2)] There must be a way to ``lift'' diffeomorphisms to act linearly on $V_\text{old}^\text{all}$. That is, to any $d\in\text{Diff}(\mathcal{M})$ there should be an associated linear map, \mbox{$d^*:V_\text{old}^\text{all}\to V_\text{old}^\text{all}$}. Moreover, the $d\mapsto d^*$ map must be continuous. For instance, in a theory about a field, \mbox{$\varphi_\text{old}:\mathcal{M}_\text{old}\to\mathcal{V}_\text{old}$}, this can done by defining $(d^*\varphi_\text{old})(p)\coloneqq \varphi_\text{old}(d(p))$. Since addition and scalar multiplication of functions happen point-wise, we are guaranteed that $d^*$ is linear.
    \item[PA3)] The theory may additionally be subject to some kinematic constraints (e.g.: smoothness, periodicity, etc.). If so, then subset of functions which are deemed kinematically possible must also form a vector space, $V_\text{old}^\text{kin}\subset V_\text{old}^\text{all}$. These will be called the theory's kinematically allowed field states, $\varphi_\text{old}\in V_\text{old}^\text{kin}$.
\end{enumerate}
Satisfying these preliminary assumptions is necessary for a spacetime theory to be a valid input to the internalization process (and, therefore, a valid input to the ISE Methodology). These assumptions guarantee that the spacetime theory under consideration has, at minimum, a certain \textit{core vectorial structure}. Importantly, however, this core vectorial structure is the \textit{minimum} structure required for the internalization process. For instance, spacetime theories with some additional algebraic structure may also be acceptable inputs to the internalization process. Indeed, for theories with non-linear dynamics (e.g.,  the \hyperlink{QKG}{\bf Quartic Klein Gordon} theory) there \textit{must} be some additionally algebraic structure present in order to even state the theory's dynamics.

How, roughly, are these preliminary assumptions used in the internalization process? As I will discuss in Ch.~\ref{ChapInternal}, internalization divorces the dynamical behavior of matter from its topological underpinnings by reducing such a spacetime theory down to this vectorial core (or, if necessary, its algebraic core). The preliminary assumptions also guarantee that at least some topological information about the old spacetime setting survives internalization (e.g., certain diffeomorphisms, $d$, understood as linear maps, $d^*$).

These are the only three assumptions required for the internalization process to begin. Whenever these preliminary assumptions are violated, internalization cannot begin and hence this dissertation's methodology is simple not applicable. But when might these preliminary assumptions be violated? For a simple example, consider a spacetime theory with a value space \mbox{$\mathcal{V}=\{\text{full},\text{empty}\}$} indicating whether a given spacetime point is full or empty. Without a suitable notion of adding and rescaling these values, the value space, $\mathcal{V}_\text{old}$, would not be a vector space.\footnote{However, one could turn this into a vector space by endowing it with a notion of addition under which $\text{full}+\text{full}=\text{empty}$. So equipped, $\mathcal{V}_\text{old}$ would then be isomorphic to the one-dimensional vector space over the finite field $\mathbb{F}_2=\{0,1\}$. This is just a technical point however, sense any such notion of addition would be in contradiction with intuitive meanings of the words ``full'' and ``empty''.} Consequently, neither  $V_\text{old}^\text{all}$ nor $V_\text{old}^\text{kin}$ would be vector spaces.

\section{Spacetime-Kinematic Compatibility (Guaranteed by Externalization)}\label{SecSKC}
As I will prove in Ch.~\ref{ChapExternal}, all spacetime theories output by externalization are guaranteed to have  spacetime-kinematic compatibility (defined below). Recall from Ch.~\ref{ChapOverview} that proving the desired \hyperlink{TRF}{\bf Topological Regularity Fact} of a given spacetime theory requires us to be able to produce this theory via externalization (see \hyperlink{D2}{\bf Desiderata \#2}). Hence, spacetime-kinematic compatibility is a necessary condition for proving this topological regularity fact of a theory. Throughout the course of this dissertation, I will show that this condition is also sufficient; I will prove the  \hyperlink{TRF}{\bf Topological Regularity Fact} for all spacetime theories with spacetime-kinematic compatibility. Thus, we can prove the topological regularity fact of a spacetime theory if and only if it has spacetime-kinematic compatibility. Therefore, we can establish a dynamics-first view of a theory's spacetime manifold if and only if this theory has spacetime-kinematic compatibility.

Given this, in the main body of this dissertation I will restrict my attention to theories with spacetime-kinematic compatibility. Importantly, however, the \hyperlink{ISE}{\bf ISE Methodology} is applicable to theories which lack spacetime-kinematic compatibility (so long as they still satisfy the above-discussed preliminary assumptions). A discussion of the ISE Methodology which does not assume spacetime-kinematic compatibility is given in Appendix~\ref{ChapISEsansSKC}. As I will discuss there, theories with spacetime-kinematic compatibility are special within the ISE methodology. Namely, a theory's spacetime manifold will survive internalization faithfully and robustly if and only if it has spacetime-kinematic compatibility.\footnote{Spacetime-kinematic compatibility being sufficient for this sort of survival is proved in Ch.~\ref{ChapInternal}. The more difficult proof of necessity is provided in Appendix~\ref{AppRFSKC}.} This is closely related to the above-discussed fact that we can only establish a dynamics-first view of a theory's spacetime if and only if that theory has spacetime-kinematic compatibility.

I have claimed above that only the first four out of five example theories have spacetime-kinematic compatibility. Consequently, for all but the last theory we can establish a dynamics-first view of its spacetime manifold. But, more generally, what scope of spacetime theories have spacetime-kinematic compatibility? As I will now discuss, the following compatibility condition is very weak. Spacetime-kinematic compatibility is defined as follows (although this definition will require a bit of unpacking). 
\begin{quote}
{\hypertarget{SKC}{\bf Definition: Spacetime-Kinematic Compatibility - }} A spacetime theory has spacetime-kinematic compatibility just in case:
\begin{enumerate}
    \item it satisfies the above-discussed \hyperlink{Prelim}{\bf Preliminary Assumptions},
    \item it is kinematically reduced, and
    \item it is kinematically homogeneous.
\end{enumerate}
\end{quote}
The term ``kinematically reduced'' is defined as follows. For any spacetime theory satisfying the above-discussed preliminary assumptions, we can define,
\begin{align}
\label{HKinDef}
H_\text{kin}&\coloneqq\{ d\in\text{Diff}(\mathcal{M}_\text{old})\,\vert\,\forall \varphi_\text{old}\in V_\text{old}^\text{kin} \ \  d^*(\varphi_\text{old})=\varphi_\text{old}\},
\end{align}
to be the theory's group of \textit{kinematically trivial} diffeomorphisms. These are the diffeomorphisms which act trivially on all kinematically allowed states.  A spacetime theory will be called \textit{kinematically reduced} if it has $H_\text{kin}=\{\openone_{\mathcal{M}_\text{old}}\}$ being the trivial group.

Being kinematically reduced is a very weak assumption. As I will now discuss, all spacetime theories which are not kinematically reduced have a certain kind of kinematic redundancy. Namely, they have distinct spacetime points which are, in a certain sense, kinematically-identical. It follows from the definition of $H_\text{kin}$ that any spacetime points, $p_1,p_2\in\mathcal{M}_\text{old}$, which are related by some $k\in H_\text{kin}$ take the same values in all kinematically allowed states:
\begin{align}\label{EquivKin}
\big(\exists k\in H_\text{kin} \ \ k(p_1)=p_2\big) \Longrightarrow \big(\forall\varphi_\text{old}\in V_\text{old}^\text{kin} \ \  \varphi_\text{old}(p_1)=\varphi_\text{old}(p_2)\big).
\end{align}
If $H_\text{kin}\neq\{\openone_{\mathcal{M}_\text{old}}\}$ is non-trivial, then it acts non-trivially somewhere. That is, there exist distinct points $p_1\neq p_2$ related by some $k\in H_\text{kin}$. It then follows that there exist distinct but kinematically-identical spacetime points.

Note that the first four spacetime theories discussed above do not have any such distinct but kinematically-identical spacetime points. From this, it follows that these theories are all kinematically reduced. Indeed, every spacetime theory which lacks distinct but kinematically-identical spacetime points is kinematically reduced. By contrast, the \hyperlink{PMNPS}{\bf Periodic Matter on a Non-Periodic Spacetime} theory is not kinematically reduced. It has non-trivial diffeomorphisms (e.g., $d:(t,x)\mapsto (t,x+L)$) which act trivially on every kinematically allowed state. Hence, this theory has $H_\text{kin}\neq\{\openone_{\mathcal{M}_\text{old}}\}$. 

Note that, as is guaranteed by Eq.~\eqref{EquivKin}, the \hyperlink{PMNPS}{\bf Periodic Matter on a Non-Periodic Spacetime} theory has distinct but kinematically-identical spacetime points (e.g., $(t,x)$ and $(t,x+L)$). One can make this theory kinematically reduced by taking its quotient over these kinematically-identical spacetime points. Doing so does, in fact, yields a kinematically reduced theory. Namely, it yields the \hyperlink{MPS}{\bf Matter on a Periodic Spacetime} theory. In general, for any spacetime theory satisfying the above-discussed preliminary assumptions, we can define a process of kinematic reduction which takes in our old spacetime theory and returns a new spacetime theory which is kinematically reduced. See Appendix~\ref{AppKinRedSKC} for details.

To summarize: Assuming that our old spacetime theory is kinematically reduced is a very weak assumption. Theories which are not kinematically reduced have spacetimes which are in excess of the theory's kinematics. This excessive spacetime can be easily removed by a process of kinematic reduction. The more substantial part of spacetime-kinematic compatibility is the assumption of kinematic homogeneity.

I will now define the term ``kinematically homogeneous'' by first defining a non-kinematic notion of homogeneity.\footnote{This non-kinematic notion of homogeneity will be discussed at length in Sec.~\ref{SecRevHomo}.} A smooth manifold, $\mathcal{M}$, is said to be \textit{homogeneous} if there exists a finite-dimensional Lie group, $G$, which acts smoothly and transitively on it.\footnote{The restriction to finite-dimensional Lie groups is not essential for any of the results of this dissertation. Moreover, allowing for infinite-dimensional Lie groups here would substantially increase the scope of theories under consideration (see footnote \ref{FnInfLie}). However, infinite-dimensional Lie groups are notoriously tricky to handle mathematically. For simplicity, I will therefore adopt this finite-dimensional definition of homogeneity.} A group action, $\theta:G\times \mathcal{M}\to \mathcal{M}$, is said to act transitively on $\mathcal{M}$ when for every pair of points, $p,q\in \mathcal{M}$, there is some $g\in G$ with $\theta(g,p)=q$. That is, a group action is transitive when it can map any point anywhere.

The intuition behind this definition of a homogeneity is that all points on such a manifold, $\mathcal{M}$, are equivalent from the perspective of the Lie group, $G$: Concretely, any point $q\in \mathcal{M}$ can be replaced by any other point $p\in \mathcal{M}$ via action by some $g\in G$. For example, $\mathbb{R}^2$ is homogeneous under rigid translations. Similarly, $S^2$ is homogeneous under rigid rotations. For an example of a non-homogeneous manifold consider $\mathcal{M}\cong \mathbb{T}^2\,\dot{\cup}\,\mathbb{R}^2$, i.e., the disjoint union of a torus and a two-dimensional plane. No smooth transformation can map a torus-point to a plane-point. This manifold is non-homogeneous because it contains two types of points (i.e., torus-points and plane-points) which cannot be related by any smooth map.\footnote{Another example of a non-homogeneous manifold is $\mathcal{M}\cong \mathbb{T}^2\#\mathbb{T}^2$, the double torus. This manifold being non-homogeneous hinges on us restricting our attention to finite-dimensional Lie groups. All Lie groups which act smoothly and transitively over the double torus are infinite-dimensional. This follows from \cite{GMostow2005}'s proof that all smooth connected manifolds with smoothly and transitively-acting finite-dimensional Lie groups have Euler characteristic, $\chi\geq0$. Note that the double torus has $\chi=-2$. Suppose, however, that we were to widen the definition of smooth homogeneous manifolds such that the smoothly and transitively-acting Lie group can be infinite-dimensional. As I will now discuss, the double torus, $\mathcal{M}\cong \mathbb{T}^2\#\mathbb{T}^2$, would then be homogeneous. By contrast, $\mathcal{M}\cong \mathbb{T}^2\,\dot{\cup}\,\mathbb{R}^2$ would still be non-homogeneous, as would every other manifold built from the disjoint union of non-diffeomorphic manifold. Notably, under this relaxed definition of homogeneity these are all of the non-homogeneous smooth manifolds in existence. Said differently, under the relaxed definition, the homogeneous smooth manifolds are exactly the smooth manifolds whose connected parts are all pair-wise diffeomorphic. In such cases, the infinite-dimensional Lie group of diffeomorphisms, $\text{Diff}(\mathcal{M})$, acts smoothly and transitively over $\mathcal{M}$. If I were to adopt this relaxed definition of homogeneity, the scope of spacetime theories which can support the DFSM view would substantially increase. The restriction to finite-dimensional Lie groups is not essential for any of the results of this dissertation. Infinite-dimensional Lie groups, however, are notoriously tricky to handle mathematically. For simplicity, I will continue on with the finite-dimensional definition of homogeneity.\label{FnInfLie}}

Spacetime-kinematic compatibility requires not only that the spacetime manifold, $\mathcal{M}_\text{old}$, be homogeneous, but also that the spacetime theory be \textit{kinematically} homogeneous. Before defining what this means, another definition is needed. For any spacetime theory satisfying the above-discussed preliminary assumptions, we can define,
\begin{align}
\label{DiffKinDef}
\text{Diff}_\text{kin}(\mathcal{M}_\text{old})&\coloneqq\{ d\in\text{Diff}(\mathcal{M}_\text{old})\,\vert\,\forall \varphi_\text{old}\in V_\text{old}^\text{kin} \ \  d^*(\varphi_\text{old})\in V_\text{old}^\text{kin}\},
\end{align}
to be the theory's group of \textit{kinematically allowed} diffeomorphisms. These are the diffeomorphisms which map kinematically allowed states onto kinematically allowed states. A spacetime theory is \textit{kinematically homogeneous} if there exists a finite-dimensional Lie group of diffeomorphisms, $H_\text{trans}\subset\text{Diff}_\text{kin}(\mathcal{M}_\text{old}),$ which: Firstly, acts transitively on $\mathcal{M}_\text{old}$; And secondly, is kinematically allowed.

All five spacetime theories discussed above are kinematically homogeneous under the obvious rigid translations. For an example of a spacetime theory without kinematic homogeneity, consider the following theory. Consider a spacetime theory set on $\mathcal{M}_\text{old}\cong \mathbb{R}^2\,\dot{\cup}\,\mathbb{R}^2$, i.e., the disjoint union of two two-dimensional planes. On one of these planes (but not the other) the theory's matter content is kinematically constrained to be periodic in both space and time with fixed periods $T$ and $L$ in some fixed coordinate system, $(t,x)$. This example helps demonstrate the difference between a spacetime manifold being homogeneous and a spacetime theory being kinematically homogeneous. Namely, despite being set on a homogeneous manifold, this theory is not kinematically homogeneous. 

To see that $\mathcal{M}_\text{old}\cong \mathbb{R}^2\,\dot{\cup}\,\mathbb{R}^2$ is homogeneous, let $H\subset\text{Diff}(\mathcal{M}_\text{old})$ be the rigid translation of each plane in the $(t,x)$ coordinate system  together with a diffeomorphism which swaps the two planes. Note that this Lie group can map any point anywhere, i.e., it acts transitively. Note, however, that if we restrict our attention to kinematically allowed diffeomorphisms, $\text{Diff}_\text{kin}(\mathcal{M}_\text{old})$, the spacetime becomes non-homogeneous. Indeed, every diffeomorphism which swaps the two planes is kinematically disallowed: Any such diffeomorphism maps some kinematically allowed state onto a kinematically disallowed state. Concretely, consider a state which is periodic on the periodic plane and non-periodic on the non-periodic plane. This state is kinematically allowed. If, however, we swap the two planes it becomes kinematically disallowed. Hence, not all points on this theory's spacetime manifold are equivalent from the perspective of $\text{Diff}_\text{kin}(\mathcal{M}_\text{old})$ (or any subgroup thereof). Indeed, in this theory there are two distinct kinds of spacetime points: those supporting periodic matter, and those supporting non-periodic matter. While this theory is set on a homogeneous manifold, it is not kinematically homogeneous.

The next three chapters will provide the mathematical details of the Internalization, Searching, and Externalization steps of the \hyperlink{ISE}{\bf ISE Methodology}, respectively. Moreover, they will discuss how this methodology satisfies \hyperlink{D1}{\bf Desiderata \#1}, \hyperlink{D2}{\bf \#2}, and \hyperlink{D3}{\bf \#3} respectively.

\chapter{Internalization and Desiderata \#1}\label{ChapInternal}
This chapter will give a complete description of the internalization process. To begin, in Sec.~\ref{SecInternalize} I will discuss how the internalization process can be applied to any spacetime theory satisfying the \hyperlink{Prelim}{\bf Preliminary Assumptions} discussed in the previous chapter. This will be followed in Sec.~\ref{SecSurvivingDynamics} by a discussion of how the theory's dynamics can survive internalization. Namely, I will demonstrate how internalization conceptually divorces the dynamical behavior of matter from any assumed topological underpinnings (as required by \hyperlink{D1}{\bf Desiderata \#1}). Importantly, however, this does not mean that internalization simply deletes all of a theory's topological information. The rest of this chapter will be spent demonstrating how a potentially significant amount of topological information about the theory's old spacetime setting survives internalization. 

Past Sec.~\ref{SecSurvivingTrans} (in all the sections describing the survival of topological information) it will be assumed that the spacetime theory being internalized has \hyperlink{SKC}{\bf Spacetime-Kinematic Compatibility}. It is important to note, however, that spacetime-kinematic compatibility is not required for the internalization process to proceed. For a discussion of the internalization process which does not assume spacetime-kinematic compatibility, see Appendix~\ref{ChapISEsansSKC}. This chapter, however, proceeds with this assumption. From this assumption it will be proved that the old theory's spacetime manifold survives internalization faithfully and robustly (definitions to come). The converse claim is proved in Appendix~\ref{AppRFSKC}. Thus, the spacetime theories whose manifold survive internalization faithfully and robustly are exactly those with spacetime-kinematic compatibility.

\section{The Internalization Process}\label{SecInternalize}
Recall from Ch.~\ref{ChapOverview} that the goal of internalization is to divorce the dynamical behavior of matter from any assumed topological substructure (see \hyperlink{D1}{\bf Desiderata \#1}). Recall also that our preliminary assumptions have guaranteed that our theory regarding the field $\varphi_\text{old}:\mathcal{M}_\text{old}\to\mathcal{V}_\text{old}$ has, at minimum, some core vectorial structure. To begin, therefore, I will demonstrate how any such theory can be reduced down to this vectorial core.

However, it is important to recall from Ch.~\ref{ChapGenerality} that the vectorial core guaranteed by our preliminary assumptions is really just the minimum structure necessary for internalization. Past this, our theory may also be equipped with additional structure, e.g., a norm or some algebraic structure. As I mentioned in Ch.~\ref{ChapGenerality}, for theories with non-linear dynamics (e.g., the \hyperlink{QKG}{\bf Quartic Klein Gordon} theory) some additional algebraic structure is needed in order to even state the theory's dynamics. In Sec.~\ref{SecSurvivingDynamics}, I will discuss briefly how the vector-based internalization process described below can be relaxed so as to also preserve these algebraic structures. For now, however, let us proceed considering theories with only this minimal vectorial structure.

Internalization is a two-step process. The first step is to reconceptualize the field state, $\varphi_\text{old}$, as a vector in some vector space. This is possible due to our preliminary assumptions, specifically PA1 and PA3. Namely, we have $\varphi_\text{old}\in V_\text{old}^\text{all}$ being in the vector space of all $\mathcal{V}_\text{old}$-values functions definable on $\mathcal{M}_\text{old}$. We also have $\varphi_\text{old}\in V_\text{old}^\text{kin}$ being in the vector space of kinematically allowed field states on $\mathcal{M}_\text{old}$.

The second step is to remove from this vector space, $V_\text{old}^\text{kin}$, any lingering attachments to the old spacetime setting. Note that while \mbox{$V_\text{old}^\text{kin}$} is a vector space, it is also much more than that. For instance, its elements \mbox{$\varphi_\text{old}\in V_\text{old}^\text{kin}$} can be evaluated at spacetime points, \mbox{$p\in\mathcal{M}_\text{old}$}. That is, $V_\text{old}^\text{kin}$ is a vector space \textit{of functions on} $\mathcal{M}_\text{old}$. In order to remove this non-vectorial structure, we applying what category theory calls a forgetful functor, $\mathcal{F}_\text{vec}$. Namely, this is a vector space isomorphism \mbox{$\mathcal{F}_\text{vec}:V_\text{old}^\text{kin}\to V_\text{neutral}$} where $V_\text{neutral}$ is some abstract vector space which is isomorphic to $V_\text{old}^\text{kin}$, namely, \mbox{$V_\text{old}^\text{kin}\cong V_\text{neutral}$}. 

But what exactly does it mean for $V_\text{neutral}$ to be an abstract vector space? This means that $V_\text{neutral}$ is \textit{only} a vector space with no additional structure. Said differently, this means that we can do with its elements $\bm{\varphi}\in V_\text{neutral}$ only those things which we could do with the elements of any other vector space. Namely, its elements can be added and rescaled (but not evaluated at spacetime points). Using these operations, we can define linear maps between $V_\text{neutral}$ and other vector spaces. That is all.\footnote{As mentioned above, in Sec.~\ref{SecSurvivingDynamics} I will discuss how some extra algebraic structure can be preserved post-internalization. So extended, internalization results in an abstract algebraic space where only strictly algebraic operations are defined.}

What objects from our old theory survive internalization? Firstly, any kinematically allowed state, $\varphi_\text{old}\in V_\text{old}^\text{kin}$, survives as an abstract vector, \mbox{$\bm{\varphi}\coloneqq\mathcal{F}_\text{vec}(\varphi_\text{old})\in V_\text{neutral}$}. Moreover, any linear map, \mbox{$L_\text{old}:V_\text{old}^\text{kin}\to V_\text{old}^\text{kin}$}, survives internalization as a linear map on $V_\text{neutral}$, namely as \mbox{$L_\text{neutral}\coloneqq\mathcal{F}_\text{vec}\circ L_\text{old}\circ\mathcal{F}_\text{vec}^{-1}$}. Since, $V_\text{neutral}$ is merely an abstract vector space, this is all that has survived: abstract linear maps, and abstract vectors.\footnote{As mentioned in the previous footnote, this can be relaxed if needed to allow for some algebraic structure to survives as well.}

In the next section, I will discuss how a theory's dynamical equations can survive internalization (even non-linear dynamics). Before this, however, allow me to introduce some notation. Throughout the rest of this dissertation I will use the following notation regarding which mathematical objects survive internalization becoming which other object. Vectors, $\varphi_\text{old}\in V_\text{old}^\text{kin}$, survive internalization as follows:
\begin{align}
\varphi_\text{old}\surint 
\bm{\varphi}\coloneqq\mathcal{F}_\text{vec}(\varphi_\text{old}) 
\end{align}
This notation, however, is not restricted to vectors. Linear maps, \mbox{$L_\text{old}:V_\text{old}^\text{kin}\to V_\text{old}^\text{kin}$}, survive internalization as follows:
\begin{align}
L_\text{old}\surint L_\text{neutral}\coloneqq\mathcal{F}_\text{vec}\circ L_\text{old}\circ\mathcal{F}_\text{vec}^{-1}
\end{align}
Indeed, this notation can also be used for vector spaces as well as groups represented on vector spaces as follows:
\begin{align}
V_\text{old}^\text{kin}&\surint
V_\text{neutral}\coloneqq\mathcal{F}_\text{vec}(V_\text{old}^\text{kin})\\
\text{GL}(V_\text{old}^\text{kin})&\surint 
\text{GL}(V_\text{neutral}).
\end{align}
The use cases of this notation will be further expanded throughout this chapter. It is important to note that in order for this notation to be applicable, the mathematical objects on either side of the arrow must be of the same type (e.g., both vector spaces, both Lie groups, both smooth manifolds, etc.). However, just because something survives internalization does not automatically mean that it does so faithfully. That is, the object appearing on the right-hand-side of one of these arrows may not be isomorphic to the object on the left-hand-side. With this notation established, let us now see how a spacetime theory's dynamics can survive internalization.

\section{Surviving Internalization: Non-Linear Dynamics}\label{SecSurvivingDynamics}
To begin, let us consider cases in which the theory's dynamical equations are linear (i.e, involving only linear operators applied to the field state, $\varphi_\text{old}\in V_\text{old}^\text{kin}$). A theory's dynamics being linear is equivalent to its space of dynamically allowed states, $V_\text{old}^\text{dyn}$, being a vector space. The dynamics of such theories survive internalization because we have that,
\begin{align}
V_\text{old}^\text{dyn}&\surint
\mathcal{F}_\text{vec}(V_\text{old}^\text{dyn}).
\end{align}
Note that because $\mathcal{F}_\text{vec}$ is a vector space isomorphism, the dynamics survives faithfully.

Before moving on to consider non-linear dynamics, let us briefly discuss an explicit example of linear dynamics surviving internalization. Consider the \hyperlink{QKG}{\bf Quartic Klein Gordon} but with $\lambda=0$ such that its dynamics simplifies to the linear equation,
\begin{align}
(\partial_t^2-\partial_x^2+M^2)\,\varphi_\text{old} = 0,
\end{align}    
for some fixed mass parameter, $M\geq0$. We have each part of this equation surviving internalization as follows:
\begin{align}
\partial_t&\surint& D_0&\coloneqq\mathcal{F}_\text{vec}\circ \partial_t\circ\mathcal{F}_\text{vec}^{-1}\\
\nonumber
\partial_x&\surint& D_1&\coloneqq\mathcal{F}_\text{vec}\circ \partial_x\circ\mathcal{F}_\text{vec}^{-1}\\
\nonumber
M^2\hat\openone_\text{old}&\surint& M^2\hat\openone_\text{neutral}&=\mathcal{F}_\text{vec}\circ M^2\hat\openone_\text{old} \circ\mathcal{F}_\text{vec}^{-1},\\
\nonumber
\varphi_\text{old}&\surint& 
\bm{\varphi}&\coloneqq\mathcal{F}_\text{vec}(\varphi_\text{old})
\end{align}
Correspondingly, we can say that the dynamical equation itself survives internalization as follows:
\begin{align}
(\partial_t^2-\partial_x^2+M^2)\,\varphi_\text{old} = 0
\surint
(D_0^2-D_1^2+M^2)\,\bm{\varphi}=0.
\end{align}
In an analogous way, the dynamics of any linear theory survives internalization.

But what about theories with non-linear dynamics? If we instead consider the 
\hyperlink{QKG}{Quartic Klein Gordon} theory with $\lambda\neq0$ then we have the following dynamics,
\begin{align}
(\partial_t^2-\partial_x^2+M^2)\,\varphi_\text{old}
+\lambda \, \varphi_\text{old}^3 = 0.
\end{align}
For this theory's dynamics to survive internalization we need the non-linear term, $\lambda \, \varphi_\text{old}^3$, to somehow survive internalization.

In general, non-linear dynamics involve the product of fields: e.g.,  $\varphi_\text{old}*_\text{old}\phi_\text{old}$ for some well-behaved product operation, $*_\text{old}$. For any such product operation, the forgetful functor, $\mathcal{F}_\text{vec}:V_\text{old}^\text{kin}\to V_\text{neutral}$, naturally induces the following product operation on a $V_\text{neutral}$:
\begin{align}
\bm{\varphi}*_\text{neutral}\bm{\phi}
\coloneqq \mathcal{F}_\text{vec}(\mathcal{F}_\text{vec}^{-1}(\bm{\varphi})*_\text{old}\mathcal{F}_\text{vec}^{-1}(\bm{\phi}))
\end{align}
where $\bm{\varphi},\bm{\phi}\in V_\text{neutral}$ are generic vectors in the $V_\text{neutral}$. Indeed, $*_\text{neutral}$ is a valid product operation defined on $V_\text{neutral}$ which could be used in writing down the theory's spacetime-neutral dynamics.

There is a slight problem however. We cannot equip $V_\text{neutral}$ with this product operation because, by definition, $V_\text{neutral}$ is an abstract vector space on which only linear operations are defined. In order to fix this issue we would need to upgrade the vector space $V_\text{neutral}$ to some algebra, $\mathcal{A}_\text{neutral}$. We can define this $\mathcal{A}_\text{neutral}$ as follows. First we upgrade $V_\text{old}^\text{kin}$ to an algebra as $\mathcal{A}_\text{old}^\text{kin}=\langle V_\text{old}^\text{kin},*_\text{old}\rangle$ by equipping it with the product, $*_\text{old}$. We can then take $\mathcal{A}_\text{neutral}$ to be any abstract algebraic space related to $\mathcal{A}_\text{old}^\text{kin}$ via some forgetful functor, $\mathcal{F}_\text{alg}:\mathcal{A}_\text{old}^\text{kin}\to\mathcal{A}_\text{neutral}$. This abstract algebraic space, $\mathcal{A}_\text{neutral}$, can now be equipped with the induced product, $*_\text{neutral}$. Following these upgrades, it is now trivial to transfer the dynamics of a non-linear theory from $\mathcal{A}_\text{old}^\text{kin}$ to $\mathcal{A}_\text{neutral}$ using $\mathcal{F}_\text{alg}$. Concretely, for the \hyperlink{QKG}{Quartic Klein Gordon} theory we have that,
\begin{align}
\varphi_\text{old}*_\text{old}\phi_\text{old}
\surint 
\mathcal{F}_\text{vec}(\varphi_\text{old})*_\text{neutral}\mathcal{F}_\text{vec}(\phi_\text{old})\coloneqq\mathcal{F}_\text{vec}(\varphi_\text{old}*_\text{old}\phi_\text{old}),
\end{align}
and hence that the dynamics survives internalization as follows:
\begin{align}
\nonumber
(\partial_t^2-\partial_x^2+M^2)\,\varphi_\text{old}
+\lambda \, \varphi_\text{old}^3 = 0
\surint
(D_0^2-D_1^2+M^2)\,\bm{\varphi}
+\lambda \, \bm{\varphi}^3 = 0,
\end{align}
where post-internalization the notation $\bm{\varphi}^3$ is short-hand for $\bm{\varphi}^3=\bm{\varphi}*_\text{neutral}\bm{\varphi}*_\text{neutral}\bm{\varphi}$.

Thus, while I have above described the internalization as being mediated by a forgetful functor, $\mathcal{F}_\text{vec}$, which only remembers vectorial structure, in order to accommodate non-linear dynamics we will instead need to use a forgetful functor, $\mathcal{F}_\text{alg}$, which remembers only algebraic structure. 

While the inclusion of such algebraic structures on top of the theory's vectorial core is allowed by the \hyperlink{ISE}{ISE Methodology}, it is \textit{mechanically irrelevant} throughout both the internalization and externalization processes. This is the case despite the fact that this additional algebraic structure may be necessary in order to state the theory's dynamics. In fact, the dynamics of the theory in question (along with any supporting algebraic structures) are both mechanically irrelevant throughout the internalization and externalization steps. In both of these steps, the dynamics and its supporting algebra merely come along for the ride, so to speak. It is only in the middle step of searching for good pre-spacetime translation operations that the theory's dynamics (and its supporting algebra) plays any role.

To be clear, the theory's dynamics can play a significant role in helping us pick good pre-spacetime translation operations to externalize.\footnote{See Ch.~\ref{ChapSearch} for concrete examples of the role that the dynamics plays in this middle searching step.} Consequently, if the theory is non-linear, then its supporting algebraic structures will be important in helping us choose between different possible spacetime settings. Thus, while it is true that the dynamics and its algebraic supports simply come along for the ride, so to speak, they ought to be seen as a mission-critical navigator rather than a mere passenger. Importantly, however, once our course has been set towards a new spacetime setting neither the theory's dynamics nor its supporting algebraic structures play any mechanical role in producing the new spacetime manifold. They are never in the driver's seat, so to speak.

This and the next chapter are focused on describing the mechanical details of internalization and externalization. For simplicity, I will there focus on theory's with only the minimal structure which is mechanically necessary to carry out this methodology (i.e., only the core vectorial structure guaranteed by the preliminary assumptions discussed in Ch.~\ref{ChapGenerality}). One brief comment will be made in Ch.~\ref{ChapExternal} regarding how additional algebraic structure is to be accommodated in the externalization process.\footnote{Specifically, see footnote \ref{FnExtAlg} in Ch.~\ref{ChapExternal}.} The theory's dynamics will then re-enter our discussion in Ch.~\ref{ChapSearch} where I discuss how it can help us negotiate between different inequivalent spacetime settings. Before that, let us investigate what topological information about the old spacetime setting survives internalization.

\section{Surviving Internalization: Spacetime Translations}\label{SecSurvivingTrans}
A generic spacetime translation operation on $\mathcal{M}_\text{old}$ maps points to points in a smooth way. That is, spacetime translations on $\mathcal{M}_\text{old}$ are just diffeomorphisms, $d\in\text{Diff}(\mathcal{M}_\text{old})$. Note that diffeomorphisms act on the field states in a linear way,
\begin{align}
d^*(\varphi_1+\varphi_2)
=d^*\varphi_1+d^*\varphi_2,
\end{align}
where $d^*(\varphi)\coloneqq\varphi\circ d$. This happens ultimately because the addition and rescaling of fields occurs point-wise. Thus, any diffeomorphism, $d$, can be seen as a linear transformation, $d^*$, acting on \mbox{$V_\text{old}^\text{all}$}, the vector space of all $\mathcal{V}_\text{old}$-valued functions definable on $\mathcal{M}_\text{old}$. The fact that diffeomorphisms can be understood to act linearly on field states is an essential part of this dissertation's methodology. Indeed, this is one of the {\hyperlink{Prelim}{\bf Preliminary Assumptions}} discussed in Sec.~\ref{SecPrelim}, specifically PA2.

As I will now discuss, only a subset of diffeomorphisms survive internalization since only the kinematically allowed subset of field states, $\varphi_\text{old}\in V_\text{old}^\text{kin}\subset V_\text{old}^\text{all}$, survive internalization. Correspondingly, only kinematically allowed diffeomorphism, $d\in \text{Diff}_\text{kin}(\mathcal{M}_\text{old})$, survives internalization.\footnote{Recall from Sec.~\ref{SecSKC} that the kinematically allowed diffeomorphism, $\text{Diff}_\text{kin}(\mathcal{M}_\text{old})$, are those that map the theory's kinematically allowed states onto each other.} They do so as follows:
\begin{align}\label{DIntMap}
d^*\surint G(d)\coloneqq\mathcal{F}_\text{vec}\circ d^*\vert_\text{kin}\circ\mathcal{F}_\text{vec}^{-1}.
\end{align}
Here \mbox{$d^*\vert_\text{kin}:V_\text{old}^\text{kin}\to V_\text{old}^\text{kin}$} is the map, \mbox{$d^*:V_\text{old}^\text{all}\to V_\text{old}^\text{all}$}, restricted act on $V_\text{old}^\text{kin}$.  

Importantly, however, it does not immediately follow from all elements of $\text{Diff}_\text{kin}(\mathcal{M}_\text{old})$ surviving internalization that the Lie group itself will survive internalization. For the Lie group, $\text{Diff}_\text{kin}(\mathcal{M}_\text{old})$, to survive internalization, it must do so \textit{as a Lie group}. That is,
\begin{align}\label{Gold}
G_\text{old}\coloneqq G(\text{Diff}_\text{kin}(\mathcal{M}_\text{old}))=\{\mathcal{F}_\text{vec}\circ d^*\vert_\text{kin}\circ\mathcal{F}_\text{vec}^{-1} \ \vert \  d\in \text{Diff}_\text{kin}(\mathcal{M}_\text{old})\},
\end{align}
must be a Lie group. Specifically, the map $G:\text{Diff}_\text{kin}(\mathcal{M}_\text{old})\to G_\text{old}$ must be a Lie group homomorphism. As I will now show, this is the case given spacetime-kinematic compatibility.\footnote{This assumption of spacetime-kinematic compatibility is relaxed in Appendix~\ref{ChapISEsansSKC}. I there prove that for any spacetime theory satisfying the \hyperlink{Prelim}{\bf Preliminary Assumptions}, the map $G:\text{Diff}_\text{kin}(\mathcal{M}_\text{old})\to G_\text{old}$ is a Lie group homomorphism. This follows from the fact that $\text{ker}(G)=H_\text{kin}$ is a normal and closed subgroup of $\text{Diff}_\text{kin}(\mathcal{M}_\text{old})$ even when $H_\text{kin}\neq \{\openone_{\mathcal{M}_\text{old}}\}$.}

It follows from the Homomorphism Theorem of group theory that $G$ will be a group homomorphism if and only if $\text{ker}(G)$ is a normal subgroup of $\text{Diff}_\text{kin}(\mathcal{M}_\text{old})$. Moreover, $G$ will be a group isomorphism if and only if 
$\text{ker}(G)=\{\openone_{\mathcal{M}_\text{old}}\}$ is the trivial group. For $G$ to be a Lie group homomorphism/isomorphism, we must additionally demand that $\text{ker}(G)$ is a closed subgroup of $\text{Diff}_\text{kin}(\mathcal{M}_\text{old})$. 

Notice that the kernel of the $G$ map is just $\text{ker}(G)=H_\text{kin}$, the kinematically trivially group defined in Eq.~\eqref{HKinDef}. If we here enforce our \hyperlink{SKC}{\bf Spacetime-Kinematic Compatibility} conditions then we have $H_\text{kin}=\{\openone_{\mathcal{M}_\text{old}}\}$ by our assumption that the old theory is kinematic reduced. The trivial group is a closed and normal subgroup of every Lie group. Hence, for all kinematically reduced spacetime theories, we have that the Lie group $\text{Diff}_\text{kin}(\mathcal{M}_\text{old})$ not only survives internalization but does so faithfully:
\begin{align}
\text{Diff}_\text{kin}(\mathcal{M}_\text{old})\surint G_\text{old}\cong \text{Diff}_\text{kin}(\mathcal{M}_\text{old}).
\end{align}
Moreover, we have every Lie subgroup thereof, $H\subset\text{Diff}_\text{kin}(\mathcal{M}_\text{old})$, surviving internalization faithfully as,
\begin{align}
H\surint G(H)\cong H,
\end{align}
for some Lie group, $G(H)$, represented on $V_\text{neutral}$.

\vspace{0.5cm}
To review: The previous three sections have introduced the internalization process which serves to conceptually divorces the dynamical behavior of matter from any assumed topological underpinnings (in line with \hyperlink{D1}{\bf Desiderata \#1}). It does so by reducing the theory down to its vectorial core (or, if necessary, its algebraic core) via a forgetful functor. As the above discussion has shown, assuming spacetime-kinematic compatibility, some diffeomorphisms (i.e., $\text{Diff}_\text{kin}(\mathcal{M}_\text{old})$) survive internalization faithfully. Hence, at least some topological information regarding the old spacetime setting, $\mathcal{M}_\text{old}$, survives internalization. The full significance of this surviving topological information will be developed in Sec.~\ref{SecSurvivingMan}. 

In light of these results, our next steps can be motivated by reflecting on \hyperlink{D2}{\bf Desiderata \#2 (D2)}: For any spacetime theory producible via externalization, after re-internalizing this theory we should always be able to then trivially re-externalize it. This chapter has shown that topological information about $\mathcal{M}_\text{old}$ can survive internalization (namely, kinematically allowed diffeomorphisms, $d\in\text{Diff}_\text{kin}(\mathcal{M}_\text{old})$). Analogously, such information about $\mathcal{M}_\text{new}$ could survive re-internalization (namely, kinematically allowed diffeomorphisms, $d\in\text{Diff}_\text{kin}(\mathcal{M}_\text{new})$). \hyperlink{D2}{\bf D2} essentially demands that $\mathcal{M}_\text{new}$ be recoverable from these surviving diffeomorphisms. This motivates us to investigate in general how smooth manifolds can be reconstructed from their diffeomorphisms. As the next section will discuss, this is possible for what mathematicians call homogeneous manifolds.

\section{Mathematical Interlude: Homogeneous Manifolds}\label{SecRevHomo}
The conditions for a smooth manifold to be considered \textit{homogeneous} have already been given in Sec.~\ref{SecSKC}: A smooth manifold, $\mathcal{M}$, is homogeneous if there exists a finite-dimensional Lie group, $G$, which acts smoothly and transitively over it.\footnote{The restriction to finite-dimensional Lie groups is not an essential part of the DFSM view. It is only imposed here for mathematical simplicity. See footnote \ref{FnInfLie} in Ch.~\ref{ChapGenerality}.} A group, $G$, acts transitively on $\mathcal{M}$ when for any two points, $p,q\in\mathcal{M}$, some $g\in G$ maps $p$ to $q$ as $\theta(g,p)=q$. The following mathematical results regarding homogeneous manifolds are of central importance to the DFSM view. 

This section will first show that the quotient of any two Lie groups is a homogeneous manifold. Following this, I will show that any homogeneous manifold is diffeomorphic to a quotient of two Lie groups. In particular, I will show how any smooth homogeneous manifold can be reconstructed up to diffeomorphism as a quotient of its diffeomorphisms. A concrete demonstration of this reconstruction will be given at the end of this section for the 2-sphere, $\mathcal{M}\cong S^2$.

\subsection{\hra Constructing Homogeneous Spaces}
Given any finite-dimensional Lie group, $G_\text{trans}$, and closed Lie subgroup thereof,
$G_\text{fix}\subset G_\text{trans}$, one can build a homogeneous manifold as follows.\footnote{These two Lie groups could come from anywhere. For instance, in Ch.~\ref{ChapExternal} these will be Lie groups represented on $V_\text{neutral}$, the spacetime-neutral vector space of kinematic possibilities. Namely, we will then have $G_\text{fix}\subset G_\text{trans}\subset \text{GL}(V_\text{neutral})$. Importantly, at this point, we do not need to have in mind any manifold upon which $G_\text{trans}$ and $G_\text{fix}$ will act. Indeed, in what follows we will build a manifold from them on which $G_\text{trans}$ will then act smoothly and transitively.} Taking the quotient of these Lie groups yields the following quotient space,
\begin{align}\label{MGDef0}
\mathcal{M}_\text{G}\coloneqq G_\text{trans}/G_\text{fix}.
\end{align}
The equivalence classes under this quotient, $[g]=g\,G_\text{fix}$, are $G_\text{fix}$ cosets. Note that the quotient between these Lie groups is, in general, not itself a Lie group. This is because the denominator, $G_\text{fix}$, may not be a normal subgroup of the numerator, $G_\text{trans}$. 

The quotient is, however, always a smooth manifold. In particular, this quotient space naturally adopts a smooth structure from the Lie group in the numerator, $G_\text{trans}$, via its quotient with $G_\text{fix}$. It follows from the Quotient Manifold Theorem that there is a unique smooth structure for $\mathcal{M}_\text{G}$ such that the quotient map, $\pi:g\mapsto [g]$, is a smooth submersion (see Theorem 1 of \cite{Zikidis}). In terms of diffeomorphisms on $\mathcal{M}_\text{G}$ this means the following: A transformation, \mbox{$d_\text{post}:\mathcal{M}_\text{G}\to \mathcal{M}_\text{G}$}, is smooth on $\mathcal{M}_\text{G}$ if and only if there exists a smooth transformation, \mbox{$d_\text{pre}:G_\text{trans}\to G_\text{trans}$}, which maps the equivalence classes onto each other in the same way,
\begin{align}
\pi\circ d_\text{pre}=d_\text{post}\circ\pi.
\end{align}
That is, some permutation of the equivalence classes, $[g]\mapsto d_\text{post}([g])$, will be smooth if and only if there is a way to implement it smoothly prior to taking the quotient, $[g]\mapsto [d_\text{pre}(g)]$.

This quotient space, $\mathcal{M}_\text{G}$, is not only a smooth manifold, but is also homogeneous with $G_\text{trans}$ acting smoothly and transitively over it. Namely, the following group action,
\begin{align}\label{gbarDef}
&\theta:G_\text{trans}\times\mathcal{M}_\text{G}\to\mathcal{M}_\text{G}\\
\nonumber
&\theta(g,[h])=\bar{g}([h])\coloneqq[g\,h].
\end{align}
acts smoothly and transitively over $\mathcal{M}_\text{G}$. This follows straightforwardly from the left-action of $G_\text{trans}$ on itself being both smooth and transitive.

Given that this group action is smooth, the above defined $\bar{g}$ maps are all diffeomorphisms on $\mathcal{M}_\text{G}$. This point should be stressed. Even if, for instance, $G_\text{trans}\subset\text{GL}(V)$ is a Lie group represented on some vector space, $V$, the above construction builds a smooth manifold, $\mathcal{M}_\text{G}$, on which each $g\in G_\text{trans}$ can be understood as a diffeomorphism, $\bar{g}\in \text{Diff}(\mathcal{M}_\text{G})$. Collecting these together we have a Lie group $\bar{g}\in\overline{G_\text{trans}}\subset\text{Diff}(\mathcal{M}_\text{G})$. In some sense, the Lie group $\overline{G_\text{trans}}$ is a representation of $G_\text{trans}$ as diffeomorphisms on $\mathcal{M}_\text{G}$. This, will be a faithful representation (i.e., $\overline{G_\text{trans}}\cong G_\text{trans}$) if and only if we have \textit{fixed-point compatibility} between $G_\text{trans}$ and $G_\text{fix}$ (defined below).\footnote{In the mathematical literature, some terminology for what I am here calling ``fixed-point compatibility between $G_1$ and $G_2$'' has already been established. One might equivalently say that ``the normal core of $G_1$ in $G_2$ is trivial''. Alternatively, one might say that ``$G_1$ is a core-free subgroup of $G_2$''.}
\begin{quote}
{\hypertarget{FPC}{\bf Definition: Fixed-Point Compatibility - }} Two groups $G_1\subset G_2$ are said to be fixed-point compatible under the following condition:
\begin{align}\label{FixedCompat}
\forall s\in G_1 \left( (\forall g\in G_2 \ \  g^{-1}s\,g\in G_1)\ \Longrightarrow \  s=e\right). 
\end{align}
That is, when only the identity, $e\in G_1$, remains in $G_1$ under conjugation by every $g\in G_2$. Equivalently, this happens exactly when $G_1$ contains no non-trivial normal subgroups of $G_2$. 
\end{quote}
Roughly, $G_\text{trans}$ will be faithfully represented as diffeomorphisms, $\overline{G_\text{trans}}$, on $\mathcal{M}_\text{G}\coloneqq G_\text{trans}/G_\text{fix}$ if and only if the quotient is in lowest terms (i.e., if no part of $G_\text{fix}$ divides neatly into $G_\text{trans}$).

\subsection{\hra Characterizing Homogeneous Spaces}
We have so far seen that the quotient of any Lie group, $G_\text{trans}$, with one of its closed Lie subgroups, $G_\text{fix}\subset G_\text{trans}$, is a homogeneous manifold,  $\mathcal{M}_\text{G}\coloneqq G_\text{trans}/G_\text{fix}$, with respect to $G_\text{trans}$. Perhaps surprisingly, these are all of the smooth homogeneous manifolds that exist (up to diffeomorphism). I will now prove this following \cite{Zikidis}.

Let us begin from a generic homogeneous manifold, $\mathcal{M}$. Because $\mathcal{M}$ is homogeneous, there exists some finite-dimensional Lie group, $G$, which acts smoothly and transitively over it. Because $G$ acts smoothly on $\mathcal{M}$ we can associate to each $g\in G$ a diffeomorphism $h=H(g)$ with $h(p)=\theta(g,p)$. Collecting these diffeomorphisms together we have a finite-dimensional Lie group of diffeomorphisms, $H_\text{trans}\subset\text{Diff}(\mathcal{M})$, which acts transitively on $\mathcal{M}$. 

I will now show that $\mathcal{M}$ is diffeomorphic to a quotient of $H_\text{trans}$ with one of its Lie subgroups, $H_\text{fix}\subset H_\text{trans}$. The first step in showing this is to choose the right equivalence relation with which to quotient $H_\text{trans}$ by. Let us next pick out any point, $p_0\in\mathcal{M}$, and define an equivalence relation, $\equiv_{p_0}$, on $H_\text{trans}$ as,
\begin{align}
h_2\equiv_{p_0} h_1 \text{ iff } h_2(p_0)=h_1(p_0).
\end{align}
Namely, $h_2$ and $h_1$ are equivalent under $\equiv_{p_0}$ if they map $p_0$ to the same place. Using this equivalence relation we can define the quotient space,
\begin{align}
\mathcal{M}_\text{H}
\coloneqq H_\text{trans}/\equiv_{p_0}.
\end{align}
Let $[h]_{p_0}$ denote the equivalence classes under $\equiv_{p_0}$ containing $h$.

It follows from the orbit-stabilizer theorem that these equivalence classes are $H_\text{fix}$-cosets as $[h]_{p_0}=h H_\text{fix}$. Here $H_\text{fix}$ is the stabilizer subgroup of $H_\text{trans}$ at $p_0$.\footnote{Note that all of $H_\text{trans}$'s stabilizer subgroups, $H_\text{fix}$, are equivalent up to conjugacy. In this sense, all choices of $p_0\in\mathcal{M}_\text{old}$ are equivalent.} Namely, $H_\text{fix}$ is the subgroup of $H_\text{trans}$ which maps $p_0$ to itself as $h(p_0)=p_0$. Rewriting our quotient in these group-theoretic terms we have,
\begin{align}\label{HomoQuotient1}
\mathcal{M}_\text{H}
=H_\text{trans}/H_\text{fix}.
\end{align}
It follows from $H_\text{trans}$ acting transitively over $\mathcal{M}$ that we have the following one-to-one correspondence between $\mathcal{M}_\text{H}$ and $\mathcal{M}$:
\begin{align}\label{PDef}
P:\mathcal{M}_\text{H}&\to \mathcal{M}\\
\nonumber
[h]_{p_0}&\mapsto h(p_0).
\end{align}
Note that $P$ sends each equivalence class, $[h]_{p_0}$, onto the point, $h(p_0)\in\mathcal{M}$, which all of its elements map $p_0$ onto. This one-to-one correspondence, $P$, gives us,
\begin{align}
\mathcal{M}\leftrightarrow \mathcal{M}_\text{H}\coloneqq H_\text{trans}/H_\text{fix}.
\end{align}
As the previous subsection has discussed this quotient space, $\mathcal{M}_\text{H}$, naturally adopts a smooth structure coming from the smooth structure of its numerator, $H_\text{trans}$, via the Quotient Manifold Theorem. It can be proved that this smooth structure on $\mathcal{M}_\text{H}$ is in fact equivalent to the smooth structure on the original manifold, $\mathcal{M}$. That is, it can be proved that the above discussed one-to-one correspondence $P:\mathcal{M}_\text{H}\to \mathcal{M}$ is smooth (see Theorem 3 of \cite{Zikidis}). Thus, we have,
\begin{align}\label{HomoRecon}
\mathcal{M}\cong \mathcal{M}_\text{H}\coloneqq H_\text{trans}/H_\text{fix}.
\end{align}
Thus, any smooth homogeneous manifold can be reconstructed up to diffeomorphism as a quotient of its diffeomorphisms. 

To close off this section, I will now prove a result which will be helpful to us later. We have above reconstructed a generic homogeneous manifold, $\mathcal{M}$, up to diffeomorphism in terms of its diffeomorphisms as $\mathcal{M}_\text{H}=H_\text{trans}/H_\text{fix}$. Just as $h\in H_\text{trans}$ can be understood as a diffeomorphism on $\mathcal{M}$, so too can it be understood as a diffeomorphism on $\mathcal{M}_\text{H}$. Namely, it can be understood as $\bar{h}\in\overline{H_\text{trans}}\subset\text{Diff}(\mathcal{M}_\text{H})$ which acts on $\mathcal{M}_\text{H}$ as $\bar{h}([g]_{p_0})=[h\,g]_{p_0}$. Given that $\mathcal{M}$ and $\mathcal{M}_\text{H}$ are isomorphic, it is natural to expect that $H_\text{trans}\subset\text{Diff}(\mathcal{M})$ and $\overline{H_\text{trans}}\subset\text{Diff}(\mathcal{M}_\text{H})$ are isomorphic. Indeed they are, as I will now prove.

As discussed above, we will have $\overline{H_\text{trans}}\cong H_\text{trans}$ if and only if we have fixed-point compatibility between $H_\text{trans}$ and $H_\text{fix}$. As I will now prove, this follows from $H_\text{trans}$ acting transitively on 
$\mathcal{M}$ and $H_\text{fix}$ being a stabilizer subgroup of $H_\text{trans}$. To prove this we must show the following:
\begin{align}\label{FixedCompatH}
\forall s\in H_\text{fix} \left((\forall h\in H_\text{trans} \ \  h^{-1}s\,h\in H_\text{fix})\ \Longrightarrow \  s=\openone_\mathcal{M}\right).
\end{align}
\textit{Proof by contrapositive}: Note that if $s\neq \openone_\mathcal{M}$ then there exists some \mbox{$q\in\mathcal{M}$} with \mbox{$s(q)\neq q$}. Since $H_\text{trans}$ acts transitively on $\mathcal{M}$ there is some $h\in H_\text{trans}$ which maps $p_0$ to $q$ as $q=h(p_0)$. For this $h$ we have \mbox{$(s\,h)(p_0)\neq q$} whereas $h(p_0)=q$. Thus we have \mbox{$(h^{-1}s\,h)(p_0)\neq p_0$} and therefore $h^{-1}s\,h\not\in H_\text{fix}$ as desired.

\subsection{\hra Reconstructing the Sphere from its Diffeomorphisms}
The previous subsection has proved that any homogeneous manifold can be reconstructed up to diffeomorphism as a quotient of its diffeomorphisms. As the above discussion has been fairly abstract, I will now provide a concrete example. Namely, I will reconstruct the 2-sphere $\mathcal{M}\cong S^2$ (i.e., the unit sphere in $\mathbb{R}^3$) in terms of its diffeomorphisms. 

Let $H_\text{trans}\subset\text{Diff}(S^2)$ be the Lie subgroup of diffeomorphisms which are also rigid rotations in $\mathbb{R}^3$, namely \mbox{$H_\text{trans}\cong\text{SO}(3)$}. Note that $H_\text{trans}$ acts smoothly and transitively on $S^2$. As such we can use it in the numerator of Eq.~\eqref{HomoRecon}. What then goes in the denominator of Eq.~\eqref{HomoRecon}? Letting $p_0=n$ be the north pole, $H_\text{fix}\subset H_\text{trans}$ is the stabilizer subgroup  of $H_\text{trans}$ which fixes the north pole. This subgroup is exactly the rotations about the z-axis. Thus we have $H_\text{fix}\cong\text{SO}(2)$.

\begin{figure}[p]
\centering
\includegraphics[width=0.95\textwidth]{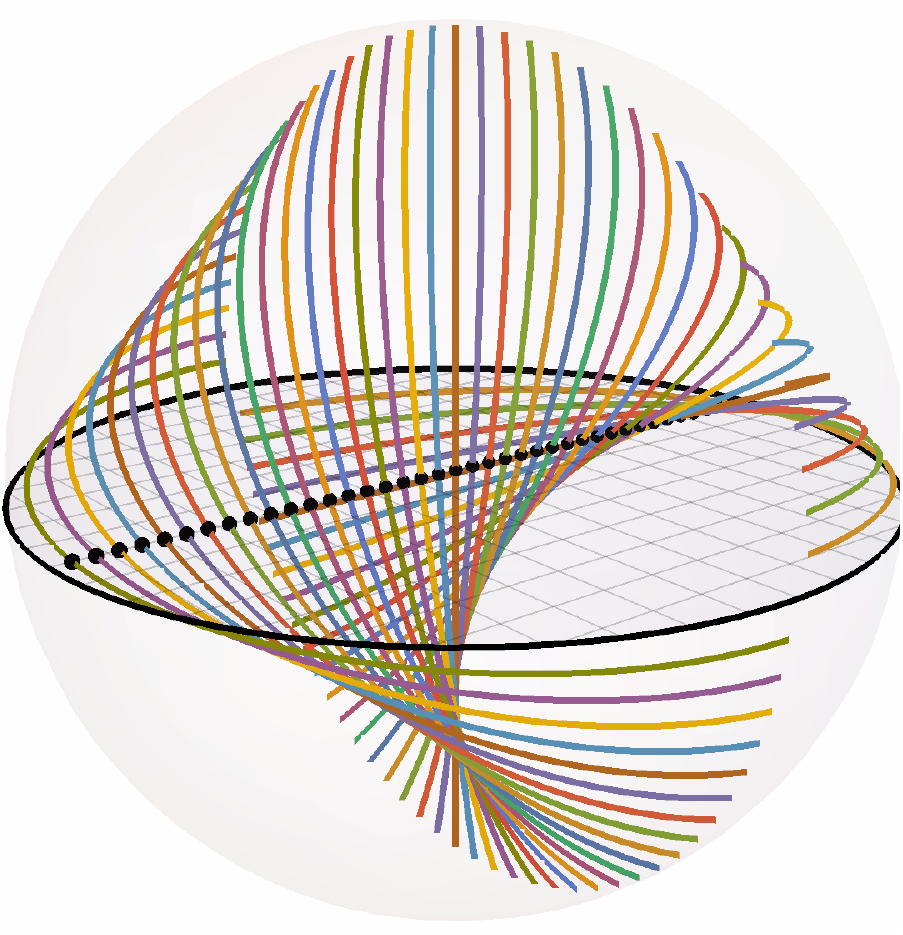}
\caption{The Lie group $\text{SO}(3)\cong\text{RP}^3$ is depicted here as a smooth manifold: namely, a ball with radius $\pi$ in $\mathbb{R}^3$ with antipodal points on its surface identified. Concretely, each point $r\in\mathbb{R}^3$ in this figure represents a rotation of $\theta=\vert r\vert$ radians about the $\hat{r}$ axis (where $\hat{r}$ is the unit vector parallel to $r$). The colored lines shown are some of the equivalence classes of $\text{SO}(3)\cong\text{RP}^3$ under $\equiv_{p_0}$ with $p_0=n$ being the north-pole. That is, rotations which are represented on the same line map the north-pole to the same place. Note that the equator itself is an equivalence class. For all equivalence classes but this one, we can pick out a unique representative on the $x,y$-plane. A selection of these are shown as black points. We can therefore take the quotient of $\text{SO}(3)$ with $\equiv_{p_0}$ by taking these points on the $x,y$-plane as our representatives and identifying all of the equatorial points. This yields a disk with its boundary compactified to a single point. Namely, this yield the smooth manifold $\text{SO}(3)/\equiv_{p_0}\cong S^2$.}\label{FigSO3}
\end{figure}

Plugging $H_\text{trans}\cong\text{SO}(3)$ and $H_\text{fix}\cong\text{SO}(2)$ into Eq.~\eqref{HomoRecon} we have,  
\begin{align}\label{SO3SO2}
\mathcal{M}_\text{H}=H_\text{trans}/H_\text{fix}\cong SO(3)/SO(2).
\end{align}
According to the results of the previous subsection we should now have $S^2\cong \mathcal{M}_\text{H}$. Let us now check this, paying careful attention to where the quotient manifold gets its smooth structure from. See Fig.~\ref{FigSO3} for an illustrated guide to how this quotient works.

In Eq.~\eqref{HomoRecon} we are supposed to regard the Lie group in the numerator, $H_\text{trans}\cong\text{SO}(3)$, as a smooth manifold: But what smooth manifold is $\text{SO}(3)$? To see this, consider first its Lie algebra $\mathfrak{so}(3)$ and note that it is three-dimensional. The group exponential from $\mathfrak{so}(3)$ to $\text{SO}(3)$ is surjective such that each element of $\text{SO}(3)$ is represented (possibly multiple times) in $\mathfrak{so}(3)$ under the group exponential. As I will now discuss, by identifying these redundancies we can see $\text{SO}(3)$ as a certain quotient space of $\mathbb{R}^3$. 

Fig.~\ref{FigSO3} illustrates how exactly $\mathfrak{so}(3)$ redundantly represents $\text{SO}(3)$. There a point, $r\in\mathbb{R}^3$, represents a rotation of $\theta=\vert r\vert$ radians about the $\hat{r}$ axis (where $\hat{r}$ is the unit vector parallel to $r$). In this way, each rotation in $\text{SO}(3)$ can represented within the region $\vert r\vert\leq \pi$ with the following redundancy: $\pi\,\hat{r}$ is identical to $-\pi\,\hat{r}$ for every $\hat{r}$. That is, a half-turn about any axis $\hat{r}$ is the same as a half-turn in the reverse direction.

For example, rotations of the sphere about the $x$-axis correspond to points on the $x$-axis in Fig.~\ref{FigSO3} (e.g., the black points). Rotating about the $x$-axis through an angle just greater than $\pi$ is equivalent to rotating in the reverse direction by an angle just less than $\pi$. Hence, the antipodal points on the $x$-axis in Fig.~\ref{FigSO3} represent the same rotation and so should be identified. This holds for every axis. Thus, as a smooth manifold $\text{SO}(3)$ is diffeomorphic to the unit ball in $\mathbb{R}^3$ with antipodal points on its surface identified. That is, as smooth manifolds we have $\text{SO}(3)\cong \text{RP}^3$, the real-projective space in three dimensions.

Note that $\text{RP}^3$ is not diffeomorphic to the manifold $\mathcal{M}=S^2$ which we are hoping to recover. Indeed, according to Eq.~\eqref{HomoRecon} the desired manifold is a quotient of this manifold by $H_\text{fix}\cong\text{SO}(2)$. The colored lines in Fig.~\ref{FigSO3} show some of these equivalence classes:\footnote{Note that only a subset of these equivalence classes are shown in Fig.~\ref{FigSO3}. Namely, only equivalence classes which contain rotations about the $x$-axis with incrementally larger values of $\theta\in[-\pi,\pi]$ are shown.} Rotations which appear on the same line map the north-pole to the same place.

Let us quickly discuss two examples of equivalence classes. Firstly, note that all rotations about the z-axis map the north-pole to the same place (they leave it fixed). Rotations about the z-axis are represented in Fig.~\ref{FigSO3} by points on the z-axis. Hence, all of the points on the z-axis are connected by a vertical line in Fig.~\ref{FigSO3}.

Secondly, note that all rotations about points on the equator through an angle of $\theta=\pi$ map the north-pole to the same place, namely the south pole. Rotations about equatorial points are represented in Fig.~\ref{FigSO3} by points on the $x,y$-plane. Rotations with $\theta=\pi$ are represented in Fig.~\ref{FigSO3} by points on the surface of the sphere. Hence, these $\theta=\pi$ equatorial rotations are represented in Fig.~\ref{FigSO3} by the sphere's equator. Notice that all of these points are connected by a circular line. This marks the fact that they all map the north-pole to the same point, namely the south-pole.

Besides these two equivalence classes, several others are also shown in Fig.~\ref{FigSO3} which smoothly transition between them. Note that each of these colored lines terminates at the boundary of the unit ball. Recall, however, that in this representation of \mbox{$\text{SO}(3)\cong\text{RP}^3$} antipodal points on the surface are to be identified. Thus, each of these lines is really a closed loop in $\text{RP}^3$. Indeed, these equivalence classes are each $H_\text{fix}\cong\text{SO}(2)$-orbits within $H_\text{trans}\cong\text{SO}{3}$.

For each equivalence class (except one) we can pick as its representative its unique member on the $x,y$-plane. For the subset of equivalence classes shown in Fig.~\ref{FigSO3}, these representative points are shown in black. The exception to this claim is the equivalence class of rotations which map the north-pole to the south-pole. This equivalence class lies entirely in the $x,y$-plane. Indeed, as discussed above, this equivalence class is represented by the points on the equator in Fig.~\ref{FigSO3}. Thus, we can take the $x,y$-plane of Fig.~\ref{FigSO3} to be our quotiented space with one slight modification: We must take the equivalence class represented by the equator to represent a single point in the quotiented space. Thus, in total we can see the quotiented space as being diffeomorphic to a disk in the $x,y$-plane with its boundary compactified to a single point. Compactifying the boundary of a disk yields the manifold $\mathcal{M}_\text{H}\cong S^2$ as desired.

\section{Surviving Internalization: Spacetime Manifold}\label{SecSurvivingMan}
Sec.~\ref{SecSurvivingTrans} showed that for a great many spacetime theories, some of the theory's diffeomorphisms faithfully survive internalization. Namely, assuming the theory satisfies the \hyperlink{Prelim}{\bf Preliminary Assumptions} discussed in Ch.~\ref{ChapGenerality} and is moreover kinematically reduced we have that,\footnote{For a discussion of internalization which does not assume spacetime-kinematic compatibility, see Appendix~\ref{ChapISEsansSKC}.}
\begin{align}
H \surint G(H)\cong H,
\end{align}
for any kinematically allowed Lie group, $H\subset\text{Diff}_\text{kin}(\mathcal{M}_\text{old})$. Recall that $G(H)\subset\text{GL}(V_\text{neutral})$ is some Lie group represented on $V_\text{neutral}$. Following this, Sec.~\ref{SecRevHomo} showed us that  any homogeneous manifold can be reconstructed up to diffeomorphism as a quotient of two Lie subgroups of its diffeomorphisms. Concretely, if the old theory's spacetime, $\mathcal{M}_\text{old}$, is homogeneous, then applying this result gives the following reconstruction:
\begin{align}\label{MoldHH}
\mathcal{M}_\text{old}\cong \mathcal{M}_\text{H}\coloneqq H_\text{trans}/H_\text{fix},
\end{align}
for some Lie groups $H_\text{trans}\subset \text{Diff}(\mathcal{M}_\text{old})$ and $H_\text{fix}\subset H_\text{trans}$. Here $H_\text{trans}$ is any finite-dimensional Lie group which acts transitively on $\mathcal{M}_\text{old}$. $H_\text{fix}$ is then any of $H_\text{trans}$'s stabilizer subgroups.

Given these results (i.e., that \textit{some} diffeomorphisms can survive internalization, and that \textit{some} manifolds can be reconstructed in terms of their diffeomorphisms) it seem conceivable that the spacetime manifolds of \textit{some} theories might also survive internalization. Concretely, the theory's manifold might survive as a quotient of its surviving diffeomorphisms. As I will prove in this section, given \hyperlink{SKC}{\bf Spacetime-Kinematic Compatibility} a theory's spacetime manifold always survives internalization. Moreover it does so faithfully (i.e., with the surviving manifold being isomorphic to the original) and robustly (i.e., up to diffeomorphism, independent of which $H_\text{trans}$ we use to reconstruct $\mathcal{M}_\text{old}$).\footnote{\hyperlink{SKC}{\bf Spacetime-Kinematic Compatibility} is not only sufficient for a theory's spacetime manifold to survive internalization faithfully and robustly, but is also necessary in order for this to happen. See Appendix~\ref{AppRFSKC} for proof.}

Consider any spacetime theory with spacetime-kinematic compatibility. This theory's spacetime manifold survives internalization as follows. From spacetime-kinematic compatibility it follows that there exists a finite-dimensional Lie group of diffeomorphisms, $H_\text{trans}\subset\text{Diff}_\text{kin}(\mathcal{M}_\text{old})$, which is kinematically allowed and acts transitively over $\mathcal{M}_\text{old}$. Using the fact that $H_\text{trans}$ acts transitively on $\mathcal{M}_\text{old}$, we can pick out one of its stabilizer subgroups, $H_\text{fix}\subset H_\text{trans}$, and reconstruct $\mathcal{M}_\text{old}$ up to diffeomorphisms as \mbox{$\mathcal{M}_\text{old}\cong \mathcal{M}_\text{H}= H_\text{trans}/H_\text{fix}$}.\footnote{See Sec.~\ref{SecRevHomo} for details.}

Next, note that (by assumption) the theory under consideration is kinematically reduced such that we have we have $\text{ker}(G)=H_\text{kin}=\{\openone_{\mathcal{M}_\text{old}}\}$. Consequently, $H_\text{trans}$ and $H_\text{fix}$ survive internalization faithfully:
\begin{align}
H_\text{trans}
&\surint G_\text{trans}\coloneqq G(H_\text{trans})\cong H_\text{trans},\\
\nonumber
H_\text{fix}
&\surint G_\text{fix}\coloneqq G(H_\text{fix})\cong H_\text{fix}
\end{align}
for some Lie groups, $G_\text{fix}\subset G_\text{trans}\subset \text{GL}(V_\text{neutral})$, represented on $V_\text{neutral}$.

In total, we therefore have the spacetime manifold surviving internalization as a quotient of its diffeomorphisms: Namely,
\begin{align}\label{ManSur1}
\mathcal{M}_\text{old}\cong\mathcal{M}_\text{H}\coloneqq \frac{H_\text{trans}}{H_\text{fix}}
\surint
\mathcal{M}_\text{G}\coloneqq \frac{G_\text{trans}}{G_\text{fix}}\cong\frac{H_\text{trans}}{H_\text{fix}}\cong \mathcal{M}_\text{old}.
\end{align}
Therefore, given spacetime-kinematic compatibility, a theory's spacetime manifold faithfully survives internalization. Moreover, it does so robustly, i.e., independent up to diffeomorphism of which $H_\text{trans}$ we have used to reconstruct $\mathcal{M}_\text{old}$. 

Thus, given \hyperlink{SKC}{\bf Spacetime-Kinematic Compatibility}, a great deal of topological information about the theory's old spacetime setting survives internalization. Indeed, it appears that, for such theories, complete information about the old spacetime setting has survived internalization. As I will discuss in Ch.~\ref{ChapExternal}, there is enough information contained in $G_\text{trans}$ and $G_\text{fix}$ to perfectly reconstruct the theory's original spacetime setting. Here one might sense a tension with \hyperlink{D1}{\bf Desiderata \#1 (D1)}: If enough topological information can survive internalization for us to be able to faithfully recover $\mathcal{M}_\text{old}$, then how can it be that we have conceptually divorced the dynamical behavior of matter from its assumed topological underpinnings? The answer to this worry lies in \hyperlink{D3}{\bf Desiderata \#3 (D3)}. While \hyperlink{D2}{\bf D2} says we can return to the original theory, \hyperlink{D3}{\bf D3} says we don't have to. In fact, it promises us that there will be a variety of spacetime settings available to us post-internalization. In this way, \hyperlink{D3}{\bf D3} provides additional support for \hyperlink{D1}{\bf D1}: While the original spacetime manifold might faithfully survive internalization, it becomes un-special in the process. (That is, it becomes un-special except insomuch as it is a uniquely good fit for the theory's dynamics and kinematics.)\footnote{See Ch.~\ref{ChapSearch}.} Further discussion of how the \hyperlink{ISE}{ISE Methodology} satisfies \hyperlink{D3}{\bf D3} (and hence, how internalization satisfies \hyperlink{D1}{\bf D1}) will be given in Ch.~\ref{ChapSearch}. Before this, Ch.~\ref{ChapExternal} will prove that externalization satisfies \hyperlink{D2}{\bf D2}.

\section{Special Properties of \texorpdfstring{$G_\text{trans}$}{Gtrans} and \texorpdfstring{$G_\text{fix}$}{Gfix}}\label{SecSpecialG}
To conclude this chapter, some special properties of the above-defined Lie groups $G_\text{trans}$ and $G_\text{fix}$ represented on $V_\text{neutral}$ should be noted. To begin, recall that we have a Lie group isomorphism, $G$, between $\text{Diff}_\text{kin}(\mathcal{M})$ and $G_\text{old}$. From this isomorphism we have both $G_\text{trans}\cong H_\text{trans}$ and $G_\text{fix}\cong H_\text{fix}$ as Lie groups. Consequently, many Lie-group-theoretic properties of $G_\text{trans}$ and $G_\text{fix}$ follow immediately:
\begin{enumerate}
    \item Both $G_\text{trans}$ and $G_\text{fix}$ are finite-dimensional Lie groups because both $H_\text{trans}$ and $H_\text{fix}$ are finite-dimensional Lie groups;
    \item $G_\text{fix}$ is a closed Lie subgroup of $G_\text{trans}$ because $H_\text{fix}$ is a closed Lie subgroup of $H_\text{trans}$;
    \item $G_\text{fix}$ and $G_\text{trans}$ have \hyperlink{FPC}{\bf Fixed-Point Compatibility} because $H_\text{fix}$ and $H_\text{trans}$ have fixed-point compatibility;\footnote{See the argument following Eq.~\eqref{FixedCompatH}.} and finally,
    \item $G_\text{trans}$ acts smoothly and transitively on $\mathcal{M}_\text{G}$ because $H_\text{trans}$ acts smoothly and transitively on $\mathcal{M}_\text{old}$.
\end{enumerate}
Past their Lie group theoretic structure, $H_\text{trans}$ and $H_\text{fix}$ also have some non-trivial relationships to $V_\text{kin}^\text{old}$, the vector space of kinematic possible fields states on $\mathcal{M}_\text{old}$. For instance, $H_\text{trans}$ is kinematically allowed. 

Post-internalization, these non-trivial relationships between $H_\text{trans}$, $H_\text{fix}$, and $V_\text{kin}^\text{old}$ become non-trivial relationships between $G_\text{trans}$, $G_\text{fix}$, and $V_\text{neutral}$. That is, $G_\text{trans}$ and $G_\text{fix}$ will have additional properties as Lie groups \textit{represented on} $V_\text{neutral}$. For later convenience, I should here note that $G_\text{trans}$, $G_\text{fix}$, and $V_\text{neutral}$ will always have injection compatibility (defined below). Motivation for the relevance of this definition will be given in Ch.~\ref{ChapExternal}.
\begin{quote}
{\hypertarget{InjCompat}{\bf Definition: Injection Compatibility - }} Let $V_\text{N}$ be a vector space over some field, $K$. Let $G_2\subset\text{GL}(V_\text{N})$ be a finite-dimensional Lie group represented on $V_\text{N}$. Let $G_1\subset G_2$ be a closed Lie subgroup of $G_2$. Let $\mathcal{M}_\text{G}\coloneqq G_2/G_1$ be the quotient manifold of $G_2$ and $G_1$. For any vector space, $\mathcal{V}_\text{G}$, let $V_\text{G}$ be the vector space of all $\mathcal{V}_\text{G}$-valued functions definable on $\mathcal{M}_\text{G}$. These Lie groups, $G_1\subset G_2\subset\text{GL}(V_\text{N})$, and the vector space, $V_\text{N}$, are said to be \textit{injection compatible} under the following condition. There exists a vector space, $\mathcal{V}_\text{G}$, over $K$ and a injective linear map, \mbox{$J:V_\text{N}\to V_\text{G}$}, with the following two properties: Firstly, we must have,
\begin{align}\label{JInjectionEq1}
&\forall g \ \forall \bm{\varphi} \ \ \Big((J\circ g)\,\bm{\varphi} = (\bar{g}^*\circ J)\,\bm{\varphi}\Big),
\end{align}
and, secondly, for all diffeomorphisms $d\in \text{Diff}(G_2)$ we must have,
\begin{align}\label{JInjectionEq2}
&\Big( \forall g \ \forall \bm{\varphi} \ \  J(\bm{\varphi})([d(g)])=J(\bm{\varphi})([g])\Big)\Longleftrightarrow \forall g \ \ [d(g)]=[g].
\end{align}
In both of these expressions $g$ ranges over $G_2$ and $\bm{\varphi}$ ranges over $V_\text{N}$.
\end{quote}

\noindent Verifying that $G_\text{trans}$ and $G_\text{fix}$ have injection compatibility with $V_\text{neutral}$ is straightforward but tedious. Explicit proof is given in Appendix~\ref{AppSpecialG}. For now, it should be noted that the conjunction of  Eq.~\eqref{JInjectionEq1} and Eq.~\eqref{JInjectionEq2} can be realized by taking $\mathcal{V}_\text{G}=\mathcal{V}_\text{old}$ and $J(\bm{\varphi})=\mathcal{F}_\text{vec}^{-1}(\bm{\varphi})\circ R$. Here $R:\mathcal{M}_\text{G}\to\mathcal{M}_\text{old}$ is the diffeomorphism relating $\mathcal{M}_\text{G}$ and $\mathcal{M}_\text{old}$.\footnote{Concretely, the $R:\mathcal{M}_\text{G}\to\mathcal{M}_\text{old}$ map acts on $[g]\in\mathcal{M}_\text{G}$ as follows. First, pick any representative $g_0\in[g]$. Recalling that $G$ is a Lie group isomorphism, we can find $h=G^{-1}(g_0)$. Recall that $H_\text{fix}$ is the stabilizer subgroup of $H_\text{trans}$ as some $p_0\in\mathcal{M}_\text{old}$. Evaluating $h$ at $p_0$ gives $R([g])=G^{-1}(g_0)(p_0)$. This $R$ map is well-defined as this does not depend on which $g_0\in[g]$ we choose.}

\vspace{0.25cm}
To review: Whenever we internalize a spacetime theory with spacetime-kinematic compatibility, its spacetime manifold faithfully and robustly survives as the quotient of two Lie groups, $G_\text{trans}$ and $G_\text{fix}$, represented on $V_\text{neutral}$.\footnote{In Appendix~\ref{AppRFSKC} I prove the converse. If a theory's spacetime manifold is to survive internalization faithfully and robustly, then the theory must have \hyperlink{SKC}{\bf Spacetime-Kinematic Compatibility}.} Moreover, as the above discussion has shown these two Lie groups have some nice properties, including \hyperlink{FPC}{\bf Fixed-Point Compatibility} and \hyperlink{InjCompat}{\bf Injection Compatibility}. As I will discuss in the next chapter, $G_\text{trans}$ and $G_\text{fix}$ having the above-discussed properties makes them what I will call \hyperlink{PSTO}{\bf Pre-Spacetime Translation Operations}. Internalizing a theory with spacetime-kinematic compatibility always produces pre-spacetime translation operations.\footnote{In Appendix~\ref{AppKinRedISE} I will prove a stronger claim: Internalizing a spacetime theory with \hyperlink{WSKC}{\bf Weak Spacetime-Kinematic Compatibility} always produces \hyperlink{PSTO}{\bf Pre-Spacetime Translation Operations}.} The existence of such Lie groups represented on $V_\text{neutral}$ will be leveraged in the externalization process. This will be the subject of the next chapter.

\chapter{Externalization and Desiderata \#2}
\label{ChapExternal}

The previous chapter has shown that, given spacetime-kinematic compatibility, a theory's spacetime manifold will faithfully and robustly survive internalization as a quotient of its surviving diffeomorphisms.\footnote{In Appendix~\ref{AppRFSKC} I prove the converse. Hence, a theory's spacetime manifold will faithfully and robustly survive internalization if and only if it has spacetime-kinematic compatibility.} Of course, knowing how a theory's spacetime manifold survives internalization is very suggestive of how externalization might be able to generate a new spacetime manifold for this theory. This chapter will give a complete description of the externalization process. To begin, Sec.~\ref{SecPSTO} will define the building blocks from which externalization constructs a new spacetime setting for our theory, namely \hyperlink{PSTO}{\bf Pre-Spacetime Translation Operations (PSTOs)}. Following this, Sec.~\ref{SecNewMan} will detail the construction of a new manifold $\mathcal{M}_\text{new}$ from these PSTOs.  As I will there discuss, externalization is the process of taking some hand-picked pre-spacetime translation operations and letting them be honest-to-goodness spacetime translations. Next, Sec.~\ref{SecInject} will discuss how externalization transplants the theory's states onto this new spacetime setting. As I will then prove, all spacetime theories producible via externalization have \hyperlink{SKC}{\bf Spacetime-Kinematic Compatibility}. 

In detailing the externalization process, this chapter will impose some constraints on both the pre-spacetime translation operations as well as how the field states are mapped onto the new spacetime. This choice of constraints is not arbitrary! In Appendix~\ref{AppD2Structure} I will prove that, given the general form of the externalization process, the specific constraints chosen here are necessary in order for the \hyperlink{ISE}{ISE Methodology} to satisfy \hyperlink{D2}{\bf Desiderata \#2 (D2)}. The chapter will finish in Sec.~\ref{SecD2} with a proof that these constraints are also sufficient for the ISE Methodology satisfies \hyperlink{D2}{\bf D2}: Given any spacetime theory producible via externalization, after we re-internalize it we can then always re-externalize it trivially. This is analogous to peeling the generalized Humean paint off of the wall and then putting it straight back on.

\section{Pre-Spacetime Translation Operations (PSTOs)}\label{SecPSTO}
Externalization begins from the spacetime-neutral vector space of kinematic possibilities, $V_\text{neutral}$. There are then three inputs to the externalization process. Firstly, we pick a finite-dimensional Lie group, \mbox{$G_\text{trans}^\text{(ext)}\subset\text{GL}(V_\text{neutral})$}, represented on $V_\text{neutral}$. Secondly, we pick a closed Lie subgroup thereof, $G_\text{fix}^\text{(ext)}\subset G_\text{trans}^\text{(ext)}$. From these two Lie groups, externalization will construct a new spacetime manifold $\mathcal{M}_\text{new}$ (see Sec.~\ref{SecNewMan}). Following this, externalization must map the theory's states and dynamics onto $\mathcal{M}_\text{new}$. This is achieved by picking a map $E:V_\text{neural}\to V_\text{new}^\text{all}$ where $V_\text{new}^\text{all}$ is the vector space of all $\mathcal{V}_\text{new}$-valued functions definable on $\mathcal{M}_\text{new}$ (see Sec.~\ref{SecInject}). Note that we might have $\mathcal{V}_\text{new}\not\cong\mathcal{V}_\text{old}$.\footnote{Technically, one might count this new value space, $\mathcal{V}_\text{new}$, as a fourth input to the externalization process. I am here considering our choice of $\mathcal{V}_\text{new}$ to be implicit within our choice of $E:V_\text{neural}\to V_\text{new}^\text{all}$.}

Before continuing on, a brief point about notation is in order. The superscripts \text{(ext)} adopted in the previous paragraph are being used to indicate that $G_\text{trans}^\text{(ext)}$ and $G_\text{fix}^\text{(ext)}$ are the Lie groups being used in the externalization process. A superscript \text{(int)} will be used when referring to the analogous but distinct objects coming from the internalization process (i.e., $G_\text{trans}^\text{(int)}$ and $G_\text{fix}^\text{(int)}$). However, as carrying around all of these \text{(ext)} superscripts is highly inconvenient, they will be dropped when context permits.

These three choices ($G_\text{trans}^\text{(ext)}$, $G_\text{fix}^\text{(ext)}$, and $E$) are firstly subject to the constraint that $G_\text{trans}^\text{(ext)}$ and $G_\text{fix}^\text{(ext)}$ be \textit{pre-spacetime translation operations} (defined below).
\begin{quote}
{\hypertarget{PSTO}{\bf Definition: Pre-Spacetime Translation Operations (PSTOs) - }} Let $V_\text{N}$ be a vector space. Let $G_2\subset\text{GL}(V_\text{N})$ be a finite-dimensional Lie group represented on $V_\text{N}$. Let $G_1\subset G_2$ be a closed Lie subgroup of $G_2$. These two groups will be called \textit{pre-spacetime translation operations}  when they satisfy the following two conditions:
\begin{enumerate}
    \item \hyperlink{FPC}{\bf Fixed-Point Compatibility} between $G_1$ and $G_2$.
    \item {\hyperlink{InjCompat}{\bf Injection Compatibility}} between $G_1$, $G_2$, and $V_\text{N}$.
\end{enumerate}
\end{quote}
The conditions which $E:V_\text{neural}\to V_\text{new}^\text{all}$ must obey will be further discussed in Sec.~\ref{SecInject}. It should be noted here, however, that the allowed choices of $E$ are, up to diffeomorphism, exactly the $J$ maps which injection compatibility guarantees the existence of.\footnote{To be explicit: The new spacetime manifold put forward by internalization is diffeomorphic to $\mathcal{M}_\text{G}=G_\text{trans}/G_\text{fix}$. Namely, it is $\mathcal{M}_\text{new}\coloneqq \mathcal{F}_\text{man}(\mathcal{M}_\text{G})$ for some forgetful functor $\mathcal{F}_\text{man}:\mathcal{M}_\text{G}\to \mathcal{M}_\text{new}$. The map $E:V_\text{neural}\to V_\text{new}^\text{all}$ is constrained to be $E(\bm{\varphi})=J(\bm\varphi)\circ\mathcal{F}_\text{man}^{-1}$ for one of the linear injectors, $J:V_\text{neutral}\to V_\text{G}$, which \hyperlink{InjCompat}{\bf Injection Compatibility} guarantees the existence of.}

Roughly put, in order for $G_\text{trans}^\text{(ext)}$ and $G_\text{fix}^\text{(ext)}$ to be pre-spacetime translation operations they must have all of the properties discussed in Sec.~\ref{SecSpecialG}. Recall from Sec.~\ref{SecSpecialG} that internalizing a spacetime theory with spacetime-kinematic compatibility always produces two Lie groups (namely, $G_\text{trans}^\text{(int)}$ and $G_\text{fix}^\text{(int)}$) which have exactly these properties. That is, internalizing such a theory guarantees us the existence of at least one set of pre-spacetime translation operations.\footnote{A more general result is proved in Appendix~\ref{AppKinRedISE}: Internalizing a spacetime theory with \hyperlink{WSKC}{\bf Weak Spacetime-Kinematic Compatibility} always produces pre-spacetime translation operations.} Thus, we are free to choose $G_\text{trans}^\text{(ext)}=G_\text{trans}^\text{(int)}$ and $G_\text{fix}^\text{(ext)}=G_\text{fix}^\text{(int)}$ as our pre-spacetime translation operations. In Sec.~\ref{SecD2}, I will prove that by making this choice we can always recover the theory's initial spacetime setting (in line with \hyperlink{D2}{\bf Desiderata \#2}).

Importantly, however, we are not forced to externalize the Lie groups given to us by internalization. We are free to search for other pre-spacetime translation operations, i.e., ones with $G_\text{trans}^\text{(ext)}\neq G_\text{trans}^\text{(int)}$ and/or $G_\text{fix}^\text{(ext)}\neq G_\text{fix}^\text{(int)}$.\footnote{To get a sense of how \hyperlink{FPC}{\bf Fixed-Point Compatibility} constrains our choice of $G_\text{trans}^\text{(ext)}$ and $G_\text{fix}^\text{(ext)}$ note the following. If $G_\text{trans}^\text{(ext)}$ is abelian, then all of its subgroups are normal subgroups. However, if $G_\text{fix}^\text{(ext)}\subset G_\text{trans}^\text{(ext)}$ is a normal subgroup then it must be trivial by fixed-point compatibility. Hence, if $G_\text{trans}^\text{(ext)}$ is abelian then we must choose $G_\text{fix}^\text{(ext)}=\{\hat\openone\}$.} This possibility will be explored in the next chapter. As I will there show, by externalizing a variety of different pre-spacetime translation operations we can achieve variety of inequivalent spacetime settings for our theory (in line with \hyperlink{D3}{\bf Desiderata \#3}).

Let us proceed now, however, with $G_\text{trans}^\text{(ext)}$ and $G_\text{fix}^\text{(ext)}$ being generic PSTOs. How, roughly, are the above-assumed properties of PSTOs used in the externalization process? Sec.~\ref{SecNewMan} will demonstrate how externalization builds a new spacetime manifold, $\mathcal{M}_\text{new}$, from these pre-spacetime translation operations. Fixed-point compatibility is a necessary and sufficient condition for $G_\text{trans}$ to be faithfully represented as diffeomorphisms on $\mathcal{M}_\text{new}$. Following this, Sec.~\ref{SecInject} will then demonstrate how externalization transplants the theory's states and dynamics onto $\mathcal{M}_\text{new}$ using the map $E:V_\text{neural}\to V_\text{new}^\text{all}$. Injection compatibility here ensures that these $G_\text{trans}$-represented-on-$\mathcal{M}_\text{new}$ diffeomorphisms act in the right way on the new field states. As mentioned above, given the general form of the externalization process, the specific constraints here placed on $G_\text{trans}$, $G_\text{fix}$ and $E:V_\text{neural}\to V_\text{new}^\text{all}$ are both necessary and sufficient for the ISE Methodology satisfies \hyperlink{D2}{\bf Desiderata \#2}. Sufficiency is proved in Sec.~\ref{SecD2} and necessity is proved in Appendix~\ref{AppD2Structure}.

\section{Externalization: Building a New Spacetime}\label{SecNewMan}
Given some pre-spacetime translation operations, $G_\text{trans}$ and $G_\text{fix}$, represented on $V_\text{ neutral}$, externalization puts forward their quotient as the theory's new spacetime manifold,
\begin{align}\label{MGdef}
\mathcal{M}_\text{new}\cong\mathcal{M}_\text{G}\coloneqq G_\text{trans}/G_\text{fix},
\end{align}
at least up to diffeomorphism.  I say here  ``up to diffeomorphism'' because the quotient manifold $\mathcal{M}_\text{G}$ contains excess structure; Its elements are sets of linear transformations on $V_\text{neutral}$ as well as manifold points on $\mathcal{M}_\text{G}$. We can remove this excess by applying a forgetful functor $\mathcal{F}_\text{man}$ to $\mathcal{M}_\text{G}$. Namely, this is a diffeomorphism  \mbox{$\mathcal{F}_\text{man}:\mathcal{M}_\text{G}\to \mathcal{M}_\text{new}$} where $\mathcal{M}_\text{new}$ is some abstract smooth manifold which is isomorphic to $\mathcal{M}_\text{G}$, i.e., with \mbox{$\mathcal{M}_\text{new}\cong \mathcal{M}_\text{G}$}. In terms of this forgetful functor, the new spacetime manifold is defined as,\footnote{Its important to note that even with an fixed abstract smooth manifold, $\mathcal{M}_\text{new}$, in mind, there are several different forgetful functors, $\mathcal{F}_\text{man}$, which one might apply when mapping $\mathcal{M}_\text{G}$ onto it. What difference do these make? Ultimately, any two forgetful functors (here, maps $\mathcal{F}_\text{man}$ and $\mathcal{F}_\text{man}'$) which we might choose here differ only by an automorphism on $\mathcal{M}_\text{new}$. Namely, we have \mbox{$\mathcal{F}_\text{man}'=d\circ\mathcal{F}_\text{man}$} for some  diffeomorphism \mbox{$d\in\text{Diff}(\mathcal{M}_\text{new})$}. Thus, while fixing an $\mathcal{F}_\text{man}$ map has helped us make things more concrete, we ought to always keep this diffeomorphism freedom in mind.}
\begin{align}\label{EqMFGG}
\mathcal{M}_\text{new}\coloneqq \mathcal{F}_\text{man}(G_\text{trans}/G_\text{fix}). 
\end{align}
What makes $\mathcal{M}_\text{new}$ an abstract smooth manifold is that it is \textit{only} a smooth manifold.\footnote{This point can be relaxed if need be. If for some reason $\mathcal{M}_\text{new}$ needs to be equipped with more structure than that of a smooth manifold, one simply needs to change how forgetful $\mathcal{F}_\text{man}$ is. This possibility for relaxation is analogous to the relaxation of $\mathcal{F}_\text{vec}$ to allow for algebraic structure discussed in Ch.~\ref{ChapInternal}.} That is, no structure or operations are well-defined on $\mathcal{M}_\text{new}$ which are not also well defined on every other smooth manifold.

It is worth reflecting momentarily on where the new spacetime, $\mathcal{M}_\text{new}$, gets its smooth structure from. As the above definition indicates, $\mathcal{M}_\text{new}$ gets its smooth structure from the quotient manifold, $\mathcal{M}_\text{G}=G_\text{trans}/G_\text{fix}$. Namely, a transformation on $\mathcal{M}_\text{new}$ is smooth if and only if its pre-image under $\mathcal{F}_\text{man}$ is a smooth transformation on $\mathcal{M}_\text{G}$. But where does $\mathcal{M}_\text{G}$ get its smooth structure from? Ultimately, it gets it from the smooth structure of its numerator, $G_\text{trans}$, via the quotient with $G_\text{fix}$.\footnote{See Sec.~\ref{SecRevHomo} for details.} But where does $G_\text{trans}$ get its smooth structure from? $G_\text{trans}$ itself has a naturally motivated smooth structure coming from it being a finite-dimensional Lie group represented on the vector space, $V_\text{neutral}$. This is the ultimate source of $\mathcal{M}_\text{new}$'s (i.e., the new spacetime manifold's) continuity and smoothness.

As I have discussed in Sec.~\ref{SecRevHomo}, the fact that $G_\text{trans}$ gives $\mathcal{M}_\text{G}$ its smooth structure (and, now, consequently, $\mathcal{M}_\text{new}$) means that its group actions are naturally represented as diffeomorphisms on $\mathcal{M}_\text{G}$ (and, now, consequently, $\mathcal{M}_\text{new}$). Indeed, because $G_\text{trans}$ and $G_\text{fix}$ have \hyperlink{FPC}{\bf Fixed-Point Compatibility} these diffeomorphisms are a faithful representation of $G_\text{trans}$. Concretely, each $g\in G_\text{trans}$ has a natural smooth action on $\mathcal{M}_\text{G}$ given by,
\begin{align}\label{DefBarg}
\bar{g}([x])=[g\, x].    
\end{align}
Collecting these into a Lie group we have $\overline{G_\text{trans}}\subset\text{Diff}(\mathcal{M}_\text{G})$. Due to fixed-point compatibility with $G_\text{fix}$ we have $\overline{G_\text{trans}}\cong G_\text{trans}$. That is, $G_\text{trans}$ is faithfully represented as diffeomorphisms on $\mathcal{M}_\text{G}$.

The same holds true for $\mathcal{M}_\text{new}$. Each $g\in G_\text{trans}$ has a natural smooth action on $\mathcal{M}_\text{new}$, namely action by the following diffeomorphism:
\begin{align}\label{Hdef}
H(g)\coloneqq\mathcal{F}_\text{man}\circ \bar{g}\circ \mathcal{F}_\text{man}^{-1}.
\end{align}
This map, $H:G_\text{trans}\to \text{Diff}(\mathcal{M}_\text{new})$, associates to every element of $G_\text{trans}$ a diffeomorphism on $\mathcal{M}_\text{new}$. Collecting these into a group we have $H_\text{trans}\coloneqq H(G_\text{trans})\subset\text{Diff}(\mathcal{M}_\text{new})$. It follows from fixed point compatibility between $G_\text{trans}$ and $G_\text{fix}$ that we have \mbox{$H_\text{trans}\cong G_\text{trans}$}. That is, $G_\text{trans}$ is faithfully represented as diffeomorphisms on $\mathcal{M}_\text{new}$. Similarly, $G_\text{fix}$ is faithfully represented by $H_\text{fix}\coloneqq H(G_\text{fix})$.\footnote{It follows from the orbit stabilizer theorem that $H_\text{fix}$ is a stabilizer subgroup of $H_\text{trans}$ at $p_0=\mathcal{F}_\text{man}([\hat\openone])$.} 

This justifies the gloss of the externalization process given above: Externalization builds a new spacetime by takes some hand-picked pre-spacetime translation operations, $G_\text{fix}\subset G_\text{trans}\subset\text{GL}(V_\text{neutral})$, and letting them be honest-to-goodness spacetime translations, $H_\text{fix}\subset H_\text{trans}\subset\text{Diff}(\mathcal{M}_\text{new})$.

\section{Externalization: Injecting Field States}\label{SecInject}
The previous section has discussed how externalization builds new spacetime manifolds, $\mathcal{M}_\text{new}$, from pre-spacetime translation operations. Past this, however, externalization must also map the theory's states and dynamics onto this new spacetime manifold. In particular, externalization requires us to specify a values space, $\mathcal{V}_\text{new}$, for the new theory as well as a map, \mbox{$E:V_\text{neural}\to V_\text{new}^\text{all}$}, with $E:\bm{\varphi}\mapsto \varphi_\text{new}$ yielding the field states of the new theory. Here $V_\text{new}^\text{all}$ is the vector space of all $\mathcal{V}_\text{new}$-valued functions definable on $\mathcal{M}_\text{new}$.

What constraints does externalization put on our choice of this \mbox{$E:V_\text{neural}\to V_\text{new}^\text{all}$} map? In short, just as $G_\text{trans}$ and $G_\text{fix}$ are constrained to be pre-spacetime translations operations, the $E$ map is constrained as follows. In order for \mbox{$E:V_\text{neural}\to V_\text{new}^\text{all}$} to be a valid input for the externalization process we must have  $E(\bm{\varphi})=J(\bm\varphi)\circ\mathcal{F}_\text{man}^{-1}$ for one of the the $J:V_\text{neutral}\to V_\text{G}$ maps which \hyperlink{InjCompat}{\bf Injection Compatibility} guarantees the existence of. After a bit of rewriting, the requirements on \mbox{$E$} are as follows. $E:V_\text{neural}\to V_\text{new}^\text{all}$ must be a linear injective map with the following two properties: Firstly, we must have,
\begin{align}\label{EqEgh*E}
\forall g\in G_\text{trans} \ \forall \bm{\varphi}\in V_\text{neutral} \ \ \big[(E\circ g)\,\bm{\varphi} &= (H(g)^*\circ E)\,\bm{\varphi}\big].
\end{align}
That is, enacting any $g\in G_\text{trans}$ on any $\bm{\varphi}\in V_\text{neutral}$ as a linear map pre-externalization ought to be equivalent to enacting the corresponding diffeomorphism, $H(g)\in\text{Diff}(\mathcal{M}_\text{new})$, post-externalization. Secondly, we must have for all $d\in \text{Diff}(\mathcal{M}_\text{new})$,
\begin{align}\label{EKinRed}
\Big(\forall \bm{\varphi}\in V_\text{neutral} \ \  (d^*\circ E)(\bm{\varphi})=E(\bm{\varphi})\Big)\Longleftrightarrow d=\openone_{\mathcal{M}_\text{new}}.
\end{align}
That is, only the trivial diffeomorphism, $d=\openone_{\mathcal{M}_\text{new}}$, acts trivially on the image of $E:V_\text{neural}\to V_\text{new}^\text{all}$.

Above, I glossed the externalization process as taking some pre-spacetime translation operations and letting them be honest-to-goodness spacetime translations. We are, however, not free to pick how these pre-spacetime translation operations act on the field states. For every $g\in G_\text{trans}$ and every $\bm{\varphi}\in V_\text{neutral}$ we already have an action specified for them, namely $g\bm{\varphi}$. Eq.~\eqref{EqEgh*E} ensures that the way we map field states onto $\mathcal{M}_\text{new}$ (namely, $E:\bm{\varphi}\mapsto\varphi_\text{new}$) respects the way that we are mapping pre-spacetime translations onto $\mathcal{M}_\text{new}$ (namely, $H:g\mapsto h$). Moreover, as I will discuss in the next subsection,  Eq.~\eqref{EKinRed} guarantees that the new theory is kinematically reduced.

As I stressed above, the constraints here placed on $G_\text{trans}$, $G_\text{fix}$ and $E:V_\text{neural}\to V_\text{new}^\text{all}$ are not arbitrary. Given the general form of the externalization process, these constraints are both necessary and sufficient for the ISE Methodology satisfies \hyperlink{D2}{\bf Desiderata \#2}. See Appendix~\ref{AppD2Structure} for proof of necessity. The next section will prove of sufficiency.

Before this, however, a few quick consequences of these constraints on $E$ should be noted. First, recall that $V_\text{neutral}$ is the spacetime-neutral vector space of kinematic possibilities. Consequently, its image under $E$, namely \mbox{$V_\text{new}^\text{kin}\coloneqq E(V_\text{neutral})$}, will be the space of kinematic possibilities for the new spacetime theory. Because $E$ is constrained to be linear and injective we have $V_\text{new}^\text{kin}\cong V_\text{neutral}$. Recall, moreover, that internalization was carried out via a linear isomorphism. Hence we have $V_\text{new}^\text{kin}\cong V_\text{neutral}\cong V_\text{old}^\text{kin}$. Note, however, that despite this we might have different value spaces for the new and old theories, $\mathcal{V}_\text{new}\not\cong \mathcal{V}_\text{old}$. The isomorphism, $V_\text{new}^\text{kin}\cong V_\text{old}^\text{kin}$, does require, however, that $\mathcal{V}_\text{new}$ and $\mathcal{V}_\text{old}$ are defined over the same field, $K$.

\section{Proofs Regarding Externalization and Desiderata \#2}\label{SecD2}
As a first step towards proving that the ISE Methodology satisfies \hyperlink{D2}{\bf Desiderata \#2}, I will first prove that every spacetime theory produced via externalization has \hyperlink{SKC}{\bf Spacetime-Kinematic Compatibility}. It should be noted that the following proof does not in any way rely on the fact that we have so far assumed that the old theory about $\varphi_\text{old}$ has spacetime-kinematic compatibility. For instance, consider the \hyperlink{PMNPS}{\bf Periodic Matter on a Non-Periodic Spacetime} theory introduced in Ch.~\ref{ChapGenerality}. This theory is not kinematically reduced and so lacks spacetime-kinematic compatibility. Despite this, as I discuss in Appendix~\ref{ChapISEsansSKC}, internalizing this theory does yield a set of PSTOs, $G_\text{fix}^\text{(int)}\subset G_\text{trans}^\text{(int)}\subset \text{GL}(V_\text{neutral})$. If we use these PSTOs for externalization, the resulting theory will have spacetime-kinematic compatibility even though the input theory did not.

\subsection{\hra Proof that Externalization Guarantees Spacetime-Kinematic Compatibility}
To begin, let us prove that the new theory is kinematically reduced. To see this, first note that as discussed above, the image of $E$ is just \mbox{$V_\text{new}^\text{kin}\coloneqq E(V_\text{neutral})$}, the vector space of kinematically possible states for the new spacetime theory. As such, demanding Eq.~\eqref{EKinRed} is equivalent to demanding that the new spacetime theory is kinematically reduced, i.e., $H_\text{kin}=\{\openone_{\mathcal{M}_\text{new}}\}$. Hence, any spacetime theory produced via externalization will be kinematically reduced.

Let us next prove that any spacetime theory produced via externalization is kinematically homogeneous. To do so, I will prove that $H_\text{trans}=H(G_\text{trans})$ is a finite-dimensional Lie group which acts transitively and smoothly on $\mathcal{M}_\text{new}$ and is moreover kinematically allowed. All but the last of these claims follows directly from the $H:G_\text{trans}\to H_\text{trans}$ map being an isomorphism. This itself follows from $G_\text{trans}$ and $G_\text{fix}$ having \hyperlink{FPC}{\bf Fixed-Point Compatibility} and $\mathcal{F}_\text{man}:\mathcal{M}_\text{G}\to\mathcal{M}_\text{new}$ being an isomorphism. 

The final claim (that $H_\text{trans}$ is kinematically allowed) follows from Eq.~\eqref{EqEgh*E}. For every $h\in H_\text{trans}$ we can find a corresponding $g\in G_\text{trans}$ with $h=H(g)$. By Eq.~\eqref{EqEgh*E}, enacting this diffeomorphism, $h$, on any kinematically allowed state is equivalent to enacting $g$ pre-externalization and then externalizing with $E$. But the $E$ map always produces kinematically allowed states by definition. Hence, $h^*$ maps the new theory's kinematically allowed states onto each other. 

This completes the proof that spacetime theories produced via externalization always have spacetime-kinematic compatibility. Let us now leverage this fact to prove that the ISE Methodology satisfies \hyperlink{D2}{\bf Desiderata \#2 (D2)}.


\subsection{\hra Proof that the ISE Methodology Satisfies Desiderata \#2}
Consider any spacetime theory with \hyperlink{SKC}{\bf Spacetime-Kinematic Compatibility} regarding a field, $\varphi_\text{old}$, set on a manifold, $\mathcal{M}_\text{old}$, with kinematically allowed states, $V_\text{old}^\text{kin}$. Such a theory can be internalized as discussed in Ch.~\ref{ChapInternal}. As I have there discussed, its spacetime manifold faithfully and robustly survives internalization:
\begin{align}
\mathcal{M}_\text{old}\cong H_\text{trans}^\text{(int)}/H_\text{fix}^\text{(int)}
&\surint \mathcal{M}_\text{G}^\text{(int)}\coloneqq G_\text{trans}^\text{(int)}/G_\text{fix}^\text{(int)}& &\text{with }\mathcal{M}_\text{G}^\text{(int)}\cong\mathcal{M}_\text{old}\\
\nonumber
H_\text{trans}^\text{(int)}
&\surint G_\text{trans}^\text{(int)}& &\text{with }G_\text{trans}^\text{(int)}\cong H_\text{trans}^\text{(int)}\\
\nonumber
H_\text{fix}^\text{(int)}
&\surint G_\text{fix}^\text{(int)}& &\text{with }G_\text{fix}^\text{(int)}\cong H_\text{fix}^\text{(int)}.
\end{align}
Recall that this internalization is mediated by a forgetful functor $\mathcal{F}_\text{vec}:V_\text{old}^\text{kin}\to V_\text{neutral}$. As I have discussed in Sec.~\ref{SecSpecialG}, the Lie groups, $G_\text{trans}^\text{(int)}$ and $G_\text{fix}^\text{(int)}$, produced in this way are always pre-spacetime translation operations.\footnote{The full proof of this claim is given in Appendix~\ref{AppSpecialG}.} Concretely, $G_\text{fix}^\text{(int)}\subset G_\text{trans}^\text{(int)}$ have \hyperlink{FPC}{\bf Fixed-Point Compatibility} with each other and \hyperlink{InjCompat}{\bf Injection Compatibility} with $V_\text{neutral}$.

We can therefore begin the externalization process by taking $G_\text{trans}^\text{(ext)}=G_\text{trans}^\text{(int)}$ and $G_\text{fix}^\text{(ext)}=G_\text{fix}^\text{(int)}$. We can next generate a new spacetime manifold from these pre-spacetime translation operations as discussed in Sec.~\ref{SecNewMan}. Doing so we construct a new manifold diffeomorphic to the old one:
\begin{align}
\mathcal{M}_\text{new}
\cong \mathcal{M}_\text{G}^\text{(ext)}\coloneqq G_\text{trans}^\text{(ext)}/G_\text{fix}^\text{(ext)}
=G_\text{trans}^\text{(int)}/G_\text{fix}^\text{(int)}\eqqcolon \mathcal{M}_\text{G}^\text{(int)}
\cong\mathcal{M}_\text{old}.
\end{align}
Thus, we have reconstructed the original theory's spacetime manifold up to diffeomorphism, $\mathcal{M}_\text{new}\cong\mathcal{M}_\text{old}$.

To map the theory's kinematically allowed states onto this new manifold we need to pick a linear injector,
\begin{align}
E(\bm{\varphi})
&=J(\bm{\varphi})\circ\mathcal{F}_\text{man}^{-1},
\end{align}
where $J$ is one of the maps whose existence is guaranteed by \hyperlink{InjCompat}{\bf Injection Compatibility}. Recalling that $G_\text{trans}^\text{(ext)}=G_\text{trans}^\text{(int)}$ and $G_\text{fix}^\text{(ext)}=G_\text{fix}^\text{(int)}$ are pre-spacetime translation operations, we are guaranteed such a map exists. Indeed, as discussed in Sec.~\ref{SecSpecialG} the map $J(\bm{\varphi})\coloneqq\mathcal{F}_\text{vec}^{-1}(\bm{\varphi})\circ R$ satisfies has the desired properties.\footnote{For a full proof of this claim, see Appendix~\ref{AppSpecialG}.} Here $R$ is the diffeomorphism which maps the smooth manifold $\mathcal{M}_\text{G}^\text{(ext)}=\mathcal{M}_\text{G}^\text{(int)}$ back onto the original manifold $\mathcal{M}_\text{old}$. With this choice of $J$ we have,
\begin{align}
E(\bm{\varphi})
&=\mathcal{F}_\text{vec}^{-1}(\bm{\varphi})\circ R',
\end{align}
where $R'=R\circ\mathcal{F}_\text{man}^{-1}$ is a diffeomorphism relating $\mathcal{M}_\text{new}$ to $\mathcal{M}_\text{old}$.\footnote{Note that a choice of value space, $\mathcal{V}_\text{new}=\mathcal{V}_\text{old}$, is here implicit in our choice of linear injector, $E$.} 

Putting the internalization and externalization steps together, we have the following: Internalization maps field state $\varphi_\text{old}\in V_\text{old}^\text{kin}$ into $V_\text{neutral}$ as $\mathcal{F}_\text{vec}:\varphi_\text{old}\mapsto\bm{\varphi}$.\footnote{Consider a spacetime theory with \hyperlink{SKC}{\bf Spacetime Kinematic Compatibility} and non-linear dynamics. For instance, consider the \hyperlink{QKG}{\bf Quartic Klein Gordon} theory introduced in Ch.~\ref{ChapGenerality}. As I have discussed in Sec.~\ref{SecSurvivingDynamics}, the vector-based internalization process which I have used here needs to be relaxed somewhat to accommodate non-linear theories. Concretely, in such cases we must make the forgetful functor, $\mathcal{F}_\text{vec}$, less forgetful instead using a different forgetful functor, $\mathcal{F}_\text{alg}$, which can remember algebraic structure as well, e.g., how to take the product of fields states. Such theories can be accommodated by simply upgrading $V_\text{neutral}$ to $\mathcal{A}_\text{neutral}$, a spacetime-neutral algebra of kinematically allowed field states.} Following this, externalization maps field states from $\bm{\varphi}\in V_\text{neutral}$ onto the new spacetime as  $E:\bm{\varphi}\mapsto\varphi_\text{new}$.\footnote{For theories with non-linear dynamics, we will here have $V_\text{neutral}$ upgraded to $\mathcal{A}_\text{neutral}$, the spacetime-neutral algebra of kinematically allowed field states (see the previous footnote). We can easily accommodate this extra algebraic structure in the externalization process. We can simply use the linear injector, $E$, to induce the necessary algebraic structure in $E(V_\text{neutral})=V_\text{new}^\text{kin}$. That is, we can upgrade $V_\text{new}^\text{kin}$ to $\mathcal{A}_\text{new}^\text{kin}$, the algebra of kinematically allowed field states on $\mathcal{M}_\text{new}$. Thus, as I claimed in Ch.~\ref{ChapInternal}, any additional algebraic structure needed for non-linear theories can just come along for the ride in both the internalization and externalization steps. It will, however, play a more important role in the searching step, discussed in the next chapter.\label{FnExtAlg}} Composing these two maps we can see how the old theory's kinematically allowed states relate to those of the new theory. Doing so, we find,
\begin{align}
\varphi_\text{new}&=E(\mathcal{F}_\text{vec}(\varphi_\text{old}))=\varphi_\text{old}\circ R'.
\end{align}
That is, internalizing and then externalizing this theory has just directly transplanted all of the field states from $\mathcal{M}_\text{old}$ to $\mathcal{M}_\text{new}$.

The proof finishes as follows: Recall that we began from a generic spacetime theory with \hyperlink{SKC}{\bf Spacetime-Kinematic Compatibility}. Using the \hyperlink{ISE}{ISE Methodology} we then internalized and externalized this theory returning right back to where we started. This proves two things: Firstly, it proves that every spacetime theory with \hyperlink{SKC}{ spacetime-kinematic compatibility} can be produced via externalization. Together with the results of the previous subsection we now have a complete characterization of the spacetime theories producible via externalization. They are exactly the theories with \hyperlink{SKC}{spacetime-kinematic compatibility}. Secondly, the above discussion has proved that all theories producible via externalization (i.e., all theories with spacetime-kinematic compatibility) can be re-internalized and re-externalized trivially. That is, we have proved that the \hyperlink{ISE}{ISE Methodology} put forward in this dissertation satisfies \hyperlink{D2}{\bf Desiderata \#2}.\footnote{Said differently, $G_\text{trans}$, $G_\text{fix}$ and $E:V_\text{neural}\to V_\text{new}^\text{all}$ being constrained as discussed in Sec.~\ref{SecPSTO} is a sufficient condition for the \hyperlink{ISE}{ISE Methodology} to satisfy \hyperlink{D2}{\bf Desiderata \#2}. Given the general form of the externalization process, these constraints are also necessary. See Appendix~\ref{AppD2Structure} for proof.}

\chapter{Searching and Desiderata \#3}\label{ChapSearch}
The previous chapter has detailed the externalization process. It was there proved that a spacetime theory is producible via externalization if and only if it has \hyperlink{SKC}{\bf Spacetime-Kinematic Compatibility}. Following this, an explicit demonstration of \hyperlink{D2}{\bf Desiderata \#2 (D2)} was performed. Namely, it was shown how any theory producible via externalization can, after being re-internalized, always be re-externalized trivially.

Externalization must be able to do more than this, however. In order to satisfy  \hyperlink{D3}{\bf Desiderata \#3 (D3)}, we ought to have a variety of \hyperlink{PSTO}{\bf Pre-Spacetime Translation Operations} available to choose from. These ought to be able to lead us to a variety of inequivalent spacetime settings. This chapter will explicitly demonstrate that the \hyperlink{ISE}{ISE Methodology} satisfies \hyperlink{D3}{\bf D3} by working through several examples in which our theory is externalized onto a different spacetime setting than it started on. As discussed in Ch.~\ref{ChapOverview} and the end of Sec.~\ref{SecSurvivingMan}, \hyperlink{D3}{\bf D3} supports \hyperlink{D1}{\bf Desiderata \#1 (D1)}. It does so by showing that while the old spacetime manifold might survive internalization, we ultimately have no special commitment to it post-internalization. (Unless, of course, it is a uniquely good fit for the theory's dynamics and kinematics.) Consequently, this chapter will also provide additional support for the claim that internalization satisfies \hyperlink{D1}{\bf D1}.

\section{In Search of New Spacetimes}
A significant result of Ch.~\ref{ChapInternal} is that internalizing any spacetime theory with \hyperlink{SKC}{\bf Spacetime-Kinematic Compatibility} yields two Lie groups $G_\text{trans}^\text{(int)}$ and $G_\text{fix}^\text{(int)}$ which are \hyperlink{PSTO}{\bf Pre-Spacetime Translation Operations}. From this, we can easily prove the following result:\footnote{Note that this result does not require the theories in question to have linear dynamics.}
\begin{quote}
{\hypertarget{ISEequiv}{\bf ISE-Equivalence - }} Consider any two spacetime theories each with spacetime-kinematic compatibility. Let $V_\text{(1)}^\text{kin}$ and $V_\text{(2)}^\text{kin}$ be these theories' vector spaces of kinematically possible field states respectively. These two spacetime theories are interchangeable by the ISE Methodology if and only if the following condition holds: There exists a linear isomorphism between $V_\text{(1)}^\text{kin}$ and $V_\text{(2)}^\text{kin}$ which moreover maps one theory's dynamical solutions onto the other's and vice versa. Such spacetime theories will be called \textit{ISE-equivalent}.
\end{quote}
The proof of this claim is simple. Internalizing either theory (1) or theory (2) yields pre-spacetime translation operations on their respective spacetime-neutral vectors spaces, $V_\text{neutral}^\text{(1)}$ and $V_\text{neutral}^\text{(2)}$. The linear isomorphism between $V_\text{(1)}^\text{kin}$ and $V_\text{(2)}^\text{kin}$ induces a linear isomorphism between $V_\text{neutral}^\text{(1)}$ and $V_\text{neutral}^\text{(2)}$. Using this, we can transfer the pre-spacetime translation operations from one theory to the other and vice-versa. Externalizing the other theory's pre-spacetime translation operation yields the other theory's spacetime setting.

This notion of \hyperlink{ISEequiv}{ISE-Equivalence} is an equivalence relation over the set of all spacetime theories with spacetime-kinematic compatibility. Note that when we search for new pre-spacetime translation operations to externalize, we are effectively searching for a new spacetime setting within one of these equivalence classes. Said differently, given any spacetime theory with spacetime-kinematic compatibility, its ISE-equivalence class contains every possible spacetime framing of this theory which maintains its spacetime-kinematic compatibility. Recall from Ch.~\ref{ChapGenerality}, that assuming spacetime-kinematic compatibility is a very weak constraint. Hence, the ISE Methodology is a powerful tool for investigating and negotiating between a wide variety of spacetime settings for a wide variety of spacetime theories. 

Two examples of ISE-equivalent theories will be discussed momentarily. Before this, however, let us see an instructive non-example.

\subsection{\hra Example: Periodic vs Non-Periodic Spacetime}
As a first example, recall the \hyperlink{PMNPS}{\bf Periodic Matter on a Non-Periodic Spacetime} theory introduced in Ch.~\ref{ChapGenerality}. As I discussed in Ch.~\ref{ChapGenerality}, this theory is not kinematically reduced and therefore does not have spacetime-kinematic compatibility. Hence, by the results of the last chapter, this spacetime theory cannot be produced via externalization. Moreover, note that this theory falls outside of the scope of the above-defined ISE-equivalence relation. It follows from this that we cannot prove the \hyperlink{TRF}{\bf Topological Regularity Fact} of this theory.  Consequently, we cannot have a dynamics-first view of its spacetime manifold. 

We can, however, find a kinematically reduced version of this theory via a process of kinematic reduction (see Appendix~\ref{AppKinRedSKC}). Intuitively, kinematic reduction produces a kinematically reduced theory by identifying all spacetime points which cannot be kinematically distinguished. Taking this reduction, we find the \hyperlink{MPS}{\bf Matter on a Periodic Spacetime} theory introduced in Ch.~\ref{ChapGenerality}. This theory is kinematically reduced and moreover has spacetime-kinematic compatibility. As such, this theory does fall within the scope of the above-defined ISE-equivalence relation. Moreover, this theory can be produced via externalization. Consequently, we can prove of this theory the \hyperlink{TRF}{topological regularity fact}. We can thereby establish a dynamics-first view of its spacetime manifold.

There is an intuitive sense in which these two theories are equivalent to each other: Indeed, there is a linear one-to-one correspondence between their kinematically allowed states which preserves their dynamics. Despite this, these two spacetime theories are not ISE-equivalent. Technically, these theories are not even not equivalent: One of them falls outside of the scope of the equivalence relation. This does not mean, however, that these two theories are not related by the ISE Methodology. Indeed, in this case, the above discussed kinematic reduction can be mediated by the ISE Methodology.\footnote{See Appendix~\ref{AppKinRedISE}.} Despite the fact that the \hyperlink{PMNPS}{\bf Periodic Matter on a Non-Periodic Spacetime} theory lacks spacetime-kinematic compatibility, internalizing it can yield Lie groups, $G_\text{trans}^\text{(int)}$ and $G_\text{fix}^\text{(int)}$, which are pre-spacetime translation operations.\footnote{As I discuss in Appendix~\ref{AppKinRedISE}, this is possible because this theory has \hyperlink{WSKC}{\bf Weak Spacetime-Kinematic Compatibility}. Roughly, the Lie groups $G_\text{trans}^\text{(int)}$ and $G_\text{fix}^\text{(int)}$ produced via internalization will be pre-spacetime translation operations if the group $H_\text{trans}$ which we use to reconstruct $\mathcal{M}_\text{old}$ is large enough to cover the orbits of $H_\text{kin}$. See Appendix~\ref{AppKinRedISE} for details.} Consequently, in this case we can skip the searching step and proceed directly to externalization with $G_\text{trans}^\text{(ext)}=G_\text{trans}^\text{(int)}$ and $G_\text{fix}^\text{(ext)}=G_\text{fix}^\text{(int)}$. Pictorially, we can in this case take the theory's generalized Humean paint off the wall, and then put it straight back on. Doing so, however, does not give us back this theory's original spacetime setting. Indeed, as discussed above nothing can return this theory back to its original spacetime setting. Rather, externalizing these pre-spacetime translation operations yields the above-discussed \hyperlink{MPS}{\bf Matter on a Periodic Spacetime} theory.

In sum: While these two spacetime theories are in some sense intuitively equivalent, they are not ISE-equivalent. They are, however, ISE-related. One theory can be transformed into the other via the ISE Methodology, but not vice versa. In this way the DFSM view rejects the \hyperlink{PMNPS}{\bf Periodic-Matter on a Non-Periodic Spacetime} theory in favor of the \hyperlink{MPS}{\bf Matter on a Periodic Spacetime} theory. The first theory has a spacetime manifold in excess of what the theory's kinematics calls for. The second theory is preferred because its spacetime framing is a better fit for the theory's kinematics.

We are now ready to see two examples of \hyperlink{ISEequiv}{\bf ISE-Equivalent} theories. In each case, we will judge which spacetime framing is a better fit for the theory's dynamics and kinematics.

\subsection{\hra Example: Locality and Fourier Duality}
As a second example, consider the following two spacetime theories regarding modified Quantum Harmonic Oscillators (QHOs) labeled (1) and (2):
\begin{quote}
\hypertarget{NLQHO}{{\bf (1) Non-Local QHO-}} Consider a spacetime theory about a field, $\varphi_\text{old}^{(1)}:\mathcal{M}_\text{old}^{(1)}\to\mathcal{V}_\text{old}^{(1)}$, with \mbox{$\mathcal{M}_\text{old}^{(1)}\cong\mathbb{R}^2$} and $\mathcal{V}_\text{old}^{(1)}\cong\mathbb{C}$. The field states are subject to the following kinematic constraint: In some fixed global coordinate system, $(t,x)$, each field must be smooth. It must also be normalized and have a well-defined Fourier transform as well as inverse Fourier transform. In this coordinate system, the field states obey the following dynamics:
\begin{align}
\nonumber
\ii\partial_t\varphi_\text{old}^{(1)}(t,x)
&=(-\partial_x^2+x^2)\varphi_\text{old}^{(1)}(t,x)+ \frac{\lambda}{2} \varphi_\text{old}^{(1)}(t,x-a)+ \frac{\lambda}{2} \varphi_\text{old}^{(1)}(t,x+a),
\end{align}
for some fixed $a>0$ and $\lambda\in\mathbb{R}$. Equivalently using the identity, $h(x+a)=\exp(a\partial_x)h(x)$, the field states obey the following dynamics:
\begin{align}
\ii\partial_t\varphi_\text{old}^{(1)}(t,x)
&=(-\partial_x^2+x^2+\lambda\text{cosh}(a\partial_x))\,\varphi_\text{old}^{(1)}(t,x),
\end{align}
This is the standard QHO plus a non-local self-coupling.
\end{quote}
and
\begin{quote}
\hypertarget{CosQHO}{{\bf (2) Cosine QHO -}} Consider a spacetime theory about a field, $\varphi_\text{old}^{(2)}:\mathcal{M}_\text{old}^{(2)}\to\mathcal{V}_\text{old}^{(2)}$, with \mbox{$\mathcal{M}_\text{old}^{(2)}\cong\mathbb{R}^2$} and $\mathcal{V}_\text{old}^{(2)}\cong\mathbb{C}$. The field states are subject to the following kinematic constraint: In some fixed global coordinate system, $(\tau,q)$, each field must be smooth. It must also be normalized and have a well-defined Fourier transform as well as inverse Fourier transform. In this coordinate system, the field states obey the following dynamics:
\begin{align}
\ii\partial_\tau\varphi_\text{old}^{(2)}(\tau,q)
&=(-\partial_q^2+q^2+\lambda\,\text{cos}(a\,q))\,\varphi_\text{old}^{(2)}(\tau,q),
\end{align}
for some $a>0$ and $\lambda\in\mathbb{R}$. This is the standard QHO plus a cosine-shaped potential.
\end{quote}
Note that while these theories' spacetimes are isomorphic $\mathcal{M}_\text{old}^{(1)}\cong\mathcal{M}_\text{old}^{(2)}$ they are nonetheless distinct, $\mathcal{M}_\text{old}^{(1)}\neq\mathcal{M}_\text{old}^{(2)}$.

Both of these theories have spacetime-kinematic compatibility. Thus, by applying the results of the previous chapter, we can internalize and then externalize these theories trivially. Said differently, they are both ISE-equivalent to themselves. Moreover, these theories are ISE-equivalent to each other. A keen reader may have noticed already that these theories are related by a Fourier transform. As such, we should be able to switch between these two spacetime settings using the ISE Methodology.

Let us see how this works moving from theory (1) to theory (2). Suppose that we internalize theory (1) using the Lie group, $H_\text{trans}^\text{(int)}\cong(\mathbb{R}^2,+)$, which implements rigid translations in the $(t,x)$ coordinate system. Note that the stabilizer group in this case is trivial, $H_\text{fix}^\text{(int)}=\{\openone_{\mathcal{M}_\text{old}}\}$. These Lie groups, $H_\text{trans}^\text{(int)}$ and $H_\text{fix}^\text{(int)}$, survive internalization faithfully becoming $G_\text{trans}^\text{(int)}\cong(\mathbb{R}^2,+)$ and $G_\text{fix}^\text{(int)}=\{\hat\openone\}$. That is, they become Lie groups represented on $V_\text{neutral}^{(1)}$. If we were to externalize this theory using this $G_\text{trans}^\text{(int)}$ and $G_\text{fix}^\text{(int)}$ we would recover the theories initial spacetime setting as discussed in Ch.~\ref{ChapExternal}. This, would be analogous to peeling the generalized Humean paint off of the wall and then putting it straight back on. Unlike in the previous example, doing so in this case returns us the theory's original spacetime setting.

We are not forced, however, to externalize in this way. We have available to us other pre-spacetime translation operations, besides the Lie group, $G_\text{trans}^\text{(int)}$, which is associated with rigid translations in the $(t,x)$ coordinate system. That is, we could instead chose to externalize a different set of pre-spacetime translation operations, $G_\text{trans}^\text{(ext)}\neq G_\text{trans}^\text{(int)}$ and/or $G_\text{fix}^\text{(ext)}\neq G_\text{fix}^\text{(int)}$. In particular, we might choose $G_\text{trans}^\text{(ext)}\cong(\mathbb{R}^2,+)$ to be the the linear maps on $V_\text{neutral}^{(1)}$ associated with translations in the $k_x$ and $t$ coordinates in Fourier space. It follow from $G_\text{trans}^\text{(ext)}\cong(\mathbb{R}^2,+)$ being abelian, that we must here pick $G_\text{fix}^\text{(ext)}=\{\hat\openone\}$ in order to satisfy \hyperlink{FPC}{\bf Fixed-Point Compatibility}. Externalizing these pre-spacetime translation operations results in a new spacetime manifold, $\mathcal{M}_\text{new}^{(1)}\cong \mathbb{R}^2$.

While we here have an isomorphism between the new and old spacetimes, $\mathcal{M}_\text{new}^{(1)}\cong \mathcal{M}_\text{old}^{(1)}$, they are nonetheless distinct manifolds as they are built from different pre-spacetime translation operations. This difference becomes more visible when we attempt to transplant the field states onto the new spacetime with a linear injection $E:V_\text{neutral}^{(1)}\to V_\text{new}^\text{all}$ satisfying Eq.~\eqref{EqEgh*E}. This equation enforces that the pre-spacetime translations $G_\text{trans}^\text{(ext)}\cong(\mathbb{R}^2,+)$ become honest-to-goodness spacetime translation operations on the new spacetime. Hence, what were rigid translations in Fourier space in the old theory must become rigid translations in spacetime in the new theory. Ultimately, this forces us to map the Fourier transform of the old field states onto the new spacetime. Doing so (and carrying theory (1)'s states and dynamics along for the ride) results in theory (2) introduced above. Thus, while theory (1) \textit{can} be externalized back onto itself, it can also be externalized onto theory (2). The same is true in reverse as well, we can go from theory (2) back to theory (1). Hence, theories (1) and (2) are ISE-equivalent.

How should we interpret this situation? 
As noted above, both of these spacetime theories have spacetime-kinematic compatibility. As such, we can prove the \hyperlink{TRF}{\bf Topological Regularity Fact} for both theory (1) and theory (2). We can therefore establish a dynamics-first view of either theory's spacetime manifold. Moreover, as the above discussion has revealed, the ISE Methodology can be used to switch between these two theories at will. In line with \hyperlink{D3}{\bf Desiderata \#3}, these inequivalent spacetime settings have been achieved by externalizing different pre-spacetime translation operations. Correspondingly, these theories can be seen as different spacetime codifications of the same dynamical behavior of matter. While in this case the two spacetime manifolds happen to be diffeomorphic, the ways in which they codify the dynamical behavior of matter are markedly different.

The competition as to which of these spacetime settings is a better codification of the dynamical behavior of matter is analogous to the competition between different law-like codifications on Lewis's Best Systems Analysis. Whereas the rough criteria there are achieving the best balance of simplicity and strength, the rough criteria here are achieving the spacetime setting which best fits with the theory's dynamics and kinematics. In the present case, we have good reason to think that the second theory is a better spacetime codification than the first theory: Namely, the theory's dynamics is local in the second spacetime setting.

\vspace{0.25cm}
Before moving on to our final example, this QHO example should be discussed from a Hilbert space perspective. As discussed in Ch.~\ref{ChapInternal}, internalization reduces a spacetime theory down to its vectorial (or, if necessary, algebraic) core. As these theories have linear dynamics, post-internalization we are left with spacetime-independent field states $\bm{\varphi}^{(1)}\in V_\text{neutral}^{(1)}$ and $\bm{\varphi}^{(2)}\in V_\text{neutral}^{(2)}$ respectively. At minimum, $V_\text{neutral}^{(1)}$ and $V_\text{neutral}^{(2)}$ are merely abstract vector spaces over $\mathbb{C}$. If the old theories have physically salient complex inner products, we can arrange for this structure to survive internalization as well (see Sec.~\ref{SecSurvivingDynamics}). In doing so, we can upgrade $V_\text{neutral}^{(1)}\to H_\text{neutral}^{(1)}$ and $V_\text{neutral}^{(2)}\to H_\text{neutral}^{(2)}$ to Hilbert spaces. Note, however, that $H_\text{neutral}^{(1)}$ and $H_\text{neutral}^{(2)}$ are not the kind of Hilbert spaces which are commonly associated with non-relativistic quantum theories such (1) and (2). The elements, $\bm{\varphi}^{(1)}=H_\text{neutral}^{(1)}$ and $\bm{\varphi}^{(2)}=H_\text{neutral}^{(2)}$, of these Hilbert space correspond to entire histories of $\varphi_\text{old}^{(1)}(t,x)$ and $\varphi_\text{old}^{(1)}(\tau,q)$ not instantaneous states. 

We can arrive at the usual Hilbert space descriptions of these theories as follows. We can pick $G_\text{trans}^\text{(ext)}\cong(\mathbb{R},+)$ to be the linear maps on $V_\text{neutral}^{(1)}$ associated with translations the $t$-coordinate in the old theory. Because $G_\text{trans}^\text{(ext)}\cong(\mathbb{R},+)$ is abelian, we must pick $G_\text{fix}^\text{(ext)}=\{\hat\openone\}$ in order to satisfy \hyperlink{FPC}{\bf Fixed-Point Compatibility}. Externalizing these pre-spacetime translation operations results in a new spacetime manifold, $\mathcal{M}_\text{new}^{(1)}\cong \mathbb{R}$, of only time. By construction, time translation in this new theory corresponds to time translation in the $(t,x)$ coordinate system of the old theory. Note that the $x$-coordinate in the old theory, however, has remained internalized.

In order to map the theory's field states onto this new spacetime we need to pick out a new value space, $\mathcal{V}_\text{new}$, and linear injector, $E:V_\text{neutral}^{(1)}\to V_\text{new}^\text{all}$. Here $V_\text{new}^\text{all}$ is the vector space of all $\mathcal{V}_\text{new}$-valued functions on $\mathcal{M}_\text{new}^{(1)}\cong \mathbb{R}$. The new theory's field states, $\varphi_\text{new}^{(1)}=E(\bm{\varphi}^{(1)})$, are at each time $t\in\mathcal{M}$ vectors in $\mathcal{V}_\text{new}$. That is, $\varphi_\text{new}^{(1)}(t)\in \mathcal{V}_\text{new}$. We can take this new value space $\mathcal{V}_\text{new}$ to be the usual Hilbert space of non-relativistic quantum mechanics. Doing so, and picking some compatible injector, $E$, we arrive at the usual Hilbert space description of theory (1).
\begin{quote}
\hypertarget{HilbertQHO}{{\bf (3) Hilbert Space Cosine QHO -}} Consider a spacetime theory about a field, $\varphi_\text{old}^{(3)}:\mathcal{M}_\text{old}^{(3)}\to\mathcal{V}_\text{old}^{(3)}$, with \mbox{$\mathcal{M}_\text{old}^{(2)}\cong\mathbb{R}$} and $\mathcal{V}_\text{old}^{(3)}\cong\mathbb{C}^\infty$. The field states are subject to the following kinematic constraint: In some fixed global coordinate system, $\tau$, each field must be smooth. The field states must also be normalized at each $\tau$. In this coordinate system, the field states obey the following dynamics:
\begin{align}
\ii\partial_\tau\varphi_\text{old}^{(3)}(\tau)
&=(\hat{p}^2+\hat{q}^2+\lambda\,\text{cos}(a\,\hat{q}))\,\varphi_\text{old}^{(3)}(\tau),
\end{align}
for some $a>0$ and $\lambda\in\mathbb{R}$. The operators $\hat{q}$ and $\hat{p}$ here are the usual canonically conjugate Hermitian operators defined on the Hilbert space, $\mathcal{V}_\text{old}^{(3)}\cong\mathbb{C}^\infty$.
\end{quote}
We now have three ISE-equivalent spacetime descriptions of this theory: one with time only, and two with differing notions of space. The ISE Methodology can be used to easy negotiate between these (and other) spacetime settings for this theory. This methodology can similarly be applied to explore various spacetime settings any theory set in a Hilbert space. For instance, recall the \hyperlink{SES}{\bf Schr\"odinger Equation on a Sphere} theory discussed in Chap.~\ref{ChapGenerality}.

\subsection{\hra Example: Discrete vs Continuous Spacetime}
Before stating the final two spacetime theories under consideration, some mathematical definitions are needed. As some of the following relates to discrete spacetime theories (i.e., lattice theories), the following matrices related to various discrete approximations of the derivative will be useful.
\begin{align}\label{BigToeplitz}
\Delta_{(1)}^2\coloneqq\text{Toeplitz}(&\dots,0,0,0,0,1,{\bf-2},1,0,0,0,0,\dots)\\
\Delta_{(2)}^2\coloneqq\text{Toeplitz}(&\dots,0,0,0,\frac{-1}{12},\frac{4}{3},{\bf \frac{-5}{2}},\frac{4}{3},\frac{-1}{12},0,0,0,\dots)\\
D^2\coloneqq\text{Toeplitz}(&\dots,\frac{2}{25},\frac{-2}{16},\frac{2}{9},\frac{-2}{4},\frac{2}{1},{\bf \frac{-2\pi^2}{6}},\frac{2}{1},\frac{-2}{4},\frac{2}{9},\frac{-2}{16},\frac{2}{25},\dots)\\
\label{Ddef}D\coloneqq\text{Toeplitz}(&\dots,\frac{-1}{5},\frac{1}{4},\frac{-1}{3},\frac{1}{2},-1,{\bf 0},1,\frac{-1}{2},\frac{1}{3},\frac{-1}{4},\frac{1}{5},\dots).
\end{align}
Toeplitz matrices are so called diagonal-constant matrices with \mbox{$[A]_{i,j}=[A]_{i+1,j+1}$}. Thus, the values listed in the above expression give the matrix's values on either side of the main diagonal. The value in the middle of the list (in bold) corresponds to the main diagonal. Past this, five values to either side of the main diagonal are given. 

Note that $\Delta_{(1)}^2$ and $\Delta_{(2)}^2$ are related to the nearest-neighbor and next-to-nearest-neighbor approximations of the second derivative. Namely,
\begin{align}
\Delta_{(1)}^2/\epsilon^2:\ \partial_x^2 f(x)
&\approx\frac{f(x+\epsilon)-2 f(x)+f(x-\epsilon)}{\epsilon^2}\\
\nonumber
\Delta_{(2)}^2/\epsilon^2:\ \partial_x^2 f(x)
&\approx\frac{-f(x+2\epsilon)+16\,f(x+\epsilon)-30\, f(x)+16\,f(x-\epsilon)-f(x+2\epsilon)}{12\,\epsilon^2}
\end{align}
These can be easily generalized to, $\Delta_{(n)}^2$, the Toeplitz matrix related to the $\text{(next-to)}^n$-nearest-neighbor approximations of the second derivative. Taking the limit $n\to\infty$ gives $D^2\coloneqq \lim_{n\to\infty}\Delta_{(n)}^2$. In this way, $D^2$ is an infinite-range discrete approximation of the second derivative. In the same way, $D$ is an infinite-range discrete approximation of the first derivative. $D^2$ is the square of $D$.

Given these definitions, we can now state the following two spacetime theories:
\begin{quote}
\hypertarget{MDS}{{\bf Matter on a Discrete Spacetime}}\footnote{In this example, space is discrete whereas time is continuous. For an example in which both space and time are discrete, see ~\cite{DiscreteGenCovPart2}.} - Consider a spacetime theory about a field, $\varphi_\text{old}:\mathcal{M}_\text{old}\to\mathcal{V}_\text{old}$, with $\mathcal{M}_\text{old}\cong\mathbb{R}\times\mathbb{Z}$ and $\mathcal{V}_\text{old}\cong\mathbb{R}$. The field states are subject to the following kinematic constraint: In some fixed global coordinate system, $(\tau,n)$, the vector of field values, $\vec{\varphi}_\text{old}(\tau)\coloneqq\{\varphi_n(\tau)\}_{n\in\mathbb{Z}}$, must have a finite $L^2$-norm at each $\tau$.

There are three variants of this theory obeying the following three dynamical equations respectively: 
\begin{align}
{\bf H1:}& \quad \frac{\d }{\d\tau}\vec{\varphi}_\text{old}(\tau)
=\alpha\,\Delta_{(1)}^2\, \vec{\varphi}_\text{old}(\tau)\\
{\bf H2:}& \quad \frac{\d }{\d\tau}\vec{\varphi}_\text{old}(\tau)
=\alpha\,\Delta_{(2)}^2\, \vec{\varphi}_\text{old}(\tau)\\
{\bf H3:}& \quad \frac{\d }{\d\tau}\vec{\varphi}_\text{old}(\tau)
=\alpha\,D^2\, \vec{\varphi}_\text{old}(\tau)
\end{align}
for some $\alpha>0$. These are the nearest neighbor (H1), next-to-nearest neighbor (H2), and infinite-range (H3) discrete heat equations respectively.
\end{quote}
\begin{quote}
\hypertarget{BMCS}{{\bf Bandlimited Matter on a Continuous Spacetime -}} Consider a spacetime theory about a  field, $\varphi_\text{old}:\mathcal{M}_\text{old}\to\mathcal{V}_\text{old}$, with $\mathcal{M}_\text{old}\cong\mathbb{R}^2$ and $\mathcal{V}_\text{old}\cong\mathbb{R}$. The field states are subject to the following kinematic constraint. In some fixed global coordinate system, $(t,x)$, each field, $\varphi_\text{old}(t,x)$, has a well-defined Fourier transform, $\tilde{\varphi}_\text{old}(\omega,k)$, with the following property: $\tilde{\varphi}_\text{old}(\omega,k)=0$ unless $\omega\in[-\Omega,\Omega]$ and $k\in[-K,K]$ for some fixed bandwidths, $\Omega>0$ and $K>0$. Moreover, the field values, $\varphi_\text{old}(t,x)$, must have a finite $L^2$-norm at each $t$.

As in the previous theory, there are three variants of this theory obeying the following three dynamical equations respectively:
\begin{align}
{\bf H1:}&\quad\partial_t\varphi_\text{old}(t,x)
=\alpha \, [2\text{cosh}(a\partial_x)-2] \ \varphi_\text{old}(t,x)\\
{\bf H2:}&\quad\partial_t\varphi_\text{old}(t,x)=\frac{\alpha}{6} [-\text{cosh}(2a\partial_x)+16\text{cosh}(a\partial_x)-15]\varphi_\text{old}(t,x)\\
{\bf H3:}&\quad \partial_t\varphi_\text{old}(t,x)
=\frac{\alpha}{a^2}\,\partial_x^2 \, \varphi_\text{old}(t,x)
\end{align}
for some $\alpha>0$. Note that H3 is the standard heat equation. H1 and H2 are similar, but featuring some non-locality at length scales or order $\sim a$.
\end{quote}
As in the QHO example, both of these theories have spacetime-kinematic compatibility. Consequently, we can prove of them the \hyperlink{TRF}{\bf Topological Regularity Fact} and thereby establish a dynamics-first view of their spacetime manifold. 

Perhaps surprisingly, there exists a linear isomorphism between these theories kinematically allowed states which preserves their dynamics. Thus, these theories are \hyperlink{ISEequiv}{\bf ISE-Equivalent}: The ISE Methodology can be used to switch between these theories at will. In line with \hyperlink{D3}{\bf Desiderata \#3}, these inequivalent spacetime settings will be achieved by externalizing different pre-spacetime translation operations.\footnote{The continuous spacial translations in the bandlimited theory, $\varphi(t,x)\mapsto \varphi(t,x+a)=\exp(a\partial_x)\varphi(t,x)$, are associated with the linear transformations, $\text{exp}(a D)$, in the discrete theory. Defined in Eq.~\eqref{Ddef}, $D$ is the best infinite-range discrete approximation of the derivative. See \cite{DiscreteGenCovPart1} for further discussion.} Correspondingly, these discrete and bandlimited theories can be seen as different spacetime codifications of the same dynamical behavior of matter. As in the QHO example, there is a competition here as to which of these spacetime settings is a better codification of the dynamical behavior of matter. 

The competition between these discrete and continuous spacetime settings is non-trivial. For instance, unlike in the previous example, the two spacetime manifolds under consideration here are topologically inequivalent ($\mathbb{R}\times\mathbb{Z}\not\cong\mathbb{R}^2$). Correspondingly, these theories have differing notions of locality which makes a straight-forward comparison on these grounds impossible. For an extended treatment this discrete-vs-continuous spacetime competition, see \cite{DiscreteGenCovPart1}.\footnote{An extended treatment of the analogous Klein Gordon equations is given in \cite{DiscreteGenCovPart2}. The discrete spacetime theories there are discrete in both space and time. Moreover, the bandlimited theories have a (non-approximate) Lorentzian symmetry. Despite what you've heard, there are perfectly Lorentzian lattice theories.} I ultimately find reason to favor the continuous spacetime framing. My analysis of this example is largely inspired by the work of the mathematical physicist Achim Kempf.\footnote{For an overview of Kempf's work see~\cite{Kempf2018} along with his other works~\cite{Kempf_1997,UnsharpKempf,Kempf2000b,Kempf2003,Kempf2004a,Kempf2004b,Kempf2006,Martin2008,Kempf_2010,Kempf2013,Pye2015,Kempf2018,KempfDistCorr}. See also~\cite{PyeThesis,Pye2022,BEH_2020}} See, for instance, \cite{Kempf_2010} entitled, ``Spacetime could be simultaneously continuous and discrete, in the same way that information can be''. It was, in fact, Kempf's work which first motivated me to look into the philosophical implications of the existence of spacetime theories with multiple inequivalent spacetime settings. 

\vspace{0.5cm}
To review: This chapter has shown the full range of spacetime theories which are interchangeable via the ISE Methodology, namely an ISE-equivalence class. These equivalence classes contain every possible spacetime framing of a theory which maintains its spacetime-kinematic compatibility. This chapter has also shown several examples of spacetime theories which are related and/or interchangeable via the ISE Methodology. In particular, two non-trivial examples of ISE-equivalent theories have been discussed. The existence of a variety of interchangeable spacetime settings for any given theory demonstrates that the \hyperlink{ISE}{ISE Methodology} satisfies \hyperlink{D3}{\bf Desiderata \#3}. Moreover, it also helps to demonstrate that \hyperlink{D1}{\bf Desiderata \#1} is satisfied. Post-internalization we no longer have any special attachment to the theory's original spacetime framing, except insomuch as it is uniquely good fit to the theory's dynamics and/or kinematics.

\chapter{Conclusion}\label{ChapConclusion}
Analogous to both Lewis's Best Systems Analysis and Huggett's Regularity Relationism, I have here put forward a dynamics-first view of the spacetime manifold (the DFSM view). At the heart of this view is the following metaphysical claim: The spacetime manifold is merely a codification of certain patterns in the dynamical behavior of matter. In particular, the spacetime manifold is a codification of the following topological regularity fact:
\begin{quote}
{\hyperlink{TRF}{\bf Topological Regularity Fact - }}
\textit{We can conceptualize the dynamical behavior of matter in a spacetime-independent way: Namely, via internalization. From this spacetime-neutral perspective we can consider various ways of picking out which pre-spacetime translation operations to externalize. One way of doing so in particular gives rise to nice geometric laws set on a smooth topological manifold.}
\end{quote}
As I discussed in Ch.~\ref{ChapInternal}, internalization conceptually divorces the dynamical behavior of matter from any assumed topological underpinnings. Namely, when applied to any spacetime theory meeting certain preliminary assumptions, internalization reduces this theory down to its vectorial (or, if necessary, algebraic) core. Perhaps surprisingly, a great deal of topological information about the old spacetime setting can survive the internalization process. Indeed, given a certain level of spacetime-kinematic compatibility, the old theory's spacetime manifold can be seen to survive internalization both faithfully and robustly. Conveniently, all of this surviving topological information is contained neatly within a set of \hyperlink{PSTO}{\bf Pre-Spacetime Translation Operations (PSTOs)}. In fact, this is exactly what pre-spacetime translation operations are: They are the minimal vectorial structure which contains complete topological information about a theory's spacetime setting. 

By externalizing the pre-spacetime translation operations delivered to us by the internalization process, one can recover the theory's original spacetime setting perfectly. Importantly, however, we also have other pre-spacetime translation operations available to us. By externalizing these, we can re-describe the same dynamical behavior of matter (understood non-topologically) in different spacetime settings. The search for good pre-spacetime translation operations is analogous to searching for good ways of bending and stretching the detached paint sheet on Huggett's Regularity Relationism. Moreover, it is analogous to the search for good ways of codifying the dynamical behavior of matter on Lewis's Best Systems Analysis. There the criteria by which codifications are judged are a balance of simplicity and strength. Here, the criteria are how well the resulting spacetime fits the theory's dynamics and kinematics.

This dissertation has developed the \hyperlink{ISE}{\bf ISE Methodology} in order to facilitate the above-discussed process of 1) internalizing the theory, 2) searching for new and improved PSTOs, and then 3) externalizing them into a new spacetime setting for our theory.
This is a powerful mathematical tools which allows us to investigate and negotiate between a wide variety of spacetime settings for a wide range of spacetime theories. Our ability to externalize into a wide range of inequivalent spacetime settings tells us something important. It gives us additional reason to believe that internalization has successfully divorced the dynamical behavior of matter from its topological underpinnings. While we can always choose to go back to the theory's original spacetime setting we do not have to. Indeed, post-internalization we no longer have any special attachment to the theory's original spacetime framing, except insomuch as it is uniquely good fit to the theory's dynamics and/or kinematics.

As I have argued in Ch.~\ref{ChapSecondVsFirst}, the primary benefit of dynamics-first views generally (of laws, geometry, spacetime, etc.) is their metaphysical sparsity. In advancing this the DFSM view, I have brought the metaphysical sparsity characteristic of dynamics-first views to the philosophy of space and time. The metaphysical commitments of the DFSM view are very light (e.g., no spacetime points contra spacetime substantivalism, and no spacetime relations contra spacetime relationalism).The only metaphysical commitments needed in support of the DFSM view is to whatever non-topological substrate which spacetime is a codification of. Moreover, unlike views of spacetime emergence, the DFSM view requires no limit taking, no regime changes, and, importantly, no new speculative physics.

In response to the DFSM view, two metaphysical questions immediately arise: Spacetime is a codification of patterns in what substrate? And, moreover, which patterns? The focus of this dissertation has been almost entirely on the second question; See the above discussion of pre-spacetime translation operations. This answer does, however, shed some light on the first question. The non-topological substrate which the spacetime is a codification of has here been modelled mathematically as being algebraic (or, at minimum, vectorial) in nature. In order to address this question, therefore, a theory of algebraic metaphysics is needed. While delivering such a theory is beyond the scope of this dissertation, some progress has here been made in that direction. In particular, this dissertation has developed the mathematics necessary to do two things: Firstly, to divorce the dynamical behavior of matter from its topological underpinnings via internalization; Secondly, to recover our intuitive descriptions of spacetime topology by codifying patterns in this non-topological substrate. Whatever this algebraic stuff/goings-on ultimately are (metaphysically speaking) the details of this recovery are suggestive of how it can be related to our usual descriptions of spacetime physics.


{\singlespacing
\setcitestyle{open={\textcolor{black}{(}},close={\textcolor{black}{)}}}
\bibliographystyle{dcu}
\bibliography{Bibliography}}


\begin{appendices}
\chapter{The ISE Methodology without Spacetime-Kinematic Compatibility}\label{ChapISEsansSKC}
Ch.~\ref{ChapInternal} - Ch.~\ref{ChapSearch} introduced the \hyperlink{ISE}{\bf ISE Methodology} with a focus on spacetime theories which satisfy \hyperlink{SKC}{\bf Spacetime Kinematic Compatibility}. This appendix will provide an expanded treatment of this methodology which does not assume spacetime kinematic compatibility. 

Before this, however, Sec.~\ref{AppSpecialG} will prove one of the key results of this dissertation (one which does assume spacetime-kinematic compatibility). Recall from Ch.~\ref{ChapInternal} that for any theory with spacetime-kinematic compatibility its spacetime manifold survives internalization faithfully and robustly. Moreover, applied to any such theory internalization yields two Lie groups, $G_\text{trans}^\text{(int)}$ and $G_\text{fix}^\text{(int)}$, represented on $V_\text{neutral}$. As I will prove in Sec.~\ref{AppSpecialG}, the Lie groups produced in this way are guaranteed to be \hyperlink{PSTO}{\bf Pre-Spacetime Translation Operations (PSTOs)}.\footnote{This result will be generalized in Sec.~\ref{AppKinRedISE} to only assume \hyperlink{WSKC}{\bf Weak Spacetime-Kinematic Compatibility}.} Recall from Ch.~\ref{ChapExternal} that the externalization process takes as input PSTOs. Thus, one option available to us (among others) is to externalize the PSTOs produced via internalization. As I proved in Ch.~\ref{ChapExternal}, doing so returns the theory to its original spacetime setting. Importantly, this fact can be leveraged to prove that the ISE Methodology satisfies \hyperlink{D2}{\bf Desiderata \#2}.

As mentioned above, the goal of this appendix is to generalize this discussion of the ISE Methodology by dropping the assumption that the old spacetime theory has spacetime-kinematic compatibility. To this end, Sec.~\ref{AppRFSKC} will answer the following questions: Can the spacetime manifold of a theory without spacetime-kinematic compatibility survive internalization? If so, under what conditions does it do so faithfully and robustly? The answers are as follows: A theory's spacetime manifold survives internalization if and only if the theory is kinematically homogeneous. For this theory's spacetime to survive internalization faithfully and robustly, it is both necessary and sufficient that the theory has spacetime-kinematic compatibility.

As this discussion will reveal, whenever a theory is kinematically homogeneous, its spacetime manifold survives internalization via a process of (partial) kinematic reduction. In Sec.~\ref{AppKinRedSKC} I will extend this notion of kinematic reduction from spacetime manifolds to spacetime theories. In particular, I will show how kinematic reduction can be used to bolster a theory's spacetime kinematic compatibility. As its name suggests, kinematic reduction produces kinematically reduced spacetime theories.\footnote{I here mean full (not partial) kinematic reduction. Applied to spacetime theories which are already kinematically reduced, this process acts as the identity.} This process can therefore be helpful in producing spacetime theories with spacetime-kinematic compatibility from theories lacking this compatibility. As I will prove, the process of kinematic reduction does not interfere with a theory's kinematic homogeneity. That is, if a spacetime theory is kinematically homogeneous before kinematic reduction, it still will be afterwards. In fact, as I will show, kinematic reduction can even enhance a theory's kinematic homogeneity.

Sec.~\ref{AppKinRedISE} will show how the process of kinematic reduction relates to the ISE Methodology. As I will prove, kinematic reduction can be achieved via the ISE Methodology given what I will call \hyperlink{WSKC}{\bf Weak Spacetime-Kinematic Compatibility}. Internalizing this theory will yield some Lie groups, $G_\text{trans}^\text{(int)}$ and $G_\text{fix}^\text{(int)}$, represented on $V_\text{neutral}$. Because this theory only has weak spacetime-kinematic compatibility, we may be outside of the scope of Sec.~\ref{AppSpecialG}. Hence, we have no proof (so far) that $G_\text{trans}^\text{(int)}$ and $G_\text{fix}^\text{(int)}$ are PSTOs. As I will prove in Sec.~\ref{AppKinRedISE}, however, given the right choice of $H_\text{trans}$, these are, in fact, PSTOs. Making this choice and externalizing these PSTOs yields the kinematic reduction of our original theory.

Finally, Sec.~\ref{AppD2Structure} will justify the constraints which $G_\text{trans}^\text{(ext)}$, $G_\text{fix}^\text{(ext)}$, and $E:V_\text{neutral}\to V_\text{new}^\text{all}$ must satisfy to be considered valid inputs to the externalization process. As I will there discuss, these constraints are necessary for the ISE Methodology to satisfy \hyperlink{D2}{\bf Desiderata \#2}.

\section{Given Spacetime-Kinematic Compatibility, \texorpdfstring{$G_\text{trans}$}{Gtrans} and \texorpdfstring{$G_\text{fix}$}{Gfix} are PSTOs}\label{AppSpecialG}
This section will prove that, given spacetime-kinematic compatibility, the Lie groups produced by internalization, $G_\text{trans}^\text{(int)}$ and $G_\text{fix}^\text{(int)}$, will always be \hyperlink{PSTO}{\bf Pre-Spacetime Translation Operations (PSTOs)}. This claim was put forward in Sec.~\ref{SecSpecialG}. I have already proved there that $G_\text{trans}$ and $G_\text{fix}$ have all of the necessary properties except for \hyperlink{InjCompat}{\bf Injection Compatibility} with $V_\text{neutral}$. Hence, to complete this proof we only need to show that there exists an value space, $\mathcal{V}_\text{G}$, and a  injective linear map, \mbox{$J:V_\text{neutral}\to V_\text{G}$}, with the following two properties:\footnote{Recall that the vector space $V_\text{G}$ is the space of all $\mathcal{V}_\text{G}$-valued functions definable over $\mathcal{M}_\text{G}=G_\text{trans}/G_\text{fix}$.} Firstly, we must have,
\begin{align}\label{AppInj1}
&\forall g \ \forall \bm{\varphi} \ \ \Big((J\circ g)\,\bm{\varphi} = (\bar{g}^*\circ J)\,\bm{\varphi}\Big),
\end{align}
and, secondly, for all diffeomorphisms, $d\in \text{Diff}(G_\text{trans})$, we must have,
\begin{align}\label{AppInj2}
&\Big( \forall g \ \forall \bm{\varphi} \ \  J(\bm{\varphi})([d(g)])=J(\bm{\varphi})([g])\Big)\Longleftrightarrow [d(g)]=[g].
\end{align}
In both of these expressions (and throughout this section) the ranges of $g\in G_\text{trans}$, $\varphi_\text{old}\in V_\text{old}^\text{kin}$, and $\bm{\varphi}\in V_\text{neutral}$ will be suppressed for notational convenience.

To find a map, $J$, satisfying Eq.~\eqref{AppInj1}, let us first identify the diffeomorphism, $R:\mathcal{M}_\text{G}\to\mathcal{M}_\text{old}$, which relates $\mathcal{M}_\text{G}$ to $\mathcal{M}_\text{old}$ as in Eq.~\eqref{ManSur1}. Concretely, this $R:\mathcal{M}_\text{G}\to\mathcal{M}_\text{old}$ map acts on points $[g]\in\mathcal{M}_\text{G}$ as follows,
\begin{align}
R([g])\coloneqq G^{-1}(g)(p_0)   
\end{align}
where $p_0\in\mathcal{M}_\text{old}$ is the point at which $H_\text{fix}$ is the stabilizer subgroup of $H_\text{trans}$. $G$ here is the isomorphism relating $H_\text{trans}$ and $G_\text{trans}$ discussed in Ch.~\ref{ChapInternal}. This is an isomorphism because the old spacetime theory is here assumed to be kinematically reduced. Note that this map is well-defined as the right-hand-side is independent of which representative of $[g]$ is used.

In addition to this diffeomorphism, $R$, relating $\mathcal{M}_\text{G}$ and $\mathcal{M}_\text{old}$ we also have another nice relationship between these smooth manifolds. Namely, the Lie group $G_\text{trans}\subset\text{GL}(V_\text{neutral})$ is faithfully represented as diffeomorphisms on both $\mathcal{M}_\text{G}$ and $\mathcal{M}_\text{old}$. Namely, we have $G_\text{trans}\cong\overline{G_\text{trans}}\subset\text{Diff}(\mathcal{M}_\text{G})$.\footnote{Recall that $\bar{g}$ acts on $\mathcal{M}_\text{G}$ as $\bar{g}([x])=[g\,x]$. The isomorphism, $G_\text{trans}\cong\overline{G_\text{trans}}$, follows from \hyperlink{FPC}{\bf Fixed-Point Compatibility} between $G_\text{trans}$ and $G_\text{fix}$.} Moreover, we have $G_\text{trans}\cong H_\text{trans}\subset\text{Diff}(\mathcal{M}_\text{old})$.\footnote{This follows from our assumption that the old spacetime theory is kinematically reduced.} Concretely, in terms in terms of the $R$ map introduced above for each $g\in G_\text{trans}$ we have $\bar{g}\in\overline{G_\text{trans}}$ and $h=G^{-1}(g)\in H_\text{trans}$ related as follows:\footnote{This is easy to confirm, Evaluating the right-hand-side at a point $[g_0]\in \mathcal{M}_\text{G}$ yields, $(R\circ \bar{g})([g_0])=G^{-1}(g\,g_0)(p_0)$. Likewise, evaluating the left-hand-side at a point $[g_0]\in \mathcal{M}_\text{G}$ yields, $(h\circ R)([g_0])=(G^{-1}(g)\,G^{-1}(g_0))(p_0)$. These are identical since $G$ is a group isomorphism.}
\begin{align}
\forall g
\ \  h\circ R=R\circ \bar{g}.
\end{align}
For any $\varphi_\text{old}\in V_\text{old}^\text{kin}$ we can apply these diffeomorphism beforehand yielding,
\begin{align}
\forall g \ \forall \varphi_\text{old}
\ \ 
\varphi_\text{old}\circ h\circ R
=\varphi_\text{old}\circ R\circ\bar{g}.
\end{align}
Using the definition of the $*$-map we then have,
\begin{align}
\forall g \ \forall \varphi_\text{old}
\ \  
(h^* \varphi_\text{old})\circ R=
\bar{g}^*(\varphi_\text{old}\circ R).
\end{align}
Recalling that we have a one-to-one correspondence between $\varphi_\text{old}\in V_\text{old}^\text{kin}$ and $\bm{\varphi}=\mathcal{F}_\text{vec}(\varphi_\text{old})\in V_\text{neutral}$ we then have,
\begin{align}
\forall g \ \forall \bm{\varphi}
\ \  
(h^*\mathcal{F}_\text{vec}^{-1}(\bm{\varphi}))\circ R = \bar{g}^*(\mathcal{F}_\text{vec}^{-1}(\bm{\varphi})\circ R).   
\end{align}
Recalling that $g$ and $h=G^{-1}(g)$ are related as $g=G(h)=\mathcal{F}_\text{vec}\circ h^*\circ \mathcal{F}_\text{vec}^{-1}$ we then have,
\begin{align}
\forall g \ \forall \bm{\varphi}
\ \  
\mathcal{F}_\text{vec}^{-1}(g\,\bm{\varphi})\circ R=\bar{g}^*(\mathcal{F}_\text{vec}^{-1}(\bm{\varphi})\circ R).   
\end{align}
This proves Eq.~\eqref{AppInj1} for $J(\bm{\varphi})=\mathcal{F}_\text{vec}^{-1}(\bm{\varphi})\circ R$. Note that implicit in this choice of $J$ is a choice of value space, $\mathcal{V}_\text{G}=\mathcal{V}_\text{old}$.

Similarly, one can prove Eq.~\eqref{AppInj2} for $J(\bm{\varphi})=\mathcal{F}_\text{vec}^{-1}(\bm{\varphi})\circ R$ as follows. Given that the old theory is kinematically reduced, we have $H_\text{kin}=\{\openone_{\mathcal{M}_\text{old}}\}$. From this it follows that for all $d_\text{old}\in\text{Diff}(\mathcal{M}_\text{old})$ we have,
\begin{align}
&\Big(\forall \varphi_\text{old} \ \   \varphi_\text{old}\circ d_\text{old}=\varphi_\text{old}\Big)\Longleftrightarrow d_\text{old}=\openone_{\mathcal{M}_\text{old}}.
\end{align}
Taking $\varphi_\text{old}=\mathcal{F}_\text{vec}^{-1}(\bm{\varphi})$ and composing with $R$ we then have,
\begin{align}
&\Big(\forall \bm{\varphi} \ \   J(\bm{\varphi})\circ R^{-1}\circ d_\text{old}\circ R=J(\bm{\varphi})\Big)\Longleftrightarrow d_\text{old}=\openone_{\mathcal{M}_\text{old}}.
\end{align}
The diffeomorphisms, $d_\text{old}$, on $\mathcal{M}_\text{old}$ are in one-to-one correspondence with diffeomorphisms, $d_\text{G}=R^{-1}\circ d\circ R$, on $\mathcal{M}_\text{G}$. Consequently, we then have that for all 
$d_\text{G}\in\text{Diff}(\mathcal{M}_\text{G})$,
\begin{align}
&\Big(\forall \bm{\varphi} \ \   J(\bm{\varphi})\circ d_\text{G}=J(\bm{\varphi})\Big)\Longleftrightarrow d_\text{G}=\openone_{\mathcal{M}_\text{G}}.
\end{align}
Evaluating these functions on $\mathcal{M}_\text{G}$ at all points $[g]\in\mathcal{M}_\text{G}$ we have,
\begin{align}
&\Big(\forall g \ \forall \bm{\varphi} \ \   J(\bm{\varphi})(d_\text{G}([g]))=J(\bm{\varphi})([g])\Big)\Longleftrightarrow d_\text{G}=\openone_{\mathcal{M}_\text{G}}.
\end{align}
Finally, recall how the smooth transformations, $d_\text{G}$, on $\mathcal{M}_\text{G}$ are defined. The map $[g]\mapsto d_\text{G}([g])$ is smooth if and only if there exists a smooth map, $d\in\text{Diff}(G_\text{trans})$, with $d_\text{G}([g])=[d(g)]$. Applying this to the above equation proves Eq.~\eqref{AppInj2} for $J(\bm{\varphi})=\mathcal{F}_\text{vec}^{-1}(\bm{\varphi})\circ R$.

\section{Robust and Faithful Manifold Survival Requires Spacetime-Kinematic Compatibility}\label{AppRFSKC}
Ch.~\ref{ChapInternal} showed that given a theory with \hyperlink{SKC}{\bf Spacetime-Kinematic Compatibility}, its spacetime manifold will faithfully survive internalization as in Eq.~\eqref{ManSur1}:
\begin{align}\label{AppManSur}
\mathcal{M}_\text{old}\cong\mathcal{M}_\text{H}\coloneqq \frac{H_\text{trans}}{H_\text{fix}}
\surint
\mathcal{M}_\text{G}\coloneqq \frac{G_\text{trans}}{G_\text{fix}}\cong\frac{H_\text{trans}}{H_\text{fix}}\cong \mathcal{M}_\text{old}.
\end{align}
Note, moreover, that given spacetime-kinematic compatibility the theory's spacetime survives internalization faithfully independently of which Lie group, $H_\text{trans}$, we have used to reconstruct $\mathcal{M}_\text{old}$. That is, $\mathcal{M}_\text{old}$ survives not only faithfully but also robustly. This section will prove the converse result: If a theory's spacetime manifold is to survive internalization faithfully and robustly then it must have spacetime-kinematic compatibility.

In order to prove this result, I will now re-trace the discussion in Ch.~\ref{ChapInternal} regarding how topological information about $\mathcal{M}_\text{old}$ can survive internalization. The key difference here is that spacetime-kinematic compatibility is not assumed.\footnote{Note, however, that the \hyperlink{Prelim}{\bf Preliminary Assumptions} discussed in Ch.~\ref{ChapGenerality} are still being assumed. This is necessary for the internalization process to even begin.} As before, individual kinematically-allowed diffeomorphisms, $d\in\text{Diff}_\text{kin}(\mathcal{M}_\text{old})$, survive internalization as follows:
\begin{align}
d^*\surint G(d)\coloneqq\mathcal{F}_\text{vec}\circ d^*\vert_\text{kin}\circ\mathcal{F}_\text{vec}^{-1}.
\end{align}
That is, the map $G:\text{Diff}_\text{kin}(\mathcal{M}_\text{old})\to G_\text{old}$ maps kinematically allowed diffeomorphisms onto linear maps on $V_\text{neutral}$. As before, if the Lie group, $\text{Diff}_\text{kin}(\mathcal{M}_\text{old})$, is to survive internalization (now, potentially unfaithfully) then $G$ must be a Lie group homomorphism. This will happen exactly when $\text{ker}(G)=H_\text{kin}$ is a closed and normal subgroup of $\text{Diff}_\text{kin}(\mathcal{M}_\text{old})$.

The following computation confirms that $H_\text{kin}\lhd \text{Diff}_\text{kin}(\mathcal{M}_\text{old})$ is a normal subgroup of $\text{Diff}_\text{kin}(\mathcal{M}_\text{old})$. As I will now show, for all $d\in\text{Diff}_\text{kin}(\mathcal{M}_\text{old})$ and all $h\in H_\text{kin}$ we have $d^{-1}\,h\,d\in H_\text{fix}$. This follows from the fact that for all $\varphi_\text{old}\in V_\text{old}^\text{kin}$ we have,
\begin{align}
(d^{-1}\,h\,d)^*(\varphi_\text{old})=
d^{-1}{}^*\,h^*\,d^*(\varphi_\text{old})=\varphi_\text{old}.
\end{align}
To see this, note that $d^*(\varphi_\text{old})$ keeps $\varphi_\text{old}$ in $V_\text{old}^\text{kin}$ where $h^*$ acts trivially. $(d^{-1})^*=(d^*)^{-1}$ then undoes the action of $d^*$. Thus we have $H_\text{kin}\lhd \text{Diff}_\text{kin}(\mathcal{M}_\text{old})$.

Moreover, $H_\text{kin}$ is a closed subgroup of $\text{Diff}_\text{kin}(\mathcal{M}_\text{old})$. To see this, first note that $H_\text{kin}=G^{-1}(\{\hat\openone\})$ is the pre-image of the singleton, $\{\hat\openone\}$, under the map \mbox{$G:\text{Diff}_\text{kin}(\mathcal{M}_\text{old})\to G_\text{old}$}. Because $G$ is continuous and the singleton is a closed set, it follows that $H_\text{kin}$ is a closed set as well. Taken together, we therefore have that:
\begin{align}
\text{Diff}_\text{kin}(\mathcal{M}_\text{old})\surint G_\text{old}\cong \text{Diff}_\text{kin}(\mathcal{M}_\text{old})/H_\text{kin}.
\end{align}
That is, up to diffeomorphism, $\text{Diff}_\text{kin}(\mathcal{M}_\text{old})$ survives internalization as a Lie group by being quotiented by $H_\text{kin}$, the theory's kinematically trivial group.

The above discussion can be easily generalized to analyze the survival of an arbitrary subgroup of the kinematically allowed diffeomorphisms, $H\subset\text{Diff}_\text{kin}(\mathcal{M}_\text{old})$. Restricting the $G$ map to this subgroup as $G\vert_H$  yields a Lie group homomorphism if and only if restricting $H_\text{kin}$ to $H$ yields a closed and normal subgroup of $H$. One can easily check that $\text{ker}(G\vert_H)=H_\text{kin}\cap H$ is both a normal and a closed subgroup of $H$. In fact, the above arguments used in the $H=\text{Diff}_\text{kin}(\mathcal{M}_\text{old})$ case apply unchanged. Thus, for every kinematically allowed Lie group, $H\subset\text{Diff}_\text{kin}(\mathcal{M}_\text{old})$, we have that,
\begin{align}
H\surint G(H)\cong H/(H_\text{kin}\cap H).
\end{align}
That is, up to diffeomorphism, $H\subset\text{Diff}_\text{kin}(\mathcal{M}_\text{old})$ survives internalization as a Lie group by being quotiented by $H_\text{kin}\cap H$, its kinematically trivial subgroup.

Next, recall from
Sec.~\ref{SecRevHomo} that any smooth homogeneous manifold can be reconstructed up to diffeomorphism as a quotient of two Lie subgroups of its diffeomorphisms. Concretely, if the old theory's spacetime, $\mathcal{M}_\text{old}$, is homogeneous, then applying this result gives the following reconstruction:
\begin{align}\label{AppMoldHH}
\mathcal{M}_\text{old}\cong \mathcal{M}_\text{H}\coloneqq H_\text{trans}/H_\text{fix},
\end{align}
for any transitively-acting finite-dimensional Lie group, $H_\text{trans}\subset \text{Diff}(\mathcal{M}_\text{old})$, and any one of its stabilizer subgroups, $H_\text{fix}\subset H_\text{trans}$. Recall from Sec.~\ref{SecRevHomo} than any such $H_\text{trans}$ and $H_\text{fix}$ together satisfy \hyperlink{FPC}{\bf Fixed-Point Compatibility}.

Given these results (i.e., that \textit{some} diffeomorphisms can survive internalization (potentially unfaithfully), and that \textit{some} manifolds can be reconstructed in terms of their diffeomorphisms) it seem conceivable that the spacetime manifolds of \textit{some} theories might also survive internalization (potentially faithfully). Concretely, the theory's manifold might survive internalization as a quotient of its surviving diffeomorphisms. 

Ch.~\ref{ChapInternal} has discussed this possibility under the assumption of spacetime-kinematic compatibility. To review: Assuming kinematic homogeneity guarantees that some such Lie group, $H_\text{trans}$, is kinematically allowed. Therefore, both $H_\text{trans}$ and its stabilizer subgroup, $H_\text{fix}$, survive internalization. Moreover, assuming that the old theory is kinematically reduced, both of these survive internalization faithfully. Consequently, the theory's spacetime survives internalization faithfully and robustly as a quotient of these surviving Lie groups. This section will prove the converse result: A theory's spacetime manifold will survive internalization faithfully and robustly only if the theory has spacetime-kinematic compatibility.

To begin this proof, let us first see that kinematic homogeneity is a necessary condition for the theory's spacetime to survive internalization simpliciter (i.e., faithfully or not). Firstly, in order for us to be able to reconstruct the old theory's spacetime manifold, $\mathcal{M}_\text{old}$, using Eq.~\eqref{AppMoldHH}, it is necessary that  $\mathcal{M}_\text{old}$ is homogeneous. In particular, there must exist some finite-dimensional Lie group of diffeomorphisms, $H_\text{trans}\subset\text{Diff}(\mathcal{M}_\text{old})$, which acts transitively on $\mathcal{M}_\text{old}$. Secondly, in order for $\mathcal{M}_\text{old}$ to survive internalization \textit{as a quotient of its surviving diffeomorphisms} the Lie group appearing in the numerator, $H_\text{trans}$, must itself survive internalization as a Lie group. That is, $H_\text{trans}$ must be kinematically allowed. Putting these two requirements together, we have the following result: For a theory's spacetime to survive internalization (faithfully or not), the theory must be kinematically homogeneous.

As I prove later in this section, being kinematically homogeneous is not only necessary but also sufficient for a theory's spacetime manifold to survive internalization (potentially unfaithfully).
\begin{quote}
{\hypertarget{CondManSur}{\bf Conditions for Manifold Survival - }} Consider a spacetime theory set on a smooth manifold, $\mathcal{M}_\text{old}$, with kinematically allowed states, $V_\text{old}^\text{kin}$. This theory's manifold, $\mathcal{M}_\text{old}$, survives internalization (potentially unfaithfully) if and only if it is kinematically homogeneous. Under this condition, the theory's spacetime manifold survives internalization as,
\begin{align}\label{ManSur0}
\mathcal{M}_\text{old}\cong \frac{H_\text{trans}}{H_\text{fix}}\surint
\mathcal{M}_\text{G}\coloneqq \frac{G_\text{trans}}{G_\text{fix}}\cong\frac{\mathcal{M}_\text{old}}{H_\text{kin}\cap H_\text{trans}}.
\end{align}
for some  Lie groups, $G_\text{trans}$ and $G_\text{fix}$, each represented on $V_\text{neutral}$. 
\end{quote}
Note that, per this claim, the way in which the theory's spacetime manifold survives internalization can depend on which Lie group, $H_\text{trans}$, we have used to reconstruct $\mathcal{M}_\text{old}$. In particular, a theory's spacetime might survive internalization faithfully for one $H_\text{trans}$ but not for another. From the above claim the following follows:
\begin{quote}
{\hypertarget{CondFRManSur}{\bf Conditions for Faithful (and Robust) Manifold Survival - }} Consider a kinematically homogeneous spacetime theory set on a smooth manifold, $\mathcal{M}_\text{old}$, with kinematically allowed states, $V_\text{old}^\text{kin}$. This theory's manifold, $\mathcal{M}_\text{old}$, survives internalization faithfully,
\begin{align}
\mathcal{M}_\text{old}\cong \frac{H_\text{trans}}{H_\text{fix}}\surint
\mathcal{M}_\text{G}\coloneqq \frac{G_\text{trans}}{G_\text{fix}}\cong\mathcal{M}_\text{old}.
\end{align}
if and only if $H_\text{kin}\cap H_\text{trans}=\{\openone_{\mathcal{M}_\text{old}}\}$. Moreover, the spacetime will survive internalization faithfully and robustly (i.e., independently of $H_\text{trans}$, up to diffeomorphism) if and only if this theory is kinematically reduced, with $H_\text{kin}=\{\openone_{\mathcal{M}_\text{old}}\}$.
\end{quote}
\textit{Proof:} The necessary and sufficient conditions for faithful survival from directly from Eq.~\eqref{ManSur0}. In order for the theory's spacetime to survives internalization robustly we must additionally demand the following: $H_\text{kin}\cap H_\text{trans}=\{\openone_{\mathcal{M}_\text{old}}\}$ for all $H_\text{trans}$. Here $H_\text{trans}$ ranges over all finite-dimensional, kinematically-allowed, transitively-acting Lie groups $H_\text{trans}\subset\text{Diff}_\text{kin}(\mathcal{M}_\text{old})$. If $H_\text{kin}$ has a trivial intersection with each such $H_\text{trans}$ then it must have a trivial intersection with their union. The union of all such $H_\text{trans}$ is $\text{Diff}_\text{kin}(\mathcal{M}_\text{old})$, the full group of kinematically allowed diffeomorphisms. Recall that $H_\text{kin}$ is a subset of $\text{Diff}_\text{kin}(\mathcal{M}_\text{old})$. Combining $H_\text{kin}\subset\text{Diff}_\text{kin}(\mathcal{M}_\text{old})$ and $H_\text{kin}\cap \text{Diff}_\text{kin}(\mathcal{M}_\text{old})=\{\openone_{\mathcal{M}_\text{old}}\}$ we have that $H_\text{kin}=\{\openone_{\mathcal{M}_\text{old}}\}$. This completes the proof.

Note that this is exactly the desired result of this section: If a theory's spacetime is to survive internalization faithfully and robustly then it must have spacetime-kinematic compatibility (i.e., it must be both kinematically reduced and kinematically homogeneous). Thus, all that remains to do is to prove the \hyperlink{CondManSur}{\bf Conditions for Manifold Survival} claim discussed above. 

The necessity of kinematic homogeneity for the theory's spacetime manifold to survive internalization, has already been established. I will now prove that kinematic homogeneity is also sufficient for spacetime survival by using this assumption to explicitly construct the surviving manifold. In the process, I will validate Eq.~\eqref{ManSur0}. 

First, note that kinematic homogeneity of the old spacetime theory guarantees the existence of a finite-dimensional Lie group, $H_\text{trans}\subset\text{Diff}(\mathcal{M}_\text{old})$, which is kinematically allowed and acts transitively over $\mathcal{M}_\text{old}$. Using the fact that $H_\text{trans}$ acts transitively on $\mathcal{M}_\text{old}$, we can pick out one of its stabilizer subgroups, $H_\text{fix}\subset H_\text{trans}$, and reconstruct $\mathcal{M}_\text{old}$ up to diffeomorphisms as \mbox{$\mathcal{M}_\text{old}\cong \mathcal{M}_\text{H}= H_\text{trans}/H_\text{fix}$}.\footnote{See Sec.~\ref{SecRevHomo} for details.}

Next, note that since (by assumption) $H_\text{trans}$ is kinematically allowed it survives internalization:
\begin{align}
H_\text{trans}
&\surint G_\text{trans}\coloneqq G(H_\text{trans})\cong H_\text{trans}/(H_\text{kin}\cap H_\text{trans}),
\end{align}
for some Lie groups $G_\text{trans}\subset \text{GL}(V_\text{neutral})$ represented on $V_\text{neutral}$. Note that because $H_\text{fix}$ is also kinematically allowed, it too survives internalization as follows:
\begin{align}
H_\text{fix}
&\surint G_\text{fix}\coloneqq G(H_\text{fix})\cong H_\text{fix}/(H_\text{kin}\cap H_\text{fix})
\end{align}
In total, we therefore have the spacetime manifold surviving internalization as a quotient of its diffeomorphisms: Namely,
\begin{align}
\mathcal{M}_\text{old}\cong\mathcal{M}_\text{H}\coloneqq \frac{H_\text{trans}}{H_\text{fix}}
\surint
\mathcal{M}_\text{G}\coloneqq \frac{G_\text{trans}}{G_\text{fix}}\cong\frac{H_\text{trans}/(H_\text{kin}\cap H_\text{trans})}{H_\text{fix}/(H_\text{kin}\cap H_\text{fix})}.
\end{align}
This proves sufficiency. I will now show that the surviving manifold $\mathcal{M}_\text{G}$ is diffeomorphic to $\mathcal{M}_\text{old}/(H_\text{kin}\cap H_\text{trans})$ as claimed in Eq.~\eqref{ManSur0}.

This quotient can be cleaned up with a bit of rewriting. For brevity, let us define $K_0\coloneqq H_\text{kin}\cap H_\text{trans}$ to be the kinematically trivial part of $H_\text{trans}$. Note that it follows from $H_\text{fix}\subset H_\text{trans}$ that we have $H_\text{kin}\cap H_\text{fix}=K_0\cap H_\text{fix}$. In these terms we have, 
\begin{align}\label{MGQuotients}
\mathcal{M}_\text{G}\cong\frac{H_\text{trans}/K_0}{H_\text{fix}/(K_0\cap H_\text{fix})}\cong\frac{H_\text{trans}/K_0}{(H_\text{fix} K_0)/K_0}\cong\frac{H_\text{trans}}{H_\text{fix} K_0}
\end{align}
The first step follows from the definition of $K_0$. The second step follows from the Isomorphism Theorem of group theory. The final step follows by cancelling the common factor of $K_0$ which is a normal subgroup of both $H_\text{trans}$ and $H_\text{fix}K_0$.

The appearance of the factors $H_\text{trans}/H_\text{fix}$ within in the right hand-side of Eq.~\eqref{MGQuotients} suggests that we may be able to understand $\mathcal{M}_\text{G}$ up to diffeomorphism directly in terms of $\mathcal{M}_\text{old}$ and $K_0$. As I will now show, this is, in fact, possible. Concretely, from Eq.~\eqref{MGQuotients} we have $\mathcal{M}_\text{G}\cong H_\text{trans}/\equiv_{FK}$ for the equivalence relation, $\equiv_{FK}$, defined below. For every $t_1,t_2\in H_\text{trans}$ we have
\begin{align}
t_1 \equiv_{FK} t_2 &\text{ iff } \exists f\in H_\text{fix} \ \exists k\in K_0 \ \ t_1=t_2\,k\,f\\
\nonumber
&\text{ iff } \exists f\in H_\text{fix} \ \exists k\in K_0 \ \ t_1=k\,t_2\,f\\
\nonumber
&\text{ iff } \exists k\in K_0 \ \ t_2^{-1}\,k^{-1}\,t_1\in H_\text{fix}\\
\nonumber
&\text{ iff } \exists k\in K_0 \ \ ( t_2^{-1}\,k^{-1}\,t_1)(p_0)=p_0\\
\nonumber
&\text{ iff } \exists k\in K_0 \ \ t_1(p_0)=k(t_2(p_0))
\end{align}
where the first step follows from $K_0$ being a normal subgroup of $H_\text{trans}$. The latter steps follow from $H_\text{fix}$ being $H_\text{trans}$'s stabilizer subgroup at $p_0$. Thus, two diffeomorphisms, $t_1,t_2\in H_\text{trans}$, are equivalent under $\equiv_{FK}$ if they map $p_0$ to places which can be mapped onto each other by some kinematically trivial diffeomorphism, $k\in K_0$. 

Recall that pre-internalization we reconstructed the spacetime as $\mathcal{M}_\text{old}\cong H_\text{trans}/\equiv_{p_0}$ using the equivalence relation $\equiv_{p_0}$ defined in Sec.~\ref{SecRevHomo}. Namely, diffeomorphisms in $H_\text{trans}$ are equivalent under $\equiv_{p_0}$ if they mapped $p_0$ to the same place. Note that equivalence under $\equiv_{p_0}$ implies equivalence under $\equiv_\text{FK}$. Indeed, $\equiv_\text{FK}$ is a courser relation than $\equiv_{p_0}$. From this it follows that every $\equiv_\text{FK}$-equivalence class is the union of $\equiv_{p_0}$-equivalence classes. Indeed we can understand each $\equiv_\text{FK}$-equivalence class as an equivalence class of $\equiv_{p_0}$-equivalence classes. Recall from Sec.~\ref{SecRevHomo} that $\equiv_{p_0}$-equivalence classes (e.g., $[h]_{p_0}$) are in one-to-one correspondence with spacetime points (e.g., $h(p_0)\in\mathcal{M}_\text{old}$). We can thus realize $\equiv_\text{FK}$ as an equivalence relation defined over $\mathcal{M}_\text{old}$ as follows:
\begin{align}\label{EquivK}
p_1 \equiv_{K_0} p_2 &\text{ iff } \exists k\in K_0 \ \ k(p_1)=p_2.
\end{align}
These equivalence classes are $K_0$-orbits. But this is just what is meant by the quotient in Eq.~\eqref{ManSur0}:
\begin{align}\label{LastEq}
\mathcal{M}_\text{G}\cong\frac{\mathcal{M}_\text{old}}{\equiv_{K_0}}=\frac{\mathcal{M}_\text{old}}{K_0}=\frac{\mathcal{M}_\text{old}}{H_\text{kin}\cap H_\text{trans}}.
\end{align}
This validates Eq.~\eqref{ManSur0} as desired.

\section{Kinematic Reduction Always Increases Spacetime-Kinematic Compatibility}\label{AppKinRedSKC}
The previous section showed that given any kinematically homogeneous spacetime theory, its spacetime manifold, $\mathcal{M}_\text{old}$, survives internalization becoming (up to diffeomorphism) a smaller manifold, $\mathcal{M}_\text{old}/K_0$. That is, $\mathcal{M}_\text{old}$ survives by being quotiented by $K_0\subset H_\text{kin}$, some subset of the theory's kinematically trivial diffeomorphisms. 

Let us call this a (partial) kinematic reduction of the theory's spacetime manifold. The word ``partial'' here refers to the fact that $K_0$ may just be a small part of $H_\text{kin}$. Notice that as $K_0$ becomes bigger, the manifold $\mathcal{M}_\text{old}/K_0$ becomes smaller. A kinematic reduction will be called ``full'' if $K_0$ is large enough that increasing its size further (towards $K_0=H_\text{kin}$) would not reduce the size of the quotient manifold any further (a more technical definition will be given momentarily). In cases of full kinematic reduction, the parenthetical ``(partial)'' will be dropped.

This section will extend this notion of (partial) kinematic reduction in two ways: Firstly, by dropping the assumption that the old theory is kinematically homogeneous; Secondly, by applying this reduction to an entire spacetime theory, not just its spacetime manifold. It is important to note that the process of kinematic reduction defined here is entirely separate from the ISE Methodology. They are, however, closely related as I will discuss in Sec.~\ref{AppKinRedISE}. 

This section will describe a process of (partial) kinematic reduction which can be applied to any spacetime theory satisfying the \hyperlink{Prelim}{\bf Preliminary Assumptions} discussed in Ch.~\ref{ChapGenerality}. As I will discuss, when applied to a spacetime theory which is not kinematically reduced, full (but not partial) kinematic reduction yields a spacetime theory which is kinematically reduced. This process can therefore be helpful in producing spacetime theories with \hyperlink{SKC}{\bf Spacetime-Kinematic Compatibility} from theories lacking this compatibility. As I will prove below, the process of kinematic reduction does not interfere with a theory's kinematic homogeneity. That is, if a spacetime theory is kinematically homogeneous before kinematic reduction, it still will be afterwards. In fact, as a coming example will show, kinematic reduction can even enhance a theory's kinematic homogeneity. 

For any spacetime theory satisfying the preliminary assumptions discussed in Ch.~\ref{ChapGenerality}, we can identify the theory's group of kinematically trivial diffeomorphisms, $H_\text{kin}$. The new spacetime manifold put forward following (partial) kinematic reduction theory is $\mathcal{M}_\text{red}\coloneqq\mathcal{M}_\text{old}/K_0$ for some subgroup $K_0\subset H_\text{kin}$.\footnote{The notation I have adopted here clashes slightly with that of the previous section. There we had $K_0= H_\text{kin}\cap H_\text{trans}$ being the kinematically trivial part of $H_\text{trans}$. Kinematic reduction is distinct from the ISE Methodology and operates without reference to any such $H_\text{trans}$. That is, $K_0$ should be thought of here as just a generic subgroup of $H_\text{kin}$.} The points, $[p]_{K_0}\in\mathcal{M}_\text{red}$, on the kinematically reduced spacetime are equivalence classes of points $p\in\mathcal{M}_\text{old}$ under the following equivalence relation,
\begin{align}
p_1 \equiv_{K_0} p_2 &\text{ iff } \exists k\in K_0 \ \ k(p_1)=p_2.
\end{align}
That is, $[p]_{K_0}$ is the orbit of $K_0$ at $p$. A kinematic reduction will be called ``full'' when we have $\mathcal{M}_\text{old}/K_0=\mathcal{M}_\text{old}/H_\text{kin}$. That is, we have a full kinematic reduction when $K_0$ is large enough that its orbits match the orbits of $H_\text{kin}$. Namely, when
\begin{align}
\forall p\in\mathcal{M}_\text{old} \ \  K_0(p)=H_\text{kin}(p). 
\end{align}
It follows from the definition of $H_\text{kin}$ that every $\varphi_\text{old}\in V_\text{old}^\text{kin}$ is constant on every $K_0$-orbit:
\begin{align}
p_1\equiv_{K_0}p_2 \Longrightarrow \big(\forall\varphi_\text{old}\in V_\text{old}^\text{kin} \ \  \varphi_\text{old}(p_1)=\varphi_\text{old}(p_2)\big).
\end{align}
Thus, for each kinematically allowed state in the old theory, $\varphi_\text{old}\in V_\text{old}^\text{kin}$, we can define a kinematically allowed state, $\varphi_\text{red}\in V_\text{red}^\text{kin}$, in the reduced theory as follows:
\begin{align}
\varphi_\text{red}([p]_{K_0})\coloneqq\varphi_\text{old}(p).    
\end{align}
This $\varphi_\text{old}\mapsto\varphi_\text{red}$ map is well defined, as it is independent of which representative $p'\in[p]_{K_0}$ we choose. Note also that this map is invertible: Knowing the value that $\varphi_\text{old}$ takes on each of its $K_0$-orbits is enough to fix it completely. Finally, note that this one-to-one correspondence $\varphi_\text{old}\leftrightarrow\varphi_\text{red}$ is linear such that we have $V_\text{old}^\text{kin}\cong V_\text{red}^\text{kin}$ as vector spaces.

We have so far identified the spacetime points and the kinematically allowed field states in the new theory. Let us next identify its smooth structure. As I discussed in Sec.~\ref{SecRevHomo}, quotient manifolds naturally inherit a smooth structure coming from their numerators. Here this means that a transformation, \mbox{$d_\text{red}:\mathcal{M}_\text{red}\to \mathcal{M}_\text{red}$}, is smooth if and only if there exists a smooth transformation, \mbox{$d_\text{old}:\mathcal{M}_\text{old}\to \mathcal{M}_\text{old}$}, which maps the $K_0$-orbits onto each other in the same way. That is, some permutation of the equivalence classes, $[p]_{K_0}\mapsto d_\text{red}([p]_{K_0})$, will be smooth if and only if there is a way to implement it smoothly prior to taking the quotient $[p]_{K_0}\mapsto [d_\text{old}(p)]_{K_0}$. 

Note that only a subset of the diffeomorphisms, $d_\text{old}\in\text{Diff}(\mathcal{M}_\text{old})$, will have a well-defined image under this quotient. Indeed, in order to survive kinematic reduction, a generic diffeomorphism, $d_\text{old}$, must respect the $K_0$-orbits collectively as follows:
\begin{align}\label{HkinConst}
\forall p,q\in\mathcal{M}_\text{old} \ \ p\equiv_{K_0} q \Leftrightarrow    d_\text{old}(p)\equiv_{K_0} d_\text{old}(q).
\end{align}
In light of this, the following definition will be useful. For any groups, $G\subset\text{Diff}(\mathcal{M}_\text{old})$ we can identify 
its subgroup which respects every $K_0$-orbit individually as,
\begin{align}
R_{K_0}(G)=\{g\in G \ \vert \ \forall p\in\mathcal{M}_\text{old} \ \ (K_0\circ g)(p)= K_0(p)\}.
\end{align}
This is the subset of $G$ which has the following property: For all $p\in\mathcal{M}_\text{old}$, the $K_0$-orbit of $g(p)$ is the same as the $K_0$-orbit of $p$. We can also identify its subgroup which respects the $K_0$-orbits collectively as,
\begin{align}
Q_{K_0}(G)=\{g\in G \ \vert \ \forall p\in\mathcal{M}_\text{old} \ \ (K_0\circ g)(p)= (g\circ K_0)(p)\}.
\end{align}
This is the subset of $G$ which has the following property: For all $p\in\mathcal{M}_\text{old}$, the $K_0$-orbit of $g(p)$ is the same as the image under $g$ of the $K_0$-orbit of $p$. Note that we have $R_{K_0}(G)\lhd Q_{K_0}(G)\subset G$. Any group, $G$, survives kinematic reduction if and only if it has $Q_{K_0}(G)=G$. Such a group survives kinematic reduction, up to diffeomorphism, as $G/R_{K_0}(G)$.

As I will now discuss, all kinematically allowed diffeomorphisms survive kinematic reduction. That is, we have $Q_{K_0}(\text{Diff}_\text{kin}(\mathcal{M}_\text{old}))=\text{Diff}_\text{kin}(\mathcal{M}_\text{old})$. To see this, recall from Sec.~\ref{AppRFSKC} that $H_\text{kin}\lhd \text{Diff}_\text{kin}(\mathcal{M}_\text{old})$ is a normal subgroup of $\text{Diff}_\text{kin}(\mathcal{M}_\text{old})$. It follows from this, that all kinematically allowed diffeomorphisms satisfy Eq.~\eqref{HkinConst}.  Intuitively, any diffeomorphism which disrespects the $K_0$-orbits (along which all kinematically allowed states are constant) is kinematically disallowed. Thus, every $d_\text{old}\in\text{Diff}_\text{kin}(\mathcal{M}_\text{old})$ has a well defined image under the quotient, namely, $d_\text{red}$ with
\begin{align}
d_\text{red}([p]_{K_0})\coloneqq [d_\text{old}(p)]_{K_0}.
\end{align}
The $d_\text{red}$ defined in this way are exactly the reduced theory's kinematically allowed diffeomorphisms, $d_\text{red}\in\text{Diff}_\text{kin}(\mathcal{M}_\text{red})$.

We are now in a position to identify the kinematically trivial diffeomorphisms in the new theory. One can easily check that we will have $H_\text{kin}^\text{(red)}\neq \{\openone_{\mathcal{M}_\text{red}}\}$ if and only if
\begin{align}
\exists d_\text{old} \ \ \ \ d_\text{old}\in H_\text{kin} \,\land\, d_\text{old}\in Q_{K_0}(H_\text{kin}) \,\land\, d_\text{old}\not\in R_{K_0}(H_\text{kin}).
\end{align}
Such a $d_\text{old}\in\text{Diff}(\mathcal{M}_\text{old})$ acts trivially on the the old theory (because $d_\text{old}\in H_\text{kin}$). Moreover, this $d_\text{old}$ survives the quotient with $K_0$ becoming a diffoeomorphism, $d_\text{red}$, on $\mathcal{M}_\text{red}$ (because $d_\text{old}\in Q_{K_0}(H_\text{kin})$). Finally, this surviving diffeomorphism is non-trivial, $d_\text{red}\neq\openone_{\mathcal{M}_\text{red}}$ (because $d_\text{old}\not\in R_{K_0}(H_\text{kin})$). Taking the logical inverse of this result we immediately have:
\begin{align}
H_\text{kin}^\text{(red)}= \{\openone_{\mathcal{M}_\text{red}}\} &\text{ if and only if } Q_{K_0}(H_\text{kin})=R_{K_0}(H_\text{kin})\\
&\text{ if and only if } \forall p\in\mathcal{M}_\text{old} \ \  K_0(p)=H_\text{kin}(p).
\end{align}
That is, (partial) kinematic reduction yields a kinematically reduced  theory exactly in the case of full kinematic reduction.

Let us briefly consider an example application. It is trivial to check that the \hyperlink{PMNPS}{\bf Periodic Matter on a Non-Periodic Spacetime} theory and the \hyperlink{MPS}{\bf Matter on a Periodic Spacetime} theory described in Ch.~\ref{ChapGenerality} are related by a full kinematic reduction. To so do, let $K_0\cong(\mathbb{R},+)\times (\mathbb{Z},+)$ be the group of continuous shifts in the $t$-coordinate together with shifts by $\pm n\,L$ in the $x$-coordinate. Note that the $K_0$ and $H_\text{kin}$-orbits here match. Hence, this is a case of full kinematic reduction, and we are guaranteed to have produced a kinematically reduced theory. Notice also, that in this example, kinematic reduction has preserved the old theory's kinematic homogeneity. As I will now prove, this always happens.

Suppose that the old theory is kinematically homogeneous. From this, it follows that there exists a finite-dimensional, kinematically allowed, transitively-acting Lie group, $H_\text{trans}\subset\text{Diff}_\text{kin}(\mathcal{M}_\text{old})$. Because this group is kinematically allowed, it has a well-defined image under the quotient.  Up to diffeomorphism this image is $H_\text{trans}/R_{K_0}(H_\text{trans})$. Moreover, this image is kinematically allowed in the new theory.\footnote{Note also that taking this quotient can only reduce the size of $H_\text{trans}$. Hence, if $H_\text{trans}$ is finite-dimensional then its image under the quotient will be too.}  Finally, note that the image of $H_\text{trans}$ under the quotient  will act transitively on $\mathcal{M}_\text{new}$. Suppose that we want to map $[p]_{K_0}$ to $[q]_{K_0}$. The transitivity of $H_\text{trans}$ on $\mathcal{M}_\text{old}$ guarantees that some $h\in H_\text{trans}$ has $h(p)=q$. The image of $h$ under the quotient maps $[p]_{K_0}$ to $[q]_{K_0}$. Thus, the (partial) kinematic reduction of a kinematically homogeneous theory is always kinematically homogeneous.

Interestingly, however, kinematic reduction can also produce a kinematically homogeneous theory beginning from a kinematically non-homogeneous theory. Consider a spacetime theory set on $\mathcal{M}_\text{old}\cong \mathbb{T}^2\,\dot\cup\,\mathbb{R}^2$, i.e., the disjoint union of a torus and a two-dimensional plane. The field states are subject to the following  kinematic constraint: In some fixed-coordinate system, $(t,x)$, for the plane, the field must be periodic in both $t$ and $x$ with fixed periods $T$ and $L$ respectively. The dynamics of this theory is unspecified. This theory is not kinematically homogeneous because its spacetime manifold is not homogeneous; No smooth map can relate a torus-point to a plane-point. Note, however, that the kinematic reduction of this theory is set on $\mathcal{M}_\text{red}\cong \mathbb{T}^2\,\dot\cup\,\mathbb{T}^2$, i.e., the disjoint union of two toruses. This new spacetime manifold is homogeneous. Moreover, the reduced spacetime theory is kinematically homogeneous. 

\section{Achieving Kinematic Reduction via the ISE Methodology}\label{AppKinRedISE}
The process of kinematic reduction described in the previous section is distinct but closely related to the \hyperlink{ISE}{\bf ISE Methodology} developed in this dissertation. As the previous section has discussed, full kinematic reduction could be used prior to applying the ISE Methodology in order to bolster a theory's spacetime-kinematic compatibility. One could also apply kinematic reduction after the ISE Methodology. However, as I proved in Ch.~\ref{ChapExternal}, every spacetime theory produced via externalization is already kinematically reduced. Hence, applying the above-described kinematic reduction process would be pointless. 

Past these sequential comparisons, this section will prove that a full kinematic reduction can be achieved via the ISE Methodology if the old spacetime theory has {\bf Weak Spacetime-Kinematic Compatibility} (defined below).
\begin{quote}
{\hypertarget{WSKC}{\bf Definition: Weak Spacetime-Kinematic Compatibility - }} Consider a spacetime theory set on a smooth manifold, $\mathcal{M}_\text{old}$, with kinematically allowed states, $V_\text{old}^\text{kin}$. This theory has weak spacetime-kinematic compatibility just in case:
\begin{enumerate}
    \item it satisfies the \hyperlink{Prelim}{\bf Preliminary Assumptions} discussed in Ch.~\ref{ChapGenerality},
    \item it is \textit{strongly kinematically homogeneous}: There exists a finite-dimensional, kinematically-allowed, transitively-acting Lie group, $H_\text{trans}\subset\text{Diff}_\text{kin}(\mathcal{M}_\text{old})$, which moreover has, 
    \begin{align}
    \forall p\in\mathcal{M}_\text{old} \ \  K_0(p)=H_\text{kin}(p),
    \end{align}
    where $K_0\coloneqq H_\text{kin}\cap H_\text{trans}$ is the kinematically trivial subgroup of $H_\text{trans}$.
\end{enumerate}
\end{quote}
As their names suggest, strong kinematic homogeneity implies kinematic homogeneity, but not vice-versa. Notice, however, that if we assume that the old theory is kinematically reduced, $H_\text{kin}=\{\openone_{\mathcal{M}_\text{old}}\}$, then strong kinematic homogeneity become equivalent to regular kinematic homogeneity. Indeed, under this assumption weak spacetime-kinematic compatibility becomes equivalent to regular spacetime-kinematic compatibility. As their names suggest, spacetime-kinematic compatibility implies weak spacetime-kinematic compatibility but not vice-versa.

The proof of this claim is structured as follows. First, notice that applying a full kinematic reduction to any theory with weak spacetime-kinematic compatibility will produce a theory with regular spacetime-kinematic compatibility. Consequently, if we then internalize this theory, we will find some Lie groups, $G_\text{trans}^\text{(red)}$ and $G_\text{fix}^\text{(red)}$, represented on a vector space, $V_\text{neutral}^\text{(red)}$. By the results of Sec.~\ref{AppSpecialG} these are pre-spacetime translation operations no matter which $H_\text{trans}^\text{(red)}$ we pick to mediate the internalization. Note that internalizing this kinematically reduced theory yields a surviving spacetime manifold, $\mathcal{M}_\text{G}^\text{(red)}\cong\mathcal{M}_\text{old}/H_\text{kin}$, independent of our choice of $H_\text{trans}^\text{(red)}$.

By contrast, we can also internalize this theory without first kinematically reducing it. As discussed in Sec.~\ref{AppRFSKC}, because this theory is kinematically homogeneous, its spacetime manifold will survive internalization (potentially unfaithfully). In the process, internalization will yield some Lie groups $G_\text{trans}^\text{(int)}$ and $G_\text{fix}^\text{(int)}$ represented on a vector space $V_\text{neutral}$. We have no reason to believe (as of yet) that these are pre-spacetime translation operations. Internalizing this theory yields a surviving spacetime manifold, $\mathcal{M}_\text{G}\cong\mathcal{M}_\text{old}/K_0$. Notice that this surviving manifold may depend on which $H_\text{trans}$ we use to mediate the internalization.

Note, however, that because we have assumed that the old spacetime theory has strong kinematic homogeneity there will exist a choice of $H_\text{trans}$ for which $\mathcal{M}_\text{G}\cong\mathcal{M}_\text{old}/K_0=\mathcal{M}_\text{old}/H_\text{kin}$. Let us take this choice of $H_\text{trans}$. Note that we then have $\mathcal{M}_\text{G}\cong \mathcal{M}_\text{G}^\text{(red)}$. Recall from the previous section that kinematic reduction gives us a linear isomorphism, $V_\text{old}^\text{kin}\cong V_\text{red}^\text{kin}$. This means that we have $V_\text{neutral}\cong V_\text{neutral}^\text{(red)}$ as well. Combining these isomorphisms, one can show that $G_\text{trans}^\text{(int)}\cong G_\text{trans}^\text{(red)}$ and $G_\text{fix}^\text{(int)}\cong G_\text{fix}^\text{(red)}$. This is sufficient to prove that $G_\text{trans}^\text{(int)}$ and $G_\text{fix}^\text{(int)}$ are, in fact, pre-spacetime translation operations. As such, they can be externalized. Doing so yields the kinematic reduction of our original spacetime theory. 

Filling in the details of this proof is tedious but straightforward. I leave this as an exercise for the reader. One important consequence of the above argument is as follows. Consider any spacetime theory with weak spacetime-kinematic compatibility. Let $H_\text{trans}$ be one of the Lie groups whose existence is guaranteed by strong kinematic homogeneity. Internalizing this theory yields two Lie groups, $G_\text{trans}^\text{(int)}$ and $G_\text{fix}^\text{(int)}$, represented on $V_\text{neutral}$. By the above discussion, these are guaranteed to be pre-spacetime translation operations. Note that this generalizes the result of Sec.~\ref{AppSpecialG} which assumes regular spacetime-kinematic compatibility.

\section{Structural Requirements on Externalization from Desiderata \#2}\label{AppD2Structure}
Ch.~\ref{ChapExternal} discussed the externalization process used in this \hyperlink{ISE}{\bf ISE Methodology}. As I discussed there, externalization takes as input two Lie groups, $G_\text{trans}$ and $G_\text{fix}$, and builds from them a new spacetime manifold, $\mathcal{M}_\text{new}$. Externalization also takes as input a map, \mbox{$E:V_\text{neutral}\to V_\text{new}^\text{all}$}, which transplants our spacetime-neutral field states onto the new spacetime as $E:\bm{\varphi}\mapsto \varphi_\text{new}$. In Ch.~\ref{ChapExternal}, I took our choices of $G_\text{trans}$, $G_\text{fix}$, and $E:V_\text{neutral}\to V_\text{new}^\text{all}$ to be constrained in some way. Concretely, $G_\text{trans}$ and $G_\text{fix}$ were constrained to be \hyperlink{PSTO}{\bf Pre-Spacetime Translation Operations}. Moreover, $E:V_\text{neutral}\to V_\text{new}^\text{all}$ was constrained to be (up to diffeomorphism) one of the $J$ maps whose existence is guaranteed by \hyperlink{InjComat}{\bf Injection Compatibility}. From these constraints, it followed that the ISE Methodology satisfies \hyperlink{D2}{\bf Desiderata \#2 (D2)}. This appendix will prove the converse result: If the \hyperlink{ISE}{\bf ISE Methodology} is to satisfy \hyperlink{D2}{\bf D2}, then $G_\text{trans}$, $G_\text{fix}$, and $E$ must be constrained as they are in Ch.~\ref{ChapExternal}.

Recall from Ch.~\ref{ChapOverview} that \hyperlink{D2}{\bf D2} requires that: ``For any spacetime theory which can be produced via externalization, after re-internalizing this theory we should always be able to be re-externalize it trivially.'' Before discussing what constraints this places on our choices of $G_\text{trans}^\text{(ext)}$, $G_\text{fix}^\text{(ext)}$, and $E$, let us clarify what exactly \hyperlink{D2}{\bf D2} requires mathematically.

What does re-internalization mean here? Re-internalization here is just the internalization process described in Ch.~\ref{ChapInternal} (and Appendix~\ref{AppRFSKC}) with its allowed inputs restricted to spacetime theories which are producible via externalization. Importantly, at this point, we have not yet characterized the spacetime theories producible via externalization as having \hyperlink{SKC}{\bf Spacetime-Kinematic Compatibility}. This result was proved in Sec.~\ref{SecD2} using certain constraints on $G_\text{trans}^\text{(ext)}$, $G_\text{fix}^\text{(ext)}$, and $E$. We are here seeking to motivate these constraints. The scope of theories producible via externalization is at this point undetermined.

What does ``re-externalized trivially'' mean? This means that we can skip over the searching step of the ISE Methodology. That is, the outputs of re-internalization should be ready-to-go as the inputs to the \mbox{(re-)externalization} process. Thus, \hyperlink{D2}{\bf D2} requires that the scope of spacetime theories producible via externalization are such that upon re-internalization they produce Lie groups, $G_\text{trans}^\text{(int)}$ and $G_\text{fix}^\text{(int)}$, which are valid inputs to the externalization process.\footnote{Importantly, at this point, we have not yet characterized the valid inputs to externalization as being pre-spacetime translation operations. In fact, we are here seeking to motivate this constraint.} As Appendix~\ref{AppRFSKC} has shown, a necessary condition for the Lie groups produced via internalization, $G_\text{trans}^\text{(int)}$ and $G_\text{fix}^\text{(int)}$, to even exist is that the input theory be kinematically homogeneous.

Finally, what does ``always'' mean here? The internalization process is unambigious given a choice of forgetful functor, $\mathcal{F}_\text{vec}$. However, as Appendix~\ref{AppRFSKC} has shown, there is a potential ambiguity in the way that the theory's spacetime manifold survives internalization, see Eq.~\eqref{ManSur0}. This is because $\mathcal{M}_\text{old}$ survives internalization by first being reconstructed up to diffeomorphism as $\mathcal{M}_\text{old}\cong H_\text{trans}/H_\text{fix}$. The Lie groups $G_\text{trans}^\text{(int)}$ and $G_\text{fix}^\text{(int)}$ produced via internalization depend on how $\mathcal{M}_\text{old}$ is reconstructed. The ``always'' appearing in \hyperlink{D2}{\bf D2} demands that $G_\text{trans}^\text{(int)}$ and $G_\text{fix}^\text{(int)}$ be valid inputs to externalization independent of which $H_\text{trans}$ we use to reconstruct $\mathcal{M}_\text{old}$.

Thus, in total, the requirements on externalization coming from \hyperlink{D2}{\bf D2} are as follows: Firstly, the valid inputs to externalization ought to be constrained such that all spacetime theories producible via externalization are kinematically homogeneous. Secondly, re-internalizing this theory ought to produce Lie groups which are themselves valid inputs for externalization (independently of which $H_\text{trans}$ mediates this re-internalization). Re-externalizing any of these Lie groups ought to lead us back to the original externalized theory.

As I will now discuss, meeting this requirement requires that we constrain $G_\text{trans}^\text{(ext)}$ and $G_\text{fix}^\text{(ext)}$ to be pre-spacetime translation operations. Moreover, $E:V_\text{neutral}\to V_\text{new}^\text{all}$ must be constrained to be (up to diffeomorphism) one of the $J$ maps whose existence is guaranteed by \hyperlink{InjComat}{\bf Injection Compatibility}. As I will now discuss, there are two types of demands coming from \hyperlink{D2}{\bf D2}: Namely, it demands that externalization and re-internalization be compatible both in terms of vectors and in terms of diffeomorphisms.


Let us begin by spelling out the vector-based compatibility required by \hyperlink{D2}{\bf D2}. Just as internalization was mediated by a forgetful functor, \mbox{$\mathcal{F}_\text{vec}^\text{(int)}:V_\text{old}^\text{kin}\to V_\text{neutral}$}, re-internalization is also mediated by a forgetful functor, \mbox{$\mathcal{F}_\text{vec}^\text{(re-int)}:V_\text{new}^\text{kin}\to V_\text{neutral}$}.\footnote{Recall from the end of Ch.~\ref{ChapInternal} that if the theory's dynamics are non-linear then this forgetful functor must be replaced with one which also remembers algebraic structures. The same allowance can be made here in terms of re-internalization.} Thus, to re-externalize after this re-internalization is to in total apply the map $E\circ \mathcal{F}_\text{vec}^\text{(re-int)}$ on $V_\text{new}^\text{kin}$. Hence, \hyperlink{D2}{\bf D2} demands that this we have $E\circ \mathcal{F}_\text{vec}^\text{(re-int)}=\hat\openone$ being the identity on $V_\text{new}^\text{kin}$. If  $E$ and $\mathcal{F}_\text{vec}^\text{(re-int)}$ are to be inverses in this way, then $E$ must be linear and injective.

Let us next discuss the diffeomorphism-based compatibility required by \hyperlink{D2}{\bf D2}. First, recall that the new theory's vector space of kinematic possibilities is defined as $V_\text{new}^\text{kin}\coloneqq E(V_\text{neutral})$, the new theory's vector space of kinematic possibilities.\footnote{As the previous paragraph has established, the $E:V_\text{neutral}\to V_\text{new}^\text{all}$ map must be linear and injective. Therefore we must have \mbox{$ V_\text{new}^\text{kin}\coloneqq E(V_\text{neutral})\cong V_\text{neutral}$}. That is, $V_\text{new}^\text{kin}$ is a vector space and is moreover isomorphic to $V_\text{neutral}$.} From this vector space, just as in the old theory, we can identify the new theory's kinematically allowed diffeomorphisms, $\text{Diff}_\text{kin}(\mathcal{M}_\text{new})$. Each of these kinematically allowed diffeomorphisms, $d\in\text{Diff}_\text{kin}(\mathcal{M}_\text{new})$, will survive re-internalization as,
\begin{align}
d&\surreint G^\text{(re-int)}(d)\coloneqq\mathcal{F}_\text{vec}^\text{(re-int)}\circ d^*\vert_\text{kin}\circ\mathcal{F}_\text{vec}^\text{(re-int)}{}^{-1}.    
\end{align}
As the Appendix~\ref{AppRFSKC} has discussed, the group of kinematically allowed diffeomorphisms will always survive internalization (albeit, potentially unfaithfully) as,
\begin{align}
\text{Diff}_\text{kin}(\mathcal{M}_\text{new})\surint G_\text{new}\cong \text{Diff}_\text{kin}(\mathcal{M}_\text{new})/H_\text{kin},
\end{align}
where $G_\text{new}\coloneqq G^\text{(re-int)}(\text{Diff}_\text{kin}(\mathcal{M}_\text{new}))$. Note that here $H_\text{kin}$ are the new theory's kinematically trivial diffeomorphisms.

\hyperlink{D2}{\bf D2} requires that externalization ought to be able to undo this re-internalization. As the above-discussion has shown, re-internalization turns certain diffeomorphisms on $\mathcal{M}_\text{new}$ (i.e., $\text{Diff}_\text{kin}(\mathcal{M}_\text{new})$) into linear maps on $V_\text{neutral}$. By contrast, externalization turns certain linear maps on $V_\text{neutral}$ (namely, $g\in G_\text{trans}$) into diffeomorphisms on $\mathcal{M}_\text{new}$ as,
\begin{align}
g&\surext H(g)\coloneqq\mathcal{F}_\text{man}\circ \bar{g}\circ \mathcal{F}_\text{man}^{-1}.
\end{align}
Thus, we can move any $g\in G_\text{trans}$ back and forth between $G_\text{trans}\subset\text{GL}(V_\text{neutral})$ and $H_\text{trans}\subset\text{Diff}(\mathcal{M}_\text{new})$ by re-internalizing and re-externalizing it (i.e., by alternatingly applying $G^\text{(re-int)}$ and $H$).

Given \hyperlink{D2}{\bf D2}'s demand that re-internalization and externalization are compatible, we must have $G^\text{(re-int)}\circ H=\openone$ being the identity on $G_\text{trans}$. This in turn implies that both $H$ and $G^\text{(re-int)}\vert_{H_\text{trans}}$ are invertible and hence both group isomorphisms. Demanding that the $H$ map be an isomorphism is equivalent to demanding that $G_\text{trans}$ and $G_\text{fix}$ have \hyperlink{FPC}{\bf Fixed-Point Compatibility}.

Demanding that $G^\text{(re-int)}\vert_{H_\text{trans}}$ is a group isomorphism is equivalent to demanding that its kernel, $H_\text{kin}\cap H_\text{trans}=\{\openone_{\mathcal{M}_\text{new}}\}$, be the trivial group. As discussed above, \hyperlink{D2}{\bf D2} demands this for all $H_\text{trans}$. Here $H_\text{trans}$ ranges over all finite-dimensional, kinematically-allowed, transitively-acting Lie groups, $H_\text{trans}\subset\text{Diff}_\text{kin}(\mathcal{M}_\text{new})$. If $H_\text{kin}$ is to have a trivial intersection with all such $H_\text{trans}$, then it must have a trivial intersection with their union. The union of all such $H_\text{trans}$ is $\text{Diff}_\text{kin}(\mathcal{M}_\text{new})$. Thus we must $H_\text{kin}\cap \text{Diff}_\text{kin}(\mathcal{M}_\text{new})=\{\openone_{\mathcal{M}_\text{new}}\}$ being the trivial group. Recall that $H_\text{kin}$ is a subset of $\text{Diff}_\text{kin}(\mathcal{M}_\text{new})$. From this it follows that we must have $H_\text{kin}=\{\openone_{\mathcal{M}_\text{new}}\}$. That is, the new theory must be kinematically reduced. Written in terms of $E$ this requirement is as follows: For all $d\in\text{Diff}(\mathcal{M}_\text{new})$,
\begin{align}\label{AppEKinRed}
\Big(\forall \bm{\varphi}\in V_\text{neutral} \ \  (d^*\circ E)(\bm{\varphi})=E(\bm{\varphi})\Big)\Longleftrightarrow d=\openone_{\mathcal{M}_\text{new}}.
\end{align}
That is, all diffeomorphisms $d\in\text{Diff}(\mathcal{M}_\text{new})$ except the identity must act non-trivially on the image of $E$, namely, \mbox{$ V_\text{new}^\text{kin}\coloneqq E(V_\text{neutral})\cong V_\text{neutral}$}.

Finally, let us return to our demand that $G^\text{(re-int)}\circ H=\openone$ be the identity on $G_\text{trans}$. Using the above-discussed requirement that $E\circ \mathcal{F}_\text{vec}^\text{(re-int)}=\hat\openone$ we can rewrite this new requirement in terms of $E$ alone as,
\begin{align}\label{AppEqEgH*E}
\forall g\in G_\text{trans} \ \forall \bm{\varphi}\in V_\text{neutral} \ \ \big[(E\circ g)\,\bm{\varphi} &= (H(g)^*\circ E)\,\bm{\varphi}\big].
\end{align}

To review: In order to satisfy \hyperlink{D2}{\bf D2}, the allowed choices for $G_\text{trans}$ and $G_\text{fix}$ must be constrained to have \hyperlink{FPC}{\bf Fixed-Point Compatibility}. Moreover, the allowed choices for $E:V_\text{neutral}\to V_\text{new}^\text{all}$ must be constrained to be linear injectors satisfying Eq.~\eqref{AppEKinRed} and Eq.~\eqref{AppEqEgH*E}. If $E$ is to satisfy these constraints then given our choices of $G_\text{trans}$ and $G_\text{fix}$ there must exist some $E$ which satisfies these constraints. That is to say, $G_\text{trans}$ and $G_\text{fix}$ must have \hyperlink{InjCompat}{\bf Injection Compatibility}. Putting these requirement together $G_\text{trans}$ and $G_\text{fix}$ must be \hyperlink{PSTO}{\bf Pre-Spacetime Translation Operations}. This guarantees the existence of the sort of maps which $E$ must be. Indeed, the constraint we must placed on $E$ is just that it is one of the linear injective maps whose existence is guaranteed by \hyperlink{InjCompat}{\bf Injection Compatibility}. In total, this proves that the constraints placed on $G_\text{trans}$, $G_\text{fix}$ and $E:V_\text{neutral}\to V_\text{new}^\text{all}$ are necessary in order for the ISE Methodology to satisfy \hyperlink{D2}{\bf Desiderata \#2}. 
\end{appendices}

\end{document}